\documentclass{ar-1col}

\usepackage{graphicx, amssymb, multirow, natbib, amsmath, upgreek, pifont,lscape}
\usepackage{color}
\usepackage{longtable, multirow}

\jname{Xxxx. Xxx. Xxx. Xxx.}
\jvol{AA}
\jyear{YYYY}
\doi{10.1146/((please add article doi))}

\begin{document}

\newcommand{\micron}{$\mu$m}

\newcommand{\RNum}[1]{\uppercase\expandafter{\romannumeral #1\relax}}

\newcommand\ion[2]{#1$\;${%
\ifx\@currsize\normalsize\small \else
\ifx\@currsize\small\footnotesize \else
\ifx\@currsize\footnotesize\scriptsize \else
\ifx\@currsize\scriptsize\tiny \else
\ifx\@currsize\large\normalsize \else
\ifx\@currsize\Large\large
\fi\fi\fi\fi\fi\fi
\RNum{#2}}\relax}

\bibliographystyle{ar-style2.bst}

\markboth{Hughes et al.}{Debris Disks}

\title{Debris Disks: Structure, Composition, and Variability}

\author{A. Meredith Hughes,$^1$ Gaspard Duch\^ene,$^{2,3}$ Brenda C.~Matthews$^{4,5}$
\affil{$^1$ Department of Astronomy, Van Vleck Observatory, Wesleyan University, 96 Foss Hill Drive, Middletown, CT 06459, USA; amhughes@wesleyan.edu}
\affil{$^2$ Astronomy Department, University of California Berkeley, Berkeley CA 94720-3411, USA}
\affil{$^3$ Univ. Grenoble Alpes, CNRS, Institut d'Astrophysique et de Plan\'etologie de Grenoble, 38000 Grenoble, France}
\affil{$^4$ Herzberg Astronomy \& Astrophysics Programs, National Research Council of Canada, 5071 West Saanich Road, Victoria, BC, V9E 2E7, Canada}
\affil{$^5$ Department of Physics \& Astronomy, University of Victoria, 3800 Finnerty Road, Victoria, BC, V8P 5C2, Canada}}

\begin{abstract}
Debris disks are tenuous, dust-dominated disks commonly observed around stars over a wide range of ages. Those around main sequence stars are analogous to the Solar System's Kuiper Belt and Zodiacal light. The dust in debris disks is believed to be continuously regenerated, originating primarily with collisions of planetesimals. Observations of debris disks provide insight into the evolution of planetary systems; the composition of dust, comets, and planetesimals outside the Solar System; as well as placing constraints on the orbital architecture and potentially the masses of exoplanets that are not otherwise detectable. This review highlights recent advances in multiwavelength, high-resolution scattered light and thermal imaging that have revealed a complex and intricate diversity of structures in debris disks, and discusses how modeling methods are evolving with the breadth and depth of the available observations.  Two rapidly advancing subfields highlighted in this review include observations of atomic and molecular gas around main sequence stars, and variations in emission from debris disks on very short (days to years) timescales, providing evidence of non-steady state collisional evolution particularly in young debris disks.  
\end{abstract}


\maketitle

\tableofcontents

\section{Introduction}

\subsection{What is a Debris Disk?}

Debris disks are both an outcome and an integral component of the formation of planetary systems. From the earliest stages of star formation to the final planetary systems, circumstellar disks of material are present, but these disks change their structure, dynamics and composition significantly over time.  In the earliest phases of star formation, a pre-main sequence star forms surrounded by a gas-rich protoplanetary (or planet-forming) disk of gas and dust.  Transition disks refer to a phase between a protoplanetary disk and a debris disk, but in this epoch, disks are still characterized by high gas-to-dust ratios and (often) continued gaseous accretion on the central star \citep{esp14b}.  Eventually the transition disk dissipates through some combination of stellar and planetary mechanisms involving accretion, photoevaporation, winds and agglomeration of large solid bodies \citep[see, e.g.,][and references therein]{wil11,wya15}.  The timescale for the decrease in emission from such disks is wavelength dependent, with most disk emission gone by 3 Myr in the near-IR and 20 Myr in the submillimeter \citep{wya15}. 

Debris disks, known also in the literature as secondary disks, are not leftover remnants of the protoplanetary or transition disks, but instead they must be continuously sustained by collisional processes (the steady state evolution of disks is typically modeled as a collisional cascade; see Sidebar) to be observable over the lifetime of the star.  Radiation forces, stellar winds, and Poynting Robertson drag forces all act to continuously clear dust grains smaller than the blowout size from orbits around the star (see Sidebar).  Thus, the detection of a debris disk indicates that planet formation processes in the system were successful in forming bodies of at least several 100s to 1000s of km in size.  Collisional cascades grind the material down to small dust grains observable in both scattered light and thermal emission.  

Protoplanetary disks and debris disks represent distinct classes of objects.  Physically, protoplanetary disks can be regarded as agglomeration dominated, while debris disks are dominated by destructive processes. Both types of disks can be detected through excess emission relative to the stellar photosphere at wavelengths that can range from the near-infrared to the (sub)millimeter. Unlike protoplanetary disks, however, for which evolution of the disks is mirrored by predictable changes in the degree and wavelengths of excess, debris disks may lack excess emission at some wavelengths simply because the disks are not radially continuous or due to the distribution of mass across the disks. Debris disk emission may also vary significantly due to perturbation of the disk material that renders them brighter or fainter, sometimes over very short timescales. Nonetheless, the presence of excess thermal emission has generally the most effective diagnostic of the presence of circumstellar dust. However, distinguishing observationally between the two classes is non-trivial. Various criteria have been proposed in the literature, but nearly all rules either have clear exceptions or rely heavily on assumptions about poorly constrained properties of the system. A system's age is not a good criterion since stellar clusters with ages of $\sim 5-15$ Myr contain coexisting examples of both protoplanetary and debris disks, and even clusters as young as 2-3 Myr show signs of incipient debris disks \citep{esp17}.
The absence of molecular gas was historically a distinguishing characteristic of debris disks, but evidence of gas is now detected toward many debris disks (see Section\,\ref{sec:gas_in_debris}). The most practical criterion seems to be optical depth: the dust in debris disks is optically thin across the electromagenetic spectrum, while the dust in protoplanetary disks is typically extremely optically thick at optical wavelengths and may remain optically thick in the inner regions even into the millimeter part of the spectrum.  The observational proxy now used most frequently to quantify the low optical depth of debris disks is the integrated fractional luminosity, $f = L_{disk} /L_*$, also often presented as $\tau = L_{IR}/L_{bol}$.  In practice, $f$ is preferred as the more physical value, but $\tau$ is more easily calculated for sparsely sampled spectral energy distributions (SEDs).  In this review, we adopt an upper bound of $\tau < 8\times10^{-3}$  to define a debris disk, selected somewhat arbitrarily to place HD~141569 right on this boundary, since it seems to be the best candidate for a truly intermediate object between the protoplanetary and debris disk phase \citep{wya15}.  

\begin{figure*}
\centering
\includegraphics[trim = 2cm 5.5cm 1.5cm 3cm,clip = true,scale=0.5]{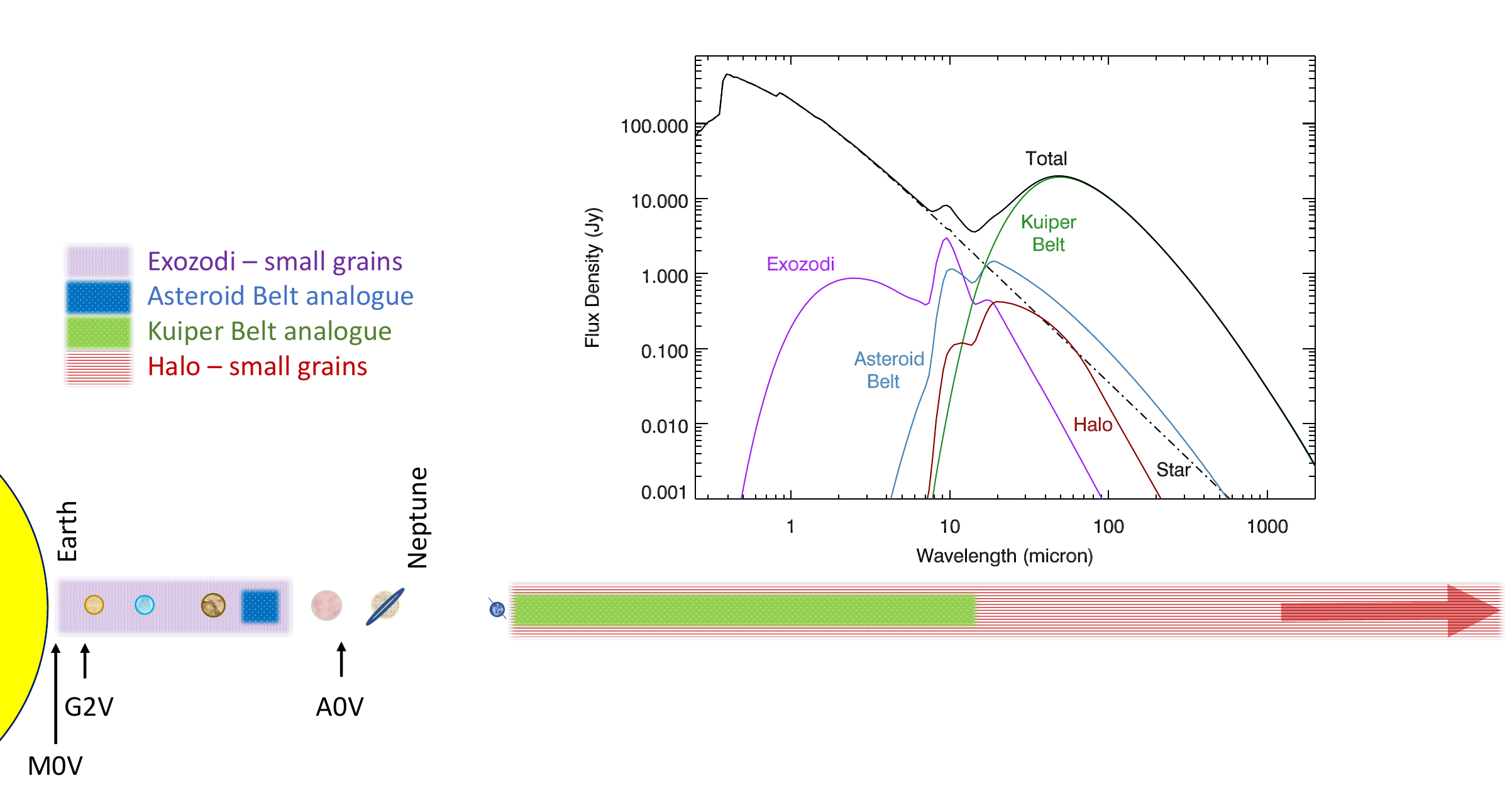}
\caption{Schematic of the potential components of a debris disk and an illustrative SED.  Depending on the system architecture, multiple planetesimal belts may be present; broad analogues to the asteroid and Kuiper belts of the Solar System are shown here. Planets are shown as a means of confining dust into separate belts, not to suggest any typical architecture. Through a steady state collisional cascade, dust is generated in the planetesimal belts down to the blowout size. Stellar radiation and/or wind can displace, or completely remove, the smallest dust grains, populating analogues of the Solar System zodiacal light and of an extended outer halo that contains small dust grains on hyperbolic orbits and is best seen in scattered light images. The black arrows indicate the equilibrium 150 K blackbody distance from the star for three spectral types: M0V, G2V and A0V.  The positions of Earth and Neptune are shown for a sense of scale for this particular disk; the radial locations and radial extent of disk components can vary dramatically from one system to another. The illustrative SED is for a model that contains each of these components, liberally adapted from the known structure of the Vega and $\beta$\,Pic debris disks. The system is built around an A0V star at 7.7 pc and with a total integrated luminosity of $\tau = 1.4 \times 10^{-3}$. Dust temperatures in the model range from 30\,K in the outer halo to as much as 2000\,K at the inner edge of the exozodiacal disk.}
\label{fig:schematic}
\end{figure*}

Our solar system contains two components in its debris disk: the zodiacal light, which is composed primarily of material from disintegrations of Jupiter-family comets in the inner Solar System \citep{nes11}, and the Edgeworth-Kuiper Belt (also known as the EKB or Kuiper Belt) consisting of material from collisions of comets between Neptune's orbit and an outer bound of approximately 50\,au.  Each of these components has $\tau \sim 10^{-7}$ \citep{vit12,nes10,rob12}.  Disks detected around other stars are thus typically referred to as Kuiper Belt analogues (typically $T<100$ K) or exozodiacal dust belts, ``exozodis" for short (typically $T>150$ K).  Extrasolar debris disks often show evidence of multiple components \citep[e.g.,][]{ken14b}, reinforcing the idea that planets are common in debris systems, since planets are a primary, but not the only, means of excluding dust from certain regions of a disk \citep[e.g.,][]{sha16}.  \textbf{Figure\,\ref{fig:schematic}} shows a schematic of a debris disk's components. Generally, the components associated with a terrestrial planet zone are referred to as the ``inner disk", characterized by warm or hot dust emission and faster collisional evolution timescales. By contrast, the ``outer disk" is characterized by cold dust in a Kuiper Belt analogue where collisional evolution timescales are longer. A radially extended halo of small grains may also be present at large radial separations from the host star.  The relative distances of the 150 K blackbody equilibrium temperature for a range of stellar spectral types is also shown on \textbf{Figure \ref{fig:schematic}}, illustrating how much more compact the terrestrial zone is for late versus early type stars. 

\begin{textbox}[t]\section{COLLISIONAL CASCADE}
The term ``collisional cascade" refers to the process by which 
planetesimals are gradually ground to sub-micron sized dust grains.  The collision of the most massive bodies ignites the cascade by producing many smaller objects that then collide with one another to create even more smaller bodies, and so on until
the smallest bodies are the most numerous even while the bulk of the mass is retained in the largest bodies. For a self-similar, steady state cascade, the differential size distribution within the cascade follows a power law such that $dN/da \propto a^{-q}$ where $a$ is the grain size and $q\approx3.5$ \citep{doh69}.
Numerical analyses support this general behavior, albeit with departures from a pure power law \citep{kri06, the07, pan12}. 
\end{textbox}

\begin{textbox}[t]\section{BLOWOUT SIZE}
The blowout size is the largest grain size in a disk that can be expelled directly by stellar pressure forces. For grains at the blowout size or smaller, the radiation pressure of a solar or earlier type star effectively transports the grains out of the disk on hyperbolic orbits (forming the halo, see \textbf{Figure\,\ref{fig:schematic}}). The luminosity of low-mass stars is too low to effectively expel grains of any size. However, their strong stellar wind can also act to remove the smallest dust grains through ram pressure \citep{str06}, thus defining a blowout size. Generally speaking, the blowout size is defined as the grain size for which the force ratio $\beta_\mathrm{forces} = \frac{F_\mathrm{rad}+F_\mathrm{wind}}{F_\mathrm{grav}} \geq \frac{1}{2}$, where $F_\mathrm{rad}$, $F_\mathrm{wind}$ and $F_\mathrm{grav}$ are radiative pressure, wind ram pressure and gravity forces, respectively. The blowout size sets a lower bound to the size distribution produced by the collisional cascade in the disk. The minimum grain size derived from observations is often larger than the blowout size, likely as a result of a change in the microphysics of grain collisions in the small grain limit \citep{paw15}.
\end{textbox}

Modern studies of planet formation strive to connect the initial conditions in a protoplanetary disk (gas and dust surface density, temperature, and velocity) to the final outcomes (exoplanet statistics), via models of planetary system formation and evolution. Historically, astronomers have tended to assume that protoplanetary disks provide information about the initial conditions, debris disks tell us about orbital evolution during the “clean-up” phase at the end of oligarchic growth, and exoplanets tell us about final configurations. The picture has become more complex recently, with mounting evidence that planets generally form quite early in the protoplanetary disk phase \citep[e.g.,][]{par15} and that gas and dust in these systems are already significantly evolved relative to interstellar medium (ISM) conditions \citep[][and references therein]{wil11}. However, it is still clear that some phases of evolution must take place during the debris disk phase.  Debris disks also provide a unique opportunity to study directly imaged planets and disks in the same system, which is much more difficult in protoplanetary disks due to their high optical depths and increased distances to their host stars. Due to their relative faintness, most debris disks currently studied lie within a few hundred parcsecs of the Sun, with the nearest debris disks just a few parsecs away.  Debris disks are also important because the secondary material of which they are composed (dust, and at least in some cases gas) can tell us about the composition of exoplanetary material, in a manner analogous to the way in which we learn about the interior of the Earth from asteroid samples and meteorites.
While the evolution of the dust in young planetary systems during the transition from the protoplanetary to debris stage is relatively well understood \citep[][and references therein]{wya08}, understanding the evolution of the gas is a more recent development in the context of debris disks. Arguably the most fundamental question is that of the origin of the gas in debris disk systems: is it primordial material, indicating that gas and dust evolve on different timescales, or is it second-generation material like the dust component, which might provide insight into the composition of icy bodies in distant planetary systems? A secondary issue is to determine the mass of the gaseous component in order to assess both its importance on the dynamics of solid bodies and the potential for late-stage gas accretion by planets. Deriving the total gas mass in a disk is much more challenging than the dust mass, as uncertain relative abundances and excitation conditions must be considered.

\subsection{Observations of Debris Disks}

In our Solar System, the zodiacal light appears brighter when observed from Earth, but the Kuiper belt would appear brighter to a distant alien astronomer due to the greater mass and larger total surface area of the dust. Neither of the Solar System's debris disk components would be detectable around a neighboring star using current technology; analogues to the Kuiper Belt are still 1-2 orders of magnitude too faint to be observed with current instrumentation around the majority of stars, and analogues to the Zodiacal dust belt are 1000x fainter than we can currently detect. All debris disks currently detected around other stars are therefore scaled-up versions of those found in our own planetary system, and we expect that debris disks are far more common than the detected fraction -- at least as common as the planetary systems that have been found to be ubiquitous throughout the galaxy \citep{win15}.

\begin{figure*}
\centering
\includegraphics[scale=0.6]{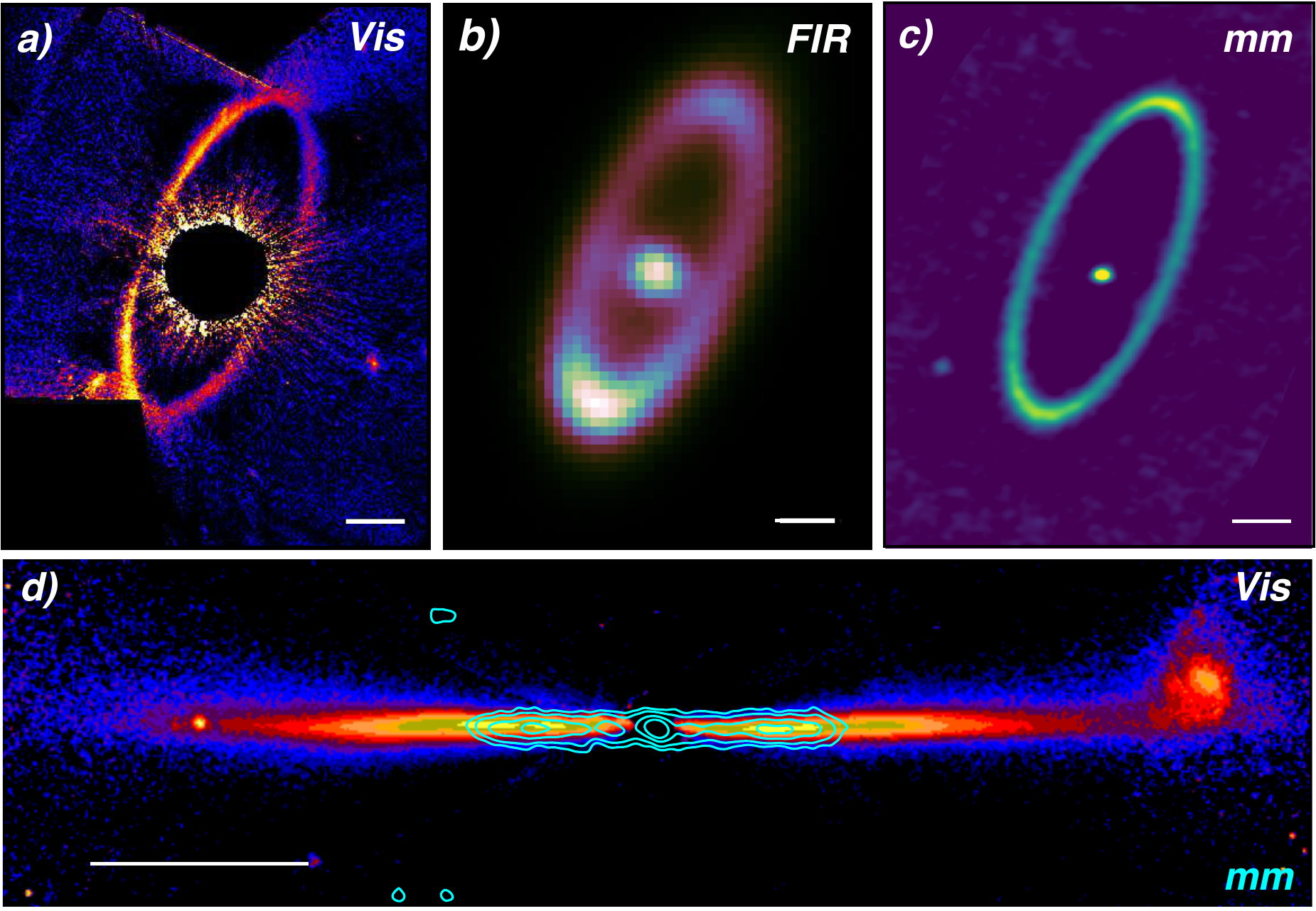}
\vspace*{-2mm}
\caption{
{\it Top:} Images of the Fomalhaut ring in {\it HST} optical scattered light \citep[panel a, from][]{kal13}, {\it Herschel} 70\,$\mu$m \citep[panel b, from][]{ack12} and ALMA 1.3\,mm thermal emission \citep[panel c, from][]{mac17}. The scale bars represent 50\,au. {\it Bottom:} Overlay of the 1.3\,mm ALMA thermal emission map (contours, from C. Daley, private communication) on the {\it HST} optical scattered light image of AU\,Mic \citep[colorscale, from][]{sch14}.
}
\label{fig:fom_aumic}
\end{figure*}

There are many different ways of observing a debris disk. The SED has been the primary diagnostic used to characterize the fractional excess luminosity of the disk over the stellar photosphere, as well as the (range of) dust temperature in the disk. While a simple blackbody excess can be fitted to sparsely sampled SEDs, often multiple components with different temperatures are required when richer datasets are available. With photometric data points in the long-wavelength (sub-millimeter) tail of the SED, a modified blackbody model that takes into account the decrease in emission/absorption efficiency of small dust grains is used instead. In this case, the long-wavelength slope of the SED is a key observable that can be directly related to the grain size distribution in the disk (see Section\,\ref{sec:dust_properties}). 
%

Images can reveal the morphology of both gas and dust emission.  Dust emission can be imaged across the electromagnetic spectrum to reveal thermal emission as well as scattered light (as illustrated in \textbf{Figure\,\ref{fig:fom_aumic}}; see also Section\,\ref{sec:outer_disk}). In thermal emission, the comparison between the observed disk radius and that expected based on the SED temperature reveals how overheated the dust really is and, thus, the size of the smallest (most populous) grains in the disk. Scattered light images indicate the disk color as well as the degree of forward scattering and linear polarization induced by scattering, again providing constraints on the properties (composition, size, porosity) of the dust grains. Overall, as imaging capabilities have improved across the electromagnetic spectrum, it has become possible to compare the spatial distribution of grains of different sizes within a single disk by making multiwavelength, spatially resolved observations of the disk (see Sidebar). 

\begin{textbox}[t]\section{OBSERVING WAVELENGTH AND GRAIN SIZE}
It is often stated that observations of dust thermal emission or scattered light are most sensitive to grains of size approximately equal to the observing wavelength. This correspondence results from a balance between two competing trends. On the one hand, the grain size distribution is heavily weighted towards small grains which thus dominate the emitting surface area. On the other hand, dust grains only emit efficiently at wavelengths shorter than their physical size, with a steep decline in emission efficiency at longer wavelengths. Therefore, at a given wavelength, only grains comparable to or larger than that wavelength can emit efficiently. As a result, the total (thermal or scattered) light output is dominated by the smallest grains capable of emitting efficiently -- namely, grains with sizes approximately equal to the wavelength of observation, if those are present in the system.  
\end{textbox}

In addition to the dust, there is occasionally a gas component that can be detected either through emission or absorption spectroscopy. While early surveys for gas around main sequence stars were impaired by poor sensitivity and low excitation temperature, which led to discouraging low detection rates \citep[e.g.,][]{zuc95}, a resurgence in interest has followed the development of large collecting area facilities like the Atacama Large Millimeter-submillimeter Array (ALMA). As a result, the study of the gas content of debris disks is now an active and rapidly developing area of study (see Section~\ref{sec:gas_in_debris}).

\subsection{Demographics: Detecting Debris Disks}
\label{sec:demographics}
Our understanding of the demographics of debris disks is driven by large surveys in the infrared where the disks are most readily detected through the presence of an infrared excess above the stellar photosphere, a signature of circumstellar dust. From their discovery with the {\it Infrared Astronomical Satellite (IRAS)} through the {\it Spitzer Space Telescope, Akari,  Wide-field Infrared Survey Explorer (WISE)} and {\it Herschel Space Observatory} missions, hundreds of stars have been targeted directly and millions of others observed in all-sky surveys to measure infrared excesses. The rates of incidence typically reported from surveys for debris signatures are in fact better termed detection rates. Comparison of detection rates in different surveys is difficult since they often probe different regions of the disk, and the likelihood of detecting a debris disk around a nearby star depends on the wavelength and sensitivity of the observation, the characterization of the photosphere, and the inherent properties of the disk and the star(s) it surrounds. \cite{wya08} and \cite{mat14c} show figures illustrating the relative detectability of disks around stars of various spectral types. Many of the results of {\it Spitzer} and {\it Herschel} surveys to date were detailed in \cite{mat14c}; here we detail primarily new analyses undertaken since that review.   

\subsubsection{Kuiper Belt Analogues}

Two large surveys of nearby stars for debris disks were undertaken by {\it Herschel}, which have yielded consistent detection rates of $\sim 17-20$\% for solar type stars from the DEBRIS \citep{sib17} and DUNES \citep{mon16b} surveys, comparable to that found at 70 \micron\ with Spitzer \citep{tri08}.  Sibthorpe et al.~further use the properties of the detected disks and the host stars of the DEBRIS survey to gauge the completeness of their detection rates, 
leading to completeness adjusted incidence rates of 36\%, 18\%, 24\% and 21\% for the F, G, K and total samples, respectively. The F star incidence rate is distinctly higher than that of G and K stars, and is in fact more comparable to that of the A star population \citep{thu14}.
This difference suggests that the traditional grouping of all FGK stars into a monolithic solar type class may not be optimal. There is further a significant difference in detection rates between early and late F stars, and a similar but less significant break within K stars.
Therefore, for samples dominated by stars in the F5 - K4 range, the variation of detection rate with spectral type is less acute than in samples that extend beyond those boundaries.  This conclusion is consistent with the finding of no significant variation with spectral type found by \cite{mon16b} with the {\it Herschel} DUNES sample and \cite{sie14} in a study of F4-K4 stars based on {\it Spitzer} and {\it Herschel} data.  To date, there remain only a few disks detected around M stars.  Whether this is due to the poorer sensitivity in terms of fractional luminosity for M stars \citep[see][and references therein]{mor14} or different progenitor properties \citep[e.g.,][]{gai17} is not yet resolved. The current detection rates are typically in the range of a few percent. The M star disk population is discussed in the review of \cite{mat14c}.

A declining detection rate with age is known and expected for debris disks \citep[e.g.,][]{mon16b}, since they collisionally evolve over time. For example, \cite{riv14} observed 19 members of the 23 Myr $\beta$\,Pic moving group with {\it Herschel} and found an excess detection rate of $\sim 50$\%. Among compiled data toward several young 20-50 Myr moving groups, \cite{moo16} find a much lower rate of detections for K stars (20\%) versus late F and G stars (56\%), though the rate for the F and G types is markedly higher than for the older populations measured for DUNES and DEBRIS. These data show that the trend in decreased detection rate as a function of spectral type is real, but it is more difficult to detect at later ages and in some spectral ranges than in others, i.e., the gradient of the change may be gradual but have some sharp transitions, particularly at later ages.

The multiplicity of a system may lower the rate of detected disks for intermediate spatial separations, but compact or very large disks are detected. Disks may be either circumprimary (or even circumsecondary) or circumbinary as dictated by the separation of
the primary and companion \citep{rod15b,mon16}.

\subsubsection{Exozodis}

Dust in the terrestrial zone, i.e., the warm dust in multiple component disks (see \textbf{Figure\,\ref{fig:schematic}}), can generally be
termed exozodiacal dust, analogous to the 270K zodiacal dust in the Solar System or to the very hot component known to exist near the Sun \citep{kim98}. Exozodis are of interest not just for insight into the underlying planetary systems, but also because their presence can significantly hinder our ability to detect planets in the habitable zone. Even Solar System levels of dust, 1000x less than the levels we can currently detect, can hinder detection of planets \citep{rob12}. We refer the reader to \cite{kra17} for a recent comprehensive review of exozodis and only briefly discuss recent studies of their occurrence rate. 

To date, there have been many surveys with space-based single dishes for warm dust
emission. These include {\it IRAS, ISO, AKARI} and more recently, {\it Spitzer} and {\it WISE}. Based on these surveys, the detection rate of warm dust was found to be quite
low, in the range of 1-2\% for young stars and up to 100$\times$ lower for older stars \citep[i.e.,][]{ken13b}.  Excesses at mid-IR wavelengths can be created by the Wein tail of cold dust belts \citep[see][and references therein]{mat14c}.  Significantly higher detection rates have been measured utilizing high spatial resolution mid-IR and near-IR interferometry.  

\cite{menn14} present detections of mid-IR excesses from the Keck
Interferometer Nuller toward 47 stars, finding detection rates above 30\% for A stars with
a global rate of 12\%. Within this sample, detection of mid-IR excesses is most common for stars earlier than
F2 that also exhibit Kuiper Belt analogues, suggesting that there is a common physical
origin for the two belts, or that the warm dust belt is sustained by the cold belt through
some mechanism. Therefore, the exozodis measured through mid-IR interferometry are consistent with the warm dust components of two-temperature belts (detected through SEDs or in single dish surveys), albeit at much higher detection rates when these components are observed with sufficient spatial resolution. Interestingly, mid-IR excesses have not been detected around stars which host hot dust detected through near-IR interferometry \citep{kra17}.

\cite{ert14} present a merged sample of 125 stars observed for near-IR excesses. Based on their data, detection rates of hot dust decline from 28\% for A-type stars, to 15\% for F stars, down to 10\% for G and K type stars, values very similar to the most recent measured
rates for Kuiper Belt analogues. The data suggest that the detected rate of hot dust increases with stellar age, the opposite of what is expected for a collisional cascade, consistent with the idea that the hot dust components are not generated by a steady state process. For A stars, no obvious correlation is seen between the presence of near-IR emission and the presence of a cold Kuiper Belt analogue; \cite{nun17} note that there is however a strong correlation for solar-type (FGK) stars.  A correlation with the presence of an outer dust belt is expected, since hot dust cannot be created in situ and therefore, it should be correlated with some outer reservoir of material from which the hot dust is drawn.  Modeling of a sample of nine hot exozodis found that the dust was located farther
from the star for higher stellar luminosities, so that the dust appears to have a consistent
temperature in each system \citep{kir17}.  The dominant source of the flux detected was thermal emission, although the models could not exclude some fraction arising from scattered light. 


Many of the systems with detected exozodis are old ($> 100$ Myr), making the origin and high frequency of the exozodis an active area of research. The short timescale required for dust depletion due to collisional evolution or radiative forces is even more acute closer to the star.  In steady state, a typically observed quantity of hot dust would survive for only a century: either the dust is replenished or there exists a means of keeping the dust in situ longer without dissipation \citep{van14}.

\cite{kra17_review} describe several mechanisms for moving
material inward in a planetary system that could assist in a build-up of dust in the inner
parts of a planetary system. For example, \cite{lis17} find evidence for hot dust
emission in ground-based spectroscopy of HR 4796A's young debris disk, which has an
outer cold component. They find that some of the excess seen in the near-IR is due to
a tenuous 850K component with evidence of organics and silicate emission. The authors
suggest that that the emission is consistent with a steady stream of dust flowing into the
sublimation zone from the disintegration of rocky cometary material.


\subsection{Scope of this review}

This review is focused on recent observational developments in understanding debris disk structure, composition, and variability.  There have been a number of notable advances in observational capability across the electromagnetic spectrum in recent years that have substantially enhanced our ability to spatially resolve debris disk structure at multiple wavelengths, yielding insight into the physical mechanisms shaping their evolution.  Recent spatially resolved observations from facilities like the {\it Hubble Space Telescope} (HST), the Gemini Planet Imager (GPI), the Spectro-Polarimetric High-contrast Exoplanet REsearch instrument (SPHERE), {\it Herschel}, and ALMA dominate the major results in this review.  We approach the review from the observational direction, while pointing out ties to theoretical work, which is intended to be in a complementary direction to other recent reviews \citep[e.g.,][]{wya08,kriv10}.  

We have attempted to select topics without comprehensive recent reviews.  For example, we have provided in this introduction only a brief update on debris disk demographics, since several recent major reviews have covered that area thoroughly \citep[][]{wya08,mat14c}, addressing questions like the incidence of debris disks around stars of different ages and masses, and how debris disk incidence as a function of time connects to planet formation processes.  We also limit the scope of this review to stars on the main sequence, since there are two comprehensive recent reviews of the rapidly developing field of debris disks around post-main sequence stars \citep{far16,ver16}.

Recently, the field has been progressing from statistics toward integrating our knowledge into understanding of the underlying planetary system and its dynamics. Therefore, in addition to addressing the overall structure (Section\,\ref{sec:outer_disk}), dust (Section~\ref{sec:dust_properties}) and gas (Section~\ref{sec:gas_in_debris}) content of debris disks, we also discuss the topic of planet-disk interaction, particularly when informed with direct imaging of both disks and planets (Section~\ref{sec:planet-disk_interaction}) and explore the emerging field of time-domain studies of debris disk properties (Section~\ref{sec:time_domain}).



\section{Outer Disk Structure}
\label{sec:outer_disk}



In this section, we review the diversity and properties of spatially resolved structure in outer debris disks, i.e., Kuiper Belt analogues. 
Recent advances in imaging capability across the electromagnetic spectrum 
have revealed details of disk structure that were previously unobservable and have begun to allow us to piece together a multiwavelength picture that connects the morphology to the underlying physics shaping the disk structure.  A spectacular example of such multiwavelength imaging of the Fomalhaut debris ring is shown in \textbf{Figure~\ref{fig:fom_aumic}}.  Catalogs of resolved images of circumstellar disks are maintained online\footnote{See http://circumstellardisks.org/ and http://www.astro.uni-jena.de/index.php/theory/catalog-of-resolved-debris-disks.html}.  

Debris disk imaging efforts stretch back across decades essentially to the moment that debris disks were discovered \citep{smi84}.  We focus on recent observational results, particularly emphasizing disks that have been imaged at a wide range of wavelengths at high ($\sim$10s of au or better) angular resolution.  {\it HST}, in particular, has been a consistent engine of high-resolution images of debris disks, and we include some recent results that make use of innovative data analysis techniques to draw out better contrast and reduce the inner working angle, the area of poor contrast near the star created by the coronograph and related star subtraction post-processing techniques.

The diversity of the spatially resolved structure in outer disks can be divided roughly into categories of radial structure, azimuthal structure (i.e., departures from axisymmetry), and vertical structure.  A summary figure illustrating examples of the different structures discussed is presented in \textbf{Figure\,\ref{fig:disk_mosaic}}.  
Some disk features do not fall neatly into any category, for example warping, which generally involves changes as a function of radius that cause discontinuities in the vertical dimension.  We review these observations with an eye towards understanding the underlying physical mechanisms driving the diversity observed in the spatially resolved structure.  

\subsection{Radial Structure}

Many debris disks are well described as a collection of rings.  Some are narrow; some are broad; and some systems have multiple rings or gaps within an otherwise broad ring.  In this section, we explore the diversity of radial structures in debris disks and discuss connections with theoretical mechanisms describing their origin.  

\begin{figure*}
\centering
\includegraphics[scale=0.75]{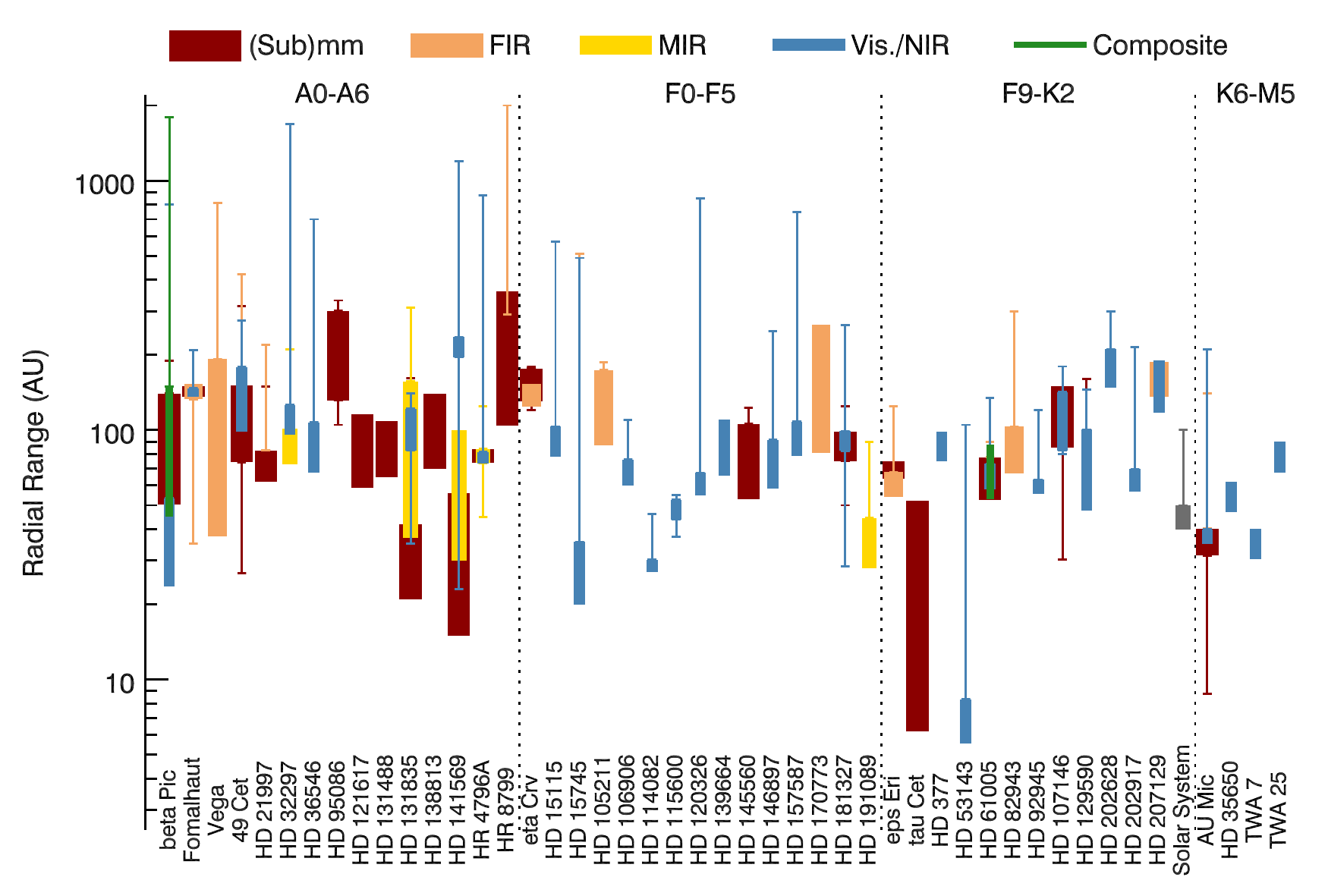}
\caption{
Radial extent of debris disks as determined in scattered light, MIR, FIR and (sub)millimeter high-resolution observations (blue, gold, orange and red bars, respectively). Green bars represent the disk structure derived from combined fits to multiple types of data. For each disk and observing methodology, the thick bar and whiskers indicate the FWHM of the inferred surface density profile and the full extent over which the disk is detected, respectively. In most cases, the latter is sensitivity-limited towards large radii, while in scattered light, the region inside the main ring is affected by severe artifacts that preclude definitive conclusions. The Solar System's Kuiper Belt is shown in gray for comparison. The observations used in generating this figure are presented in Part\,A of the Supplementary Materials.
\label{fig:ring_width}}
\end{figure*}

\textbf{Figure\,\ref{fig:ring_width}} summarizes the radial extent of all of the debris disks with well-resolved observations (i.e., the angular resolution is small compared to the disk angular extent) at a variety of different wavelengths. The tendency of scattered light observations to yield radially narrower structures is likely due to a selection effect (surface brightness contrast is enhanced in  narrow ring) and possible methodology biases (post-processing methods to suppress contaminating starlight often partially suppress disk brightness and sharpen existing structures). Also, scattered light images are limited by the inner working angle within which artifacts from subtracting the bright star obscure disk structure.
For instance, the inner edge of the solid bar for two disks (HD\,15745 and HD\,53143) marks the inner working angle rather than a true inner disk edge, since the inner radius is not resolved. Similarly, our detection of the outermost region of each disk is sensitivity-limited and thus the full extent of the disks is likely underestimated.
While these systems are all far dustier than our own Solar System's Kuiper belt, the radial extent is comparable: the classical Kuiper Belt has an inner radius of 40\,au and a width of 10\,au, while the scattered belt extends hundreds of au farther in radius \citep{bar08}. 

There are no obvious trends in disk radius or disk width as a function of stellar spectral type. Studies of the temperature of the dust from the SED have either revealed weak trends towards a constant temperature regardless of spectral type \citep[e.g.][]{mor11}, possibly due to ice lines setting the disk location, or a slight correlation between dust temperature and spectral type suggestive of an alternative mechanism \citep[e.g.][]{bal13,paw14}, but the spatially resolved data have so far not yielded similar trends (although the sample sizes are smaller).  

In general, the search for trends between disk size and stellar parameters like spectral type or age is used to distinguish between different proposed mechanisms for the dust generation that result in different patterns of radial rings.  Briefly, the proposed mechanisms tend to divide into the following categories: delayed or self-stirring \citep[e.g.,][]{ken08,ken10}, which predicts that dust is generated at larger radii with increasing time; stirring by a planet \citep[e.g.,][]{mus09}, in which the radius of the disk can be related to the mass, semi-major axis, and eccentricity of the stirring planet; or explanations related to ice lines or cometary sublimation \citep[e.g.,][]{mor11,jur98}.  There are also explanations that tend more towards stochasticity, but rely on rare events to initiate the collisional cascade, for example a massive collision or stellar flyby \citep[e.g.,][]{wya02,ken02} and therefore are unlikely to explain the majority of the extremely commonly observed debris disk population.  For a  more detailed description of proposed stirring mechanisms, we refer to \citet{wya08} and references therein.  

As \textbf{Figure\,\ref{fig:ring_width}} shows, the maximum radial extent of debris disks tends to be largest for the most luminous stars and smallest for the least luminous, although in most cases the large outer radius is a scattered light (or IR) feature easily understood as a ``halo'' of small grains blown from the parent planetesimal belt by radiation pressure or a stellar wind.  These extended halos are common around A star debris disks, with spectacular examples imaged around Vega and HR\,8799 \citep{su05,su09}.  The phenomenon is not limited to A stars, however; for example, the disk around the M star AU\,Mic also exhibits an extended scattered light halo reaching to hundreds of au distance from the star \citep{kal04}, while the large grains traced by millimeter-wavelength thermal emission are confined within 40\,au \citep[][see also \textbf{Figure\,\ref{fig:fom_aumic}}]{wil12}.  
This configuration was explained by the model of \citet{str06} in which a ``birth ring'' of parent planetesimals initiate a collisional cascade, the smallest grains of which are then blown out into an extended halo.  The surface brightness profile of the halo then reflects the importance of collisions relative to Poynting-Robertson drag.  This birth ring paradigm has also been observed in $\beta$\,Pic \citep{wil11b}, with a number of other systems exhibiting extended scattered light haloes as well.  A few canonical narrow ring systems, including Fomalhaut and HR\,4796A \citep{kal13, sch18}, also show faint extended features that may indicate that the phenomenon is widespread at a variety of contrast levels.  Another interesting observation is that at least some disks have tenuous dust haloes {\it inside} the parent belt, which may be populated by grains that spiral inward due to Poynting-Robertson drag \citep{loh12,sch16c,ren18}.



\begin{figure*}
\centering
\includegraphics[scale=0.45]{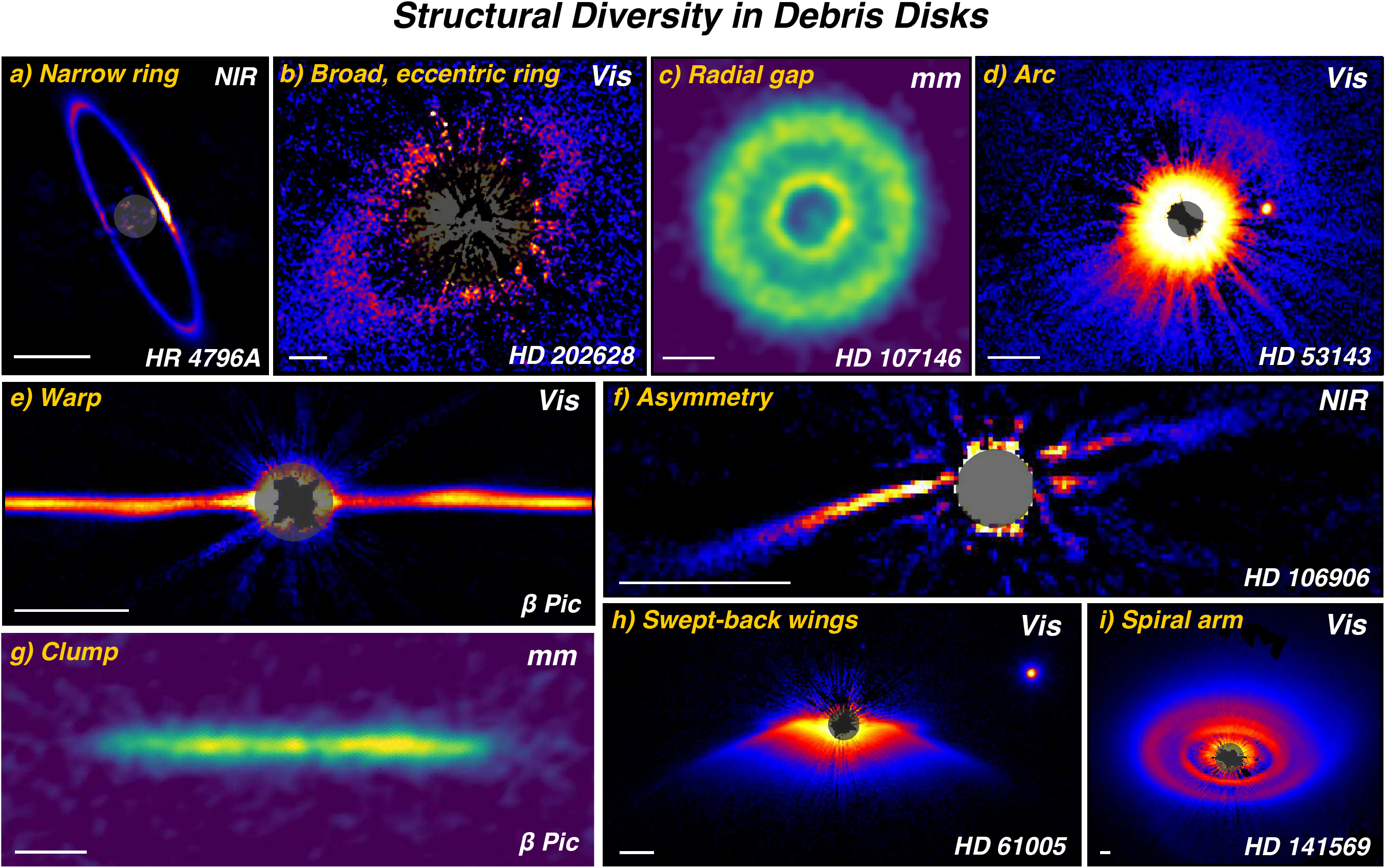}
\vspace*{-2mm}
\caption{
Mosaic of scattered light and millimeter thermal emission for eight systems illustrating the range of asymmetries observed in debris disks. The scale bars represent 50\,au. In some cases, smoothing and high-pass filtering has been applied to the data to better emphasize substructures. Data are from \citet[][panel g]{den14}, \citet[][panels d and h]{sch14},  \citet[][panel e]{apa15}, \citet[][panel f]{kal15}, \citet[][panel i]{kon16c}, \citet[][panel b]{sch16b}, \citet[][panel a]{mil17b} and S.~Marino (private communication, panel c).
}
\label{fig:disk_mosaic}
\end{figure*}

Aside from the halo phenomenon, narrow rings (in this case meaning rings with $\Delta R/R \lesssim 0.5$) are generally quite common, with particularly spectacular examples found around the A stars Fomalhaut \citep[e.g.,][and see \textbf{Figure\,\ref{fig:fom_aumic}}]{kal05} and HR\,4796A \citep[e.g.,][see also panel a of \textbf{Figure\,\ref{fig:disk_mosaic}}]{per15}, as well as the F star HD\,181327 \citep[e.g.,][]{sch14}.  There are also examples of radially broad debris disks, even when traced by large grains impervious to radiation pressure at millimeter wavelengths, including those around the A stars HR\,8799 \citep{hug11,boo16} 49\,Ceti \citep{hug17}, the G stars $\tau$\,Ceti \citep{mac16} and HD\,107146 \citep[][see also panel c of \textbf{Figure\,\ref{fig:disk_mosaic}}]{ric15}, and the M star AU\,Mic \citep{mac13}.  

The question of what makes a ring narrow or broad is still a fairly open one.  The broad rings around AU\,Mic and HD\,107146 both exhibit surface brightness profiles that increase steeply with radius at millimeter wavelengths \citep{mac13,ric15}, consistent with predictions for self-stirred debris disks \citep[which include planetesimals but not necessarily any planets,][]{ken02,ken10}, although the timescale required to assemble Pluto-size planetesimals and initiate a collisional cascade at the outer boundaries of these disks is longer than the age of the systems \citep{ken08}. The 49\,Ceti disk, on the other hand, exhibits a deficit of millimeter emission at small radii, with a surface density profile that generally decreases with radius beyond $\sim$100\,au, extending to radii of more than 300\,au.  Given similar difficulties with reconciling the age of the system with the timescale for self-stirring at such large radii, it is likely that at least some parts of the disk are stirred by planets or other mechanisms \citep{hug17}. Narrow rings, on the other hand, may require multiple planets shepherding the inner and outer radii of the ring \citep[e.g.,][]{bol12}, truncation by external perturbers \citep[e.g.,][]{nes17} or they may be confined by mechanisms related to the interaction between gas and dust \citep[e.g.,][]{lyr13}.  



Substructure in the radial dust distribution is now becoming visible as high-contrast imaging techniques improve.  For example, the disk around HD~\,07146 contains a broad and smooth ring at optical wavelengths \citep{sch14}, but is much broader and exhibits a statistically significant gap in its radial brightness distribution at millimeter wavelengths \citep[][see also panel c of \textbf{Figure\,\ref{fig:disk_mosaic}}]{ric15}.  Similarly, the disk around HD\,131835 
shows clear evidence for multiple rings \citep{fel17}, as does that around 
HD\,120326
\citep{bon17}. A related phenomenon, veering into the territory of departures from axisymmetry that will be discussed below, is that of arcs and spiral arms, which have been imaged in the disks around HD\,141569 \citep[e.g.,][]{kon16c} and HD\,53143 \citep[][see panel d of \textbf{Figure\,\ref{fig:disk_mosaic}}]{sch14}.  HD\,141569 may be unique in representing a truly transitional object between the protoplanetary and debris disk phases, which might explain why it is so unusual in exhibiting a spiral pattern (see panel i in \textbf{Figure\,\ref{fig:disk_mosaic}}), but clearly arcs and concentric rings are a relatively common phenomenon in debris disk systems.  At this point the data quality is not sufficient to distinguish between a broad disk punctuated by narrow gaps and multiple concentric narrow rings in most cases, though recent high-resolution ALMA observations of the HD\,107146 system are extremely suggestive of a gap-like structure that retains dust at a reduced level within the gap (see panel c of  \textbf{Figure\,\ref{fig:disk_mosaic}}, S.~Marino, private communication).  Such a broad and shallow gap is likely to be more consistent with planet-sculpting scenarios than gas-dust interactions, and in debris systems these gaps are quite easy to open and might be sensitive to very low-mass planets \citep{ric15}.  


As a final example of radial substructure, we must consider the presence of warps in debris disks.  The iconic example within this category is the disk around $\beta$\,Pic, which exhibited a clearly warped inner disk in early scattered light imaging \citep[][see also panel e of \textbf{Figure\,\ref{fig:disk_mosaic}}]{hea00,gol06} that pointed the way to the planet $\beta$\,Pic\,b \citep[][discussed in more detail in Section~\ref{sec:planet-disk_interaction}]{lag09}. Similar structures have been observed in the disks around AU\,Mic \citep{wan15} and HD\,110058 \citep{kas15}.  Interestingly, modeling suggests that a single planet may produce only transient warps, and may not be able to explain all of the observed features of the $\beta$\,Pic system, indicating that multiple planets or other mechanisms may be necessary to sustain warped structures \citep[e.g.,][and references therein]{lee16}.  




\subsection{Departures from Axisymmetry}

With the central star providing an essentially spherically symmetric radiation field and acting as a point mass, departures from axisymmetry are of particular interest because they require something beyond the star-disk system to break the axisymmetry.  One tempting interpretation is gravitational sculpting by a planet, but other interpretations have also been proposed for various phenomena, including interactions with the ISM, gravitational perturbations from stellar flybys or companions, instabilities produced by interactions between gas and dust, and recent planetesimal collisions.  Here we review several categories of non-axisymmetric structure in debris disks.   \\

\subsubsection{Eccentricity}

A number of debris disks exhibit subtle deviations from axisymmetry in the form of eccentricity, such as Fomalhaut \citep[][see \textbf{Figure\,\ref{fig:fom_aumic}}]{kal05}.  Nearly all narrow rings have been found to have a stellocentric offset.
Measured eccentricities range from $\sim$0.06 \citep[HR\,4796A, HD\,15115,][]{rod15,sai15,mil17b}  to $\sim$0.2 \citep[HD\,106906,][see panels b and f of \textbf{Figure\,\ref{fig:disk_mosaic}}]{kal15}. Eccentricities smaller than $\sim$0.05 cannot routinely be measured with current instrumentation.


A detailed look at the flux distribution in an eccentric disk provides a wealth of information about the grains and their dynamics.  In scattered light observations at short wavelengths, debris disks exhibit a brightness enhancement at pericenter due to the steep $r^{-2}$ dropoff in flux with distance from the central star; the closest point to the star (pericenter) therefore glows the brightest.  At longer wavelengths, by contrast, thermal emission in an optically thin disk is proportional to the product of temperature (which decreases as $r^{-0.5}$ with distance from the star), surface density, and opacity. In the mid- to far-infrared, the higher temperature at pericenter can result in ``pericenter glow" \citep[e.g.,][]{wya99}, whereas in the submillimeter, the accumulation of dust grains on eccentric orbits at apocenter leads to an opposite brightness asymmetry, called ``apocenter glow'' \citep[e.g.,][]{wya05,pan16}. Both pericenter and apocenter glow are observed in the Fomalhaut ring at different wavelengths (see panels b and c of \textbf{Figure\,\ref{fig:fom_aumic}}).

The variation of the apo-to-peri flux ratio as a function of wavelength can be related, in the context of steady-state models of grain populations, to properties of the grain size distribution like its power-law slope $q$ and the long-wavelength slope of the grain emissivity $\beta$ \citep{pan16}. 
Recent high-resolution ALMA imaging of the Fomalhaut system have provided the first conclusive observational evidence for apocenter glow and have been used to place stringent constraints on the power law slope of the grain size distribution \citep{mac17}.  Detailed models of the scattered light from the HD\,61005 disk perplexingly derive a density that is two times higher at pericenter than apocenter \citep{olo16}.  

The underlying cause of eccentric rings in debris disks is more difficult to pinpoint.  The categories are similar to those proposed as the confinement mechanism for narrow rings, although eccentricity requires mechanisms that break the intrinsic symmetry of the star-disk system, i.e., eccentric planets, dust-gas interactions and recent catastrophic giant impacts. 
A potential way of distinguishing between a recent collision and a more steady-state planet shepherding scenario is via the long-wavelength slope of the spectral energy distribution, since impacts are predicted to exhibit specific particle size distributions \citep[e.g.][]{lei12}.  \\

\subsubsection{Swept-back Wings}

Another departure from axisymmetry 
is the asymmetric, swept-back features occasionally associated with dust rings in debris disks.  The prototypical example is HD\,61005, also poetically named ``The Moth'' due to the appearance of the swept-back ``wings'' that dominate its scattered light images (see panel h of \textbf{Figure\,\ref{fig:disk_mosaic}}).  Also included in this category are HD\,32297 and HD\,15115 and two other possible members are the asymmetric edge-on disks around HD\,111520 \citep{dra16} and HD\,30447 \citep{sou14}.
\citet{sch14} also note that systems like Fomalhaut and HR\,4796A that exhibit low signal-to-noise ratio (SNR) evidence of faint extended features beyond a narrow dust ring may indicate that these systems are simply the most spectacular examples of a common phenomenon, perhaps due to their edge-on configuration.  As a class, in addition to their large-scale asymmetries these systems feature a distinct ring component -- which may be eccentric as in the case of HD\,61005 \citep{bue10} or apparently circular as in the case of HD\,15115 \citep{maz14} -- and color gradients that trace the structural asymmetries in scattered light \citep[e.g.][]{deb08,deb09}. The degree of asymmetry of these systems in early millimeter interferometry mapping is weak if present at all \citep[e.g.,][]{man08,olo16}, but higher resolution mapping with ALMA is crucially needed to assess whether large dust grains trace the asymmetry observed in scattered light or are primarily confined to the parent-body ring.

The most popular explanation for this morphology has so far been interactions with the ISM.  Two similar models based on interaction between disk particle and ISM gas under different configurations have been proposed to explain the observed  structures \citep{man09,deb09}.
Another torque-based method of breaking the symmetry, proposed by \citet{esp16}, involves an eccentric, inclined planet sculpting the dust in the HD\,61005 system. Independently, \citet{lee16} developed a unifying model of debris disk morphology that invokes planets to explain a wide range of morphological features including needles and wings like those observed in HD\,15115 and HD\,61005.  To provide yet another interpretation, \citet{maz14} speculate that a recent collision like that described by \citet{jac14} might plausibly explain the morphology of HD\,15115.  At this time, no definitive explanation for the observed structure exists, although higher-resolution imaging at long wavelengths that spatially resolves the locations of the large grains is one promising avenue of investigation, since ISM interactions generally preferentially manifest in the small grains whereas dynamical scenarios are more likely to affect the large grains as well.  \\

\subsubsection{Clumps}

Clumpy structure in debris disks has long been a predicted consequence of resonant interactions between planets and debris dust \citep[e.g.,][]{lio99,oze00}. The prediction is reasoned largely on the basis of analogy with Neptune and the resonant structure it induces on Kuiper Belt objects in our Solar System.  The predicted orbital period of a dusty clump is super-Keplerian, orbiting with the period of the planet inducing the resonance rather than at the orbital period of the more distant KBOs themselves.  Such features are most readily observed at long wavelengths that trace large dust grains, since the radiation pressure that affects the orbits of small grains can smooth over and erase clumpy structure \citep{wya06}. 


While initially there appeared to be clumpy structure in low-SNR millimeter-wavelength maps of several systems, including Vega and $\epsilon$\,Eridani
\citep{wil02,gre05}, follow-up observations have generally not robustly recovered the clumpy structure.  There are different explanations for different sources: in the case of Vega a combination of low SNR and positioning of the features at the edge of the primary beam where noise is difficult to characterize likely led to an overestimate of the significance of the reported clumpy structures 
\citep{hug12}.  In $\epsilon$\,Eri, by contrast, while it is still not clear whether the features are recovered 
\citep{boo17} or not 
\citep{cha16}, it is clear that background galaxies account for at least some of the previously reported clumpy structure, since the high proper-motion system recently passed in front of an unusually strong concentration of galaxies bright in the submillimeter \citep{gre14}.

At this stage, the only system with an unambiguous departure from axisymmetry in the form of a ``clump'' of millimeter emission on one side is the debris disk around $\beta$\,Pic\footnote{AU\,Mic also exhibits clumpiness in scattered light imaging, which is discussed in Section~\ref{sec:time_domain}.}, which exhibits a relatively weak asymmetry in dust continuum emission 
and a stronger asymmetry 
in CO emission \citep[][see also panel g of \textbf{Figure\,\ref{fig:disk_mosaic}}]{mat17}.
Unfortunately, since $\beta$\,Pic is viewed edge-on it is impossible to study the true azimuthal variation of the dust emission.  In general, even with high-sensitivity imaging by ALMA, $\beta$\,Pic has so far been the primary exception to the rule that debris disks tend to be extremely azimuthally smooth at millimeter wavelengths.  
A robust approach to determining the role of low-SNR features is to subtract an axisymmetric model and examine the residuals (in the visibility domain, for interferometric data), but the role of background galaxies is much more difficult to determine without long-term observations that monitor changes in structure due to both orbital motion in the system and proper motion of the system across the sky.  

The low levels of non-axisymmetry do not rule out the presence of planets. Despite the earlier predictions of resonant structure, recent sophisticated modeling of debris disk evolution that takes into account the role of collisions shows that resonant structure can be washed out by collisions even for the large grains that dominate the emission at millimeter wavelengths \citep{kuc10,nes15}.  Another implication of these models is that in systems with lower dust densities where the collision rate is proportionally lower, collisions may be less efficient at washing out the signature of resonant interactions.  \citet{sha15} present a fast, semi-analytic method for generating images of such (collisionless) disks for comparison with data.  Asymmetries are also expected to be more common in the terrestrial planet-forming regions, where velocities are large and destructive collisions between planetary embryos more frequent \citep[e.g.,][]{ray09}.  
Another potential method of distinguishing between planet and non-planet mechanisms is studying the alignment of the disk's spin axis with the parent star's rotation axis, which can be connected to planetary obliquity studies.  While this field is still in its infancy, some initial investigations show that the rotation axes of disks and stars are mostly well aligned \citep[see, e.g.,][]{ken13,gre14}.


\subsection{Vertical Structure}

The vertical structure of debris disks can be a revealing probe of the physical mechanisms shaping debris disk structure. Unfortunately, it is often very difficult to measure as the result of the degeneracy between the radial and vertical structure in optically thin disks seen at intermediate inclinations.

The most favorable, albeit still imperfect, case for measuring the vertical structure is an edge-on, radially narrow disk, although in scattered light images PSF subtraction can complicate the interpretation of the disk width. There have been several measurements of vertical structure in debris disks in this regime. The vertical structure inferred from image modeling is generally expressed as an $h/r$ ratio, where $h$ is the scale height (typically Gaussian $\sigma$ or HWHM of a Lorentzian or other distribution) and $r$ is the disk radius. An interesting apparent trend, from the few disks with sufficient resolution to make such a determination, is that these systems exhibit a projected FWHM perpendicular to the disk major axis that is constant or decreases with radius close to the star, and then switches direction to increase with radius far from the star, possibly with the outer edge of the parent planetesimal belt falling somewhere near the inflection point. This is best seen in the case of AU\,Mic and $\beta$\,Pic \citep{gra07,kal95}.
The behavior inside of the parent body projected radius likely stems from the combination of a radially broad parent body belt with a (more or less) forward-scattering phase function. The sharp increase outside of this radius is an effect of the radial-vs-vertical degeneracy applied to prominent haloes in these systems.

The most reliable estimates of the scale height are obtained for systems in which the scattered light has been modeled within a framework that takes into account the phase function, rather than systems in which only the projected FWHM on the sky is recorded, since the projected FWHM is degenerate with the inclination of the system. Such modeling has been carried out for a half dozen disks so far \citep[e.g.,][]{sai15, olo16, cur17}, with $h/r$ values ranging between 0.02 and 0.12 and a median of 0.06. 

The vertical structure of debris disks has often been interpreted as a measurement of dynamical excitation -- indeed, it is the most direct way of probing the system's velocity dispersion, and therefore the processes and masses of the bodies sculpting it.  The large bodies stirring the collisional cascade gravitationally excite the small dust grains, increasing their eccentricity and inclination through collisions that bring the bodies into equilibrium with impact velocities comparable to the escape velocity of the largest bodies.  It is therefore possible to relate the disk thickness directly to the mass of the largest bodies stirring the collisional cascade \citep[e.g.,][]{the07,qui07}.  

However, \citet{the09} demonstrates that the vertical structure at optical and infrared wavelengths should exhibit a substantial ``natural" scale height of $0.04\pm0.02$ due to radiation pressure from the star that can excite the eccentricities and inclinations of the small grains that dominate the scattered light images, even in the absence of large bodies. 
Therefore, caution must be exercised in interpreting these scale heights as probes of the dynamical state of host debris disks.  

The high angular resolution and sensitivity available with ALMA provides an opportunity to circumvent this problem by measuring scale heights at longer wavelengths, where the grains that dominate the thermal emission are large enough to be effectively impervious to the radiation pressure that produces the natural scale height at optical and infrared wavelengths.  There is even some evidence from multiwavelength dynamical modeling of the SED and resolved images that the AU Mic system should display a smaller scale height at millimeter wavelengths than at shorter wavelengths \citep{sch15}.  Self-consistent modeling of the size-dependent velocity distribution in the collisional cascade by \citet{pan12} allows vertical scale height measurements at a particular (millimeter) wavelength to be connected with the dynamical state of the debris disk in a robust way.

\section{Dust properties}
\label{sec:dust_properties}

In this section, we discuss how observations of debris disks are used to place constraints on some key dust properties, such as composition and size distribution. The morphology and SED of a debris disk are jointly determined by its geometry and by the properties of the dust grains it contains. While some results are solidly established, we also emphasize some issues that are currently open but where progress is possible in the future.

\subsection{From observations to dust properties}

The absorption, emission and scattering properties of a dust grain depend on its physical characteristics (size, shape, composition, porosity). These quantities can be modeled parametrically or with theoretically grounded models. While the latter are preferable to illuminate the origin of the debris disk phenomenon and its role in the evolution of planetary systems, they are generally more challenging to implement and often rely on assumptions that are difficult to verify. We first present a brief overview of both types of models before describing how physical parameters qualitatively affect the appearance of debris disks.

\subsubsection{General methodology}

The thermal emission of debris disks is determined by the balance between the heating (absorption) and cooling (emission) of dust grains. The fact that small dust grains are imperfect long-wavelength emitters results in hotter-than-blackbody dust temperatures, as well as in a long-wavelength SED tail that is steeper than the Rayleigh-Jeans law. Analyses of debris disk SEDs generally adopt the modified blackbody model to represent this behavior (see Sidebar).

\begin{textbox}[t]\section{MODIFIED BLACKBODY MODELS}
The fact that small dust grains are imperfect long-wavelength emitters  is most often described with a two-parameter opacity law, whereby the perfect blackbody emission is multiplied either by a $\lambda^{-\beta}$ term for $\lambda \geq \lambda_0$ \citep{bac93} or by $1 - \mathrm{exp}(-(\lambda_0/\lambda)^\beta)$ \citep{wil04}. The latter formalism asymptotically converges to the same power law at the longest wavelengths as the former while avoiding its sharp break around $\lambda_0$.
\end{textbox}

A scattered light image provides constraints on several quantities: albedo, color, scattering phase function, polarizability. The latter two quantities measure the dependence of scattered intensity and linear polarization, respectively, on scattering angle. A resolved surface brightness map can therefore be used, under the assumption of axisymmetric volume density, to derive the dust scattering phase function. While the dust albedo can in principle be used to derive the grain composition, it relies on knowledge of the Bond albedo (averaged over all scattering angles), which is not accessible from our fixed vantage point. On the other hand, the scattered light color of a disk relative to its parent star is essentially a differential albedo measurement, and is therefore more directly informative about the dust properties.

The most common analysis tool for scattering phase functions is the analytical Henyey-Greenstein prescription, characterized by the asymmetry parameter $g$ (see Sidebar). One practical complication is that estimating the sign of $g$ requires identifying which side of the disk is in front of the parent star, which is normally assumed to be the brightest as a result of preferential forward scattering, but can be ambiguous in some cases \citep[e.g.,][]{per15}. Other approaches involve using Mie theory for spherical grains or numerical approximations applicable to complex grain shapes \citep[e.g.,][]{min16}.

\begin{textbox}[t]\section{HENYEY-GREENSTEIN SCATTERING PHASE FUNCTION}
The most commonly used analytical prescription for the scattering phase function is the Henyey-Greenstein formalism \citep{hen41}, which is described by a single parameter, the asymmetry parameter $g$, which is the intensity-averaged cosine of the scattering angle. In this prescription, the scattered intensity varies as $I(\theta) = \frac{1}{4\pi} \frac{1 - g^2}{[1+g^2-2g \cos \theta]^{3/2}}$, where $\theta$ is the scattering angle. Isotropic scattering corresponds to $g = 0$, while forward and back scattering are characterized by positive and negative values of $g$, respectively. While it is easy to implement, this formalism does not rely on any physical model of scattering. Furthermore, empirically determined scattering phase function generally have a significantly different shape, which can at best be approximated by the linear combination of several distinct Henyey-Greenstein phase functions \citep[e.g.,][for the zodiacal light]{hon85}. Interpreting scattered light images of astrophysical objects with this formalism can therefore be misleading, as discussed here for debris disks.
\end{textbox}

Because starlight is unpolarized whereas scattered light is polarized, imaging polarimetry with instruments such as GPI and SPHERE provides a powerful analysis tool. The simplest parametric model for polarizability is Rayleigh scattering, even though it is unlikely to be physically appropriate for the micron-sized (and larger) dust grains found in debris disks. As a result, a combination of Rayleigh polarizability and Henyey-Greenstein phase function is sometimes adopted.

While each of the approaches discussed above provides important clues regarding the dust properties of a debris disk, they all suffer from ambiguities and potential biases. Critically, while parametric models are easy to implement, they do not rely on a physical framework and their interpretation in terms of dust properties is not unique. A physically-grounded, multi-wavelength approach that combines several methodologies is required to obtain a complete picture of the dust properties. Such analyses are based on Mie theory, as this is the only theoretical framework that provides a self-consistent estimate of the absorption, scattering and emission cross-sections at all wavelengths. The key parameters in such a model include the minimum grain size, the power law index of the grain size distribution, the dust composition, and the grain porosity, although some of these may be fixed to default values for computational expediency.

\subsubsection{Qualitative behavior}

\paragraph{Dust composition and porosity}

In thermal emission, the dust composition is most readily probed through the presence and shape of solid state spectral features. Most prominent is a collection of silicate features that are observed at wavelengths ranging from 10 to 70\,$\mu$m, although these are only excited for grains larger than $\approx10\,\mu$m and heated to a minimum of 150--200\,K \citep[except for the 69\,$\mu$m forsterite feature, which can be present at lower temperatures, e.g.,][]{dev12}. Thus, studies of the silicate mineralogy are limited to small grains in the warm and hot components of debris disks. Within these constraints, many species of silicates can be identified in both amorphous and crystalline forms, with the latter typically displaying sharper features. Besides silicates, many non-silicate species have spectral features in the mid-infrared, albeit usually weak and/or very broad ones. High SNR spectra, together with dedicated spectral decomposition codes \citep[e.g.,][]{olo12, lis12}, can thus be used to infer their presence in debris disks. On the other hand, dust composition does not significantly affect the modified blackbody $\beta$ parameter \citep{dra06}. 

In scattered light, dust composition primarily influences the surface brightness and color of debris disks. However, even the difference between low-albedo carbon-rich dust and high-albedo silicate- and ices-rich species is nearly impossible to establish since the albedo is degenerate with the total mass of scatterers for optically thin disks. On the other hand, certain species have markedly distinct colors, such as the organic compounds responsible for red colors 
observed for many Solar System minor bodies. Finally, high grain porosity can be inferred from (polarized) scattered light imaging, as the first order effect of porosity is to reduce the effective size of dust grains. In other words, larger porous grains have scattering properties that are qualitatively similar to those of (smaller) compact grains of the same mass and composition \citep[e.g.,][]{gra07, she09, kir13}. 

\paragraph{Grain size distribution}

Assuming a power law grain size distribution and an arbitrarily large maximum grain size, the grain size distribution's slope $q$ is directly encoded in the modified blackbody parameter $\beta$ : the shallower the size distribution, the shallower the long-wavelength end of the SED (see Sidebar). A second constraint on this slope is provided by detailed analysis of the mid-infrared silicate emission, as micron-size grains have weaker and broader spectral features than smaller grains. Thus, spectral decomposition enables a determination of the relative number of grains of different sizes, thereby providing a constraint on $q$ in the small grain limit. The $\lambda_0$ modified blackbody parameter is also related to the size distribution. A standard interpretation is that $\lambda_0 / 2\pi$ represents the size of a ``characteristic" dust grain in the system, i.e., an average over the entire grain size distribution weighted by mass and emissivity. Given the typical values of $q$, this quantity is close to, but possibly somewhat larger than, the minimum grain size. Similarly, the smallest detectable dust grains, which are the hottest, dominate the thermal emission at any given wavelength. Thus, a comparison between the spatially resolved inner disk radius and the radius expected from the SED temperature measures the degree of overheating and provides a direct constraint on the minimum grain size.

\begin{textbox}[t]\section{GRAIN SIZE DISTRIBUTION AND SUB-MILLIMETER SPECTRAL INDEX}
\citet{dra06} showed that a dust population characterized by a power law size distribution of index $q$ will emit thermally a spectrum whose long-wavelength behavior is a power law of index $\beta = (q-3)\,\beta_s$ if a single small dust grain ($a \ll \lambda$) has a long wavelength emissivity that follows a power law of index $\beta_s$. This underlying behavior is observed in laboratory experiments and predicted theoretically, for instance from Mie theory calculations. Since most astronomical dust compositions have very similar values of $\beta_s$ (in the 1.5--2 range), the $\beta$ parameter is directly connected to the grain size distribution power law index.  The parameter $q$ can then be calculated from the observed (sub)millimeter spectral index $\alpha_\mathrm{mm}$ using the relation $q = \frac{\alpha_\mathrm{mm} - \alpha_\mathrm{Pl}}{\beta_s}+3$, where $\alpha_\mathrm{Pl}$ is the spectral index of the Planck function between the two measured wavelengths given the inferred temperature of the disk.  In the Rayleigh-Jeans limit $\alpha_\mathrm{Pl} = 2$, but for low-temperature disks observed at submillimeter wavelengths a more precise correction factor is sometimes necessary \citep[e.g.,][]{ric15}.  
\end{textbox}

Only grains that are roughly micron-sized contribute to scattered light images. Grains much smaller than the observing wavelength are expected to scatter nearly isotropically while larger grains display a strong preference for forward scattering. Furthermore, small dust grains scatter in the Rayleigh regime and disks dominated by such grains will be bluer than their parent star, whereas large grains have roughly neutral scattering colors. Given the steep slope of the size distribution, analysis of the scattering phase function and dust colors therefore provides a constraint on the minimum grain size. Within the context of Mie scattering, the scattering phase function and polarizability curves are expected to be strongly sensitive to the minimum grain size and to the slope of the grain size distribution in the small grain regime.

\subsection{Observed dust properties in debris disks}

\subsubsection{Dust composition}

{\it Spitzer} spectroscopy of large sample of debris disks has revealed that their dust composition is dominated by standard silicates \citep[e.g.,][]{olo12, mit15}. In addition, a broad range of crystallinity fractions, from $\lesssim1$\% to $\approx95$\%, is derived. This illustrates the intrinsic diversity of dust properties in warm debris disks, but also reveals some degree of dust processing, either through annealing or violent collisions, as the crystalline fraction in the interstellar medium does not exceed a few percent \citep[e.g.,][]{li07}. Beyond silicates, the list of species whose presence has been proposed in some systems includes water ice, amorphous carbon, metal sulfides, silica, SiO and organic materials \citep[e.g.,][]{lis12}. However, the interpretation of these analyses is subject to several caveats, such as nearly featureless spectra for some species, the possibility of non-unique solutions, and underlying assumptions about the structure of individual grains. \citet{leb16} point out that, in the case of the $\eta$\,Crv disk, the latter issue is often significant, thus implying that detailed compositional conclusions reached from such analyses should be treated with caution.

Scattered light imaging has not yet proven powerful in uniquely constraining the dust composition of debris disks. For instance, the red optical-to-near-infrared colors of the best-studied HR\,4796A ring have been proposed by \cite{deb08b} as indicative of the presence of organic compounds similar to those found on Solar System minor bodies. However, similar colors can also be reproduced with porous grains primarily made of silicates, amorphous carbon and possibly water ice \citep{koh08}, emphasizing the difficulties inherent to a color-based approach to constraining dust properties. Furthermore, whenever data are available at $\lambda \geq 3\,\mu$m, disk colors become essentially neutral irrespective of whether the disk is bluer or redder than the star at shorter wavelengths \citep{rod12, rod15}. This behavior is consistent for most species. For instance, abundance fractions of water ice of up to $\approx50\%$ are as consistent with observed disk colors as water-free compositions in at least some systems \citep[e.g.,][]{rod15}. No definitive evidence for water ice, such as the detection of its strong 3--3.4\,$\mu$m solid-state feature, has been obtained to date. 

While simultaneous fits to the SED and scattered light images often assume astronomical silicates as the default pure dust composition for simplicity, some studies consider a wide range of possible mixtures. The inclusion of other components, such as carbon-bearing species (either amorphous carbon or organics) and water ice,  generally improve the quality of the fit \citep[e.g.][]{leb12, olo16}. Similarly, porosities of several tens of percent are often necessary to reproduce the disk observations, although constraints are sometimes rather weak due to degeneracies \citep[e.g.,][]{rod15}. A 90\% porosity was also inferred from the high degree of optical linear polarization observed in the AU\,Mic disk \citep{gra07}. The $\beta$\,Pic disk is the exception to the rule, as compact grains are favored by global modeling \citep{bal16}.

\subsubsection{Grain size distribution}

\begin{figure*}
\centering
\includegraphics[scale=0.7]{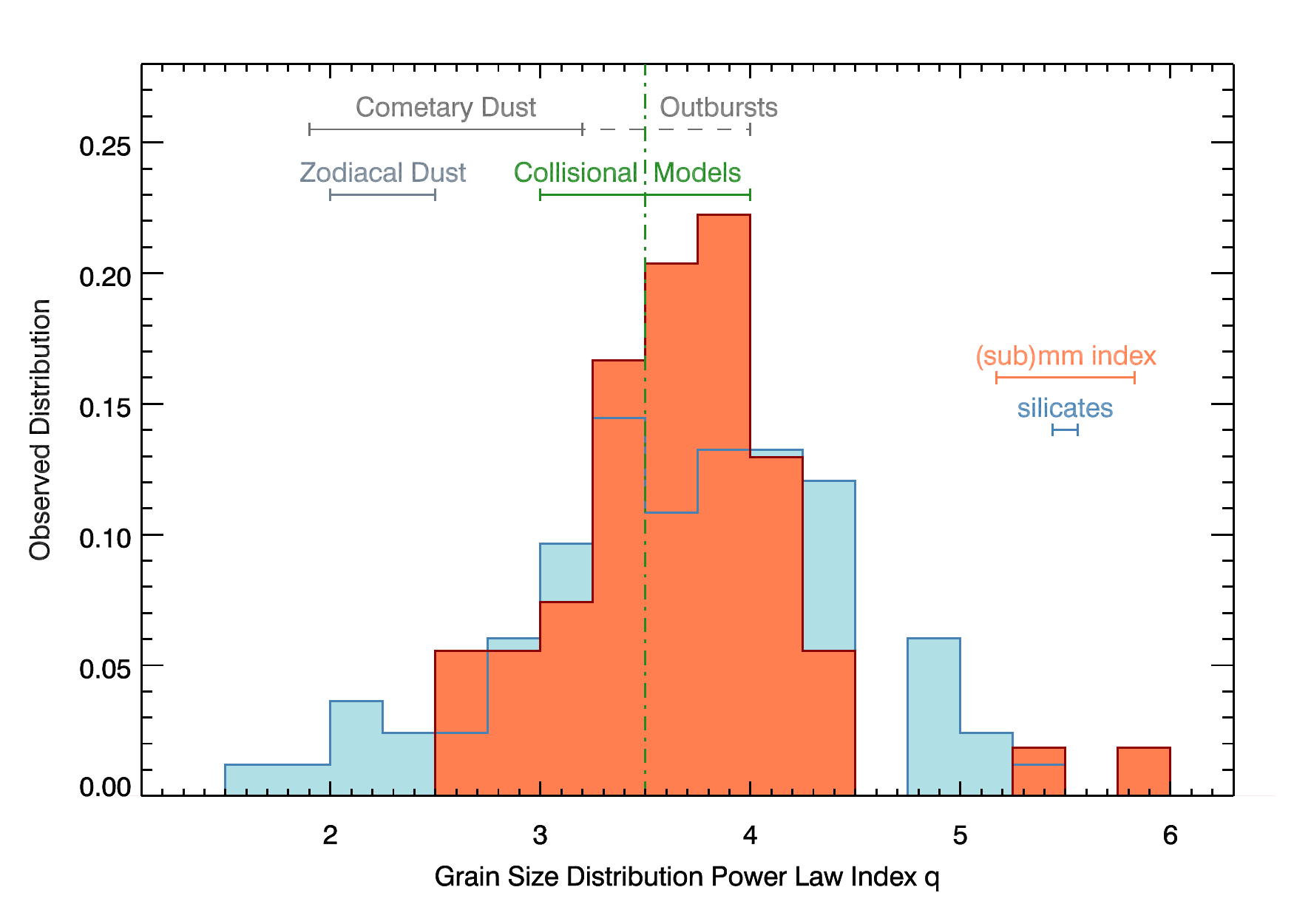}
\caption{
Inferred grain size power law index for debris disks based on the (sub)mm spectral index \cite[orange histogram;][]{mat07, mat15b, don13, olo13, mar14, mar17, paw14, mac16b} and the spectral shape of the mid-infrared silicate features \citep[blue histogram;][]{mit15}. The median uncertainty associated with both methods is represented by the colored errorbars. The ranges of derived power law indices for zodiacal \citep{lei76, gru85} and cometary \citep{hor06, gic12, ish14, ful16} dust are indicated by gray errorbars at the top of the figure. The vertical dot-dashed line marks the theoretical prediction from \citet{doh69} while the green errorbar spans the range of predictions from modern theoretical and numerical models of collisional cascades \citep{pan05, gas12, pan12}.}
\label{fig:q_dist}
\end{figure*}

\paragraph{Power law index} Large (sub)mm surveys have sampled the long-wavelength end of the SED of debris disk systems and have revealed typical values of 0.5--1 for the modified blackbody parameter $\beta$ \citep[e.g.,][]{hol17, sib17}. In turn, this implies a grain size distribution power law index in the $q\approx3.2$--4.2 range with a handful of exceptions that show $q\lesssim3$ or $q\gtrsim5$ \citep[e.g.,][]{ric15,mac16}. 
Similarly, spectral analyses of the mid-infrared silicate feature yielded inferred slopes of the grain size distribution with a typical value of $q=3.5$--4.5, albeit with a range from $q\approx2$ to $q\approx5$ that is broader than expected given quoted statistical uncertainties \citep{mit15}. \textbf{Figure\,\ref{fig:q_dist}} illustrates the range of derived grain size distribution slope from both methods. Because they are sensitive to different ranges of grain sizes, this confirms that a single power law is a reasonable approximation of the actual distribution in debris disks, even though they probe different regions in the disk (i.e., cold dust for long-wavelength thermal emission and warm and hot dust for silicate emission).

The inferred range for $q$ in debris disks is in good agreement with the predictions of steady-state collisional models, both from the classical \citet{doh69} model and from more modern models \citep[][and references therein]{mar17}. On the other hand, estimates for the power law index associated with zodiacal dust are markedly shallower, $q\approx 2$--2.5, and studies of cometary dust also generally yield similar power law indices (see \textbf{Figure\,\ref{fig:q_dist}}). One notable exception to this rule is comet 67P/Churyumov-Gerasimenko, whose size distribution steepened from $q\approx2$ to $q\approx3.7$ around periastron passage \citep{ful16}. The shallower distributions are generally explained by the size-dependence of drag and radiative forces, through which the smallest dust grains are rapidly removed from the system \citep[e.g.,][]{ued17}. Only freshly produced size distributions, such as enhanced mass-loss episodes associated with periastron passage for comets or through discrete collisions \citep{gic12}, are therefore expected to match the predictions of collisional models. By analogy, this suggests that the dust population in cold debris disk components is constantly fed by new collisions.

\paragraph{Minimum grain size} The minimum grain size inferred from the modified blackbody parameter $\lambda_0$ typically range from a few to a few tens of microns \citep{hol17, sib17}. Similarly, spectral analyses of the mid-infrared silicate feature suggest minimum sizes ranging from $\approx2$ to $\approx20\,\mu$m \citep{mit15}. Finally, the observation that the actual size of debris disks as measured in thermal emission maps is larger than the prediction from pure blackbody emission by a factor of up to 10 times \citep[e.g.,][]{mor16} indicates minimum grain sizes in the 1--10\,$\mu$m range \citep{paw15}. Not only are these estimates in general agreement with predicted blowout size, but the minimum grain size is found to be positively correlated with the stellar luminosity. However, the dependency of the minimum grain size with stellar luminosity is significantly shallower than would be expected from simple blowout calculations, suggesting more complicated physics \citep{paw15}. 

Constraints on the minimum grain size from scattered light are not as tight at this point. Most notably, the blue scattered light colors of the AU\,Mic disk throughout the optical and near-infrared \citep[][and references therein]{lom17} is consistent with the idea that the low-luminosity star cannot expel small dust grains through radiation pressure.
However, some debris disks surrounding earlier-type stars, which should have large minimum grain sizes, have also been found to have blue colors in the optical and/or near-infrared \citep[e.g., HD\,61005;][]{esp16}. Conversely, red disks \citep[e.g., HD\,107146 and $\beta$\,Pic;][]{ard04, gol06} are found around stars that represent a range of spectral types which partially overlaps with that of blue disk hosts, indicating that the minimum grain size is not the only factor setting a disk's color. Polarization measurements in debris disks, available for a half-dozen systems, also yield modest constraints. The observed maximum polarization fraction is in the 20-50\% range, with no clear wavelength dependence in the cases where this could be probed \citep{tam06}. In most cases where it can be determined, the polarizability curve peaks at roughly 90$^\circ$ scattering angle, consistent with both model calculations spanning diverse grain size distributions and with observed polarizability curves for Solar System dust \citep[e.g.,][]{lam86, had07}. In the case of AU\,Mic, this curve is consistent with a sub-micron minimum grain size \citep{gra07}. On the other hand, the unusual polarizability curve observed for the HR\,4796A ring, with a peak polarization at a scattering angle of $\approx$50$^\circ$ rather than 90$^\circ$ \citep{per15}, is consistent with dust grains which behave in the Fresnel regime \citep{she09}. This implies a minimum grain size of at least 5--10\,$\mu$m in size, in reasonable agreement with the conclusions reached from the disk color and phase function, as well as with the blowout size for such an intermediate-mass star.


\subsubsection{Scattering phase function}

Assuming axisymmetric disk models and applying the Henyey-Greenstein formalism, estimates of the asymmetry parameter have been obtained for over two dozen debris disks (see Part\,A of the Supplementary Materials). There is considerable scatter in the resulting values, from $g=0$ to $g=0.85$. While this is often interpreted as evidence for small (isotropic) or large (strongly forward scattering) minimum grain sizes, the fact that it does not correlate with wavelength, disk colors, or stellar properties suggests that there is another physical effect at play. In particular, knowledge of the phase function is affected by our viewing geometry, which limits the range of scattering angles that can be probed. Interestingly, there is a clear positive correlation between the fitted $g$ values and the breadth of scattering angles probed by the observations (see Part\,B of the Supplementary Materials). This correlation suggests that the scattering phase function of debris disk dust does not obey a simple Henyey-Greenstein function. Instead, it can be generally characterized by a shallow slope around 90$^\circ$, thus yielding low $g$ value for nearly face-on disks, coupled with a strong forward scattering peak that is only sampled for disks seen close to edge-on ($i \gtrsim 70^\circ$). This trend is also qualitatively consistent with the 2- or 3-parameter composite Henyey-Greenstein models used to represent scattering off zodiacal dust \citep{hon85}.

\begin{figure*}
\centering
\includegraphics[scale=0.7]{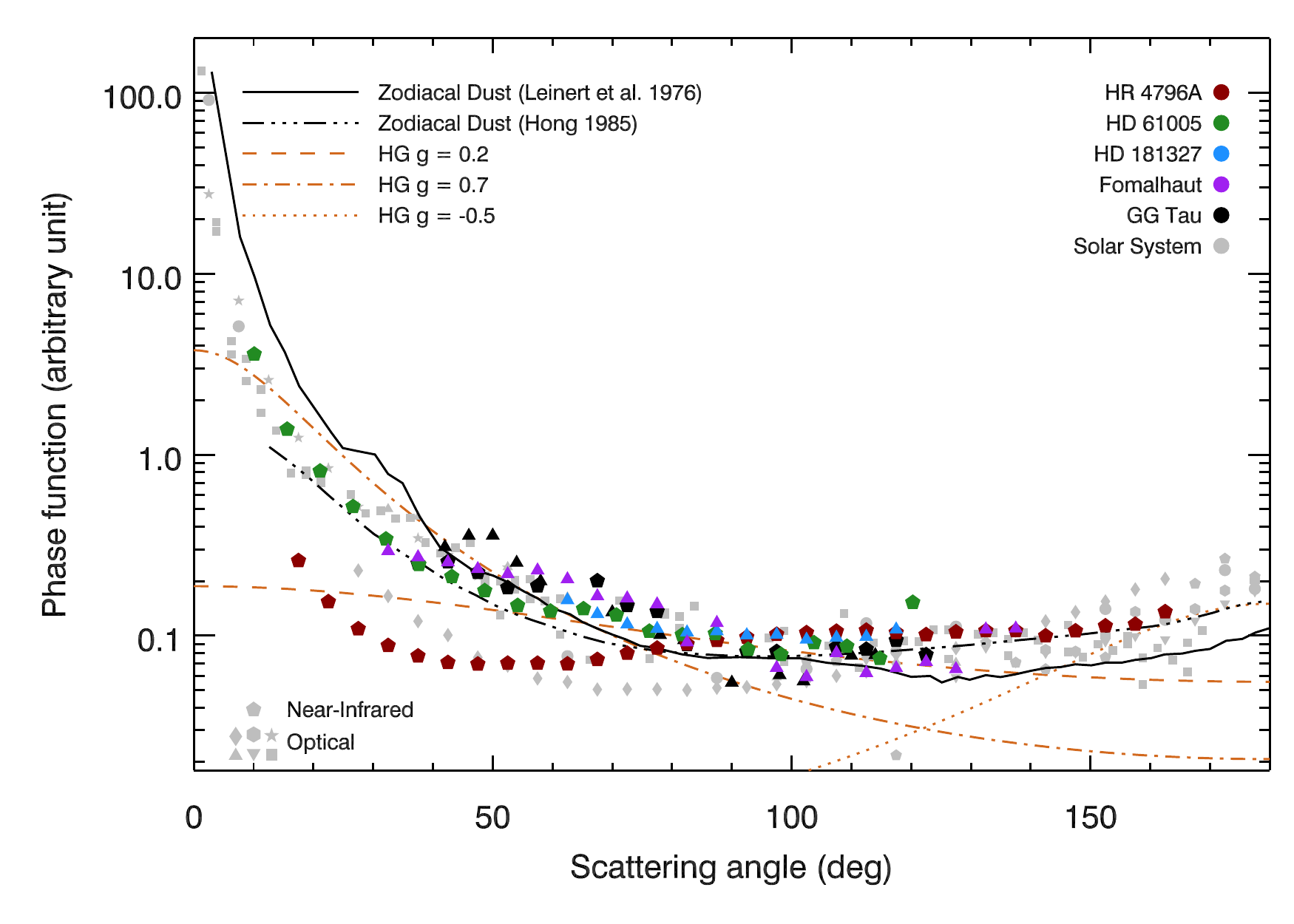}
\caption{
Observed scattering phase function for various debris disks \citep[colored symbols;][]{kal13, sta14, olo16, mil17b}, the GG\,Tau protoplanetary ring \citep[black symbols;][]{mcc02, kri05b}, and several dust populations in the Solar System (gray symbols and black curves; see Part\,C of the Supplementary Materials). For consistency, we have assumed that the brighter region of each debris disk corresponds to the front side of the disk \citep[for the case of Fomalhaut, see][]{leb09}. Near-infrared observations are shown as pentagons whereas other symbols indicate visible light observations. The orange curves display representative Henyey-Greenstein phase functions. 
}
\label{fig:dust_scatt}
\end{figure*}

A more agnostic approach consists of the recent trend of empirically determining the phase function in debris disks. \textbf{Figure\,\ref{fig:dust_scatt}} compares all observed phase functions of debris disks to those measured in the Solar System and in a protoplanetary disk. None of the empirical phase functions matches a simple Henyey-Greenstein function. Furthermore, with the notable exception of the HR\,4796A system (and comet 67P/Churyumov-Gerasimenko in the Solar System), the overall shape of all phase functions are similar, with a strong forward scattering peak, a shallow slope at intermediate scattering angles and, when probed, a significant back scattering peak. The similarity of the debris disk and zodiacal dust phase functions has been pointed out by \cite{gra07} and \cite{ahm09}. Since these environments are unlikely to be characterized by the same dust populations, this suggests that there is a nearly universal phase function for circumstellar dust. Indeed, even volcanic ash and Sahara sand particles have a very similar phase function \citep{hed15}.

As studies of cometary dust have shown, it is possible to reproduce this generic phase function using Mie theory and assuming minimum grain sizes larger than 1\,$\mu$m and high degrees of porosity \citep[e.g.][]{gry04}. Numerical determinations of the phase function for more complex grains, such as aggregates of small spheres and fractal grains also match well the generic phase function \citep[][and references therein]{min16, taz16}. This suggests that debris disk dust particles share a similarly porous structure to cometary dust, whereas the detailed dust composition only plays a secondary role in setting the phase function. Even the unusual phase function observed in the HR\,4796A system is consistent with scattering off porous aggregates, albeit with a larger minimum size of 20--30\,$\mu$m \citep{mil17b}. Based on these calculations, it is worth emphasizing that a larger minimum grain size typically results in a shallower phase function around 90$^\circ$ scattering angle, contrary to naive expectations, which further emphasizes the danger in interpreting the degree of forward scattering in terms of minimum grain size. 

\subsection{Implications and perspectives}

As discussed above, the availability of increasingly higher quality observations of debris disks in recent years has led to a number of robust conclusions about their dust content. The dust composition is dominated by silicates, albeit with non-negligible amounts of refractories and ices, much like dust in the Solar System. Individual grains are characterized by an aggregate-type structure with significant degrees of porosity. The overall shape of the size distribution is in reasonable agreement with the prediction of collisional models, with a $q\approx3.5$ power law index and a minimum grain size that is 3--5 times the blowout size. Finally, scattering of debris disk dust grains, which deviates strongly from the often used Henyey-Greenstein formalism, is remarkably similar to that observed in Solar System dust populations, with much less dispersion between systems than was predicted based on the diversity of grain size distribution and composition.

Despite these significant steps forward, our current understanding of dust in debris disks still suffers from significant shortcomings. Simultaneous analysis of the thermal and scattered light properties of debris disks generally results in serious tension. One illustration of such shortcomings is the HD\,181327 disk, for which \citet{sch06} found that no power law size distribution can simultaneously reproduce the disk scattered light colors, its scattering phase function and the system's SED. Similarly, the apparently low scattered light surface brightness, hence low albedo, of many debris disks \citep[e.g.,][]{kri10} is inconsistent with usual Mie scattering models. On the other hand, while the observed scattering phase function of debris disks can be explained with Mie scattering and/or assuming aggregate grains, the fact that most dust populations share a nearly universal phase function was unexpected.

As the quality of the observations has dramatically improved in recent years, from mid-infrared {\it Spitzer} spectra to full far-infrared coverage of the SED with {\it Herschel} and high-resolution images with ALMA, HST, GPI and SPHERE, so now must our models of debris disks. In particular, it is now necessary to critically review the assumptions on which current models are built and to adopt more sophisticated models as necessary. For instance, it is becoming increasingly clear that debris disks deviate from simple axisymmetry. Comparison of high-resolution thermal emission and scattered light observations may also reveal that each method is probing a physically distinct dust population (see \textbf{Figure\,\ref{fig:ring_width}}), contrary to standard assumptions. This distinction is already hinted at in multiple systems, for example through observations of scattered light color gradients and predominantly blue outer haloes \citep[e.g.,][]{man09} and the observed steepening of the scattering phase function with stellocentric distance in the outer halo of HD\,181327 \citep{sta14}. Both of these observations indicate that dust grains located outside of their parent body belt are predominantly submicron, as expected from the fact that such grains are more sensitive to radiation pressure and, thus, more easily set on high-eccentricity orbits.

The commonplace assumption that the size distribution follows a simple power law also will need to be revisited. Collisional models predict significant departures from this simple form \citep[e.g.,][]{kri06, the07, paw15}, especially for grains up to 10 times larger than blowout size, which dominate most scattering and thermal emission from debris disks. If this is indeed the case, it is likely that current constraints on the grain size distribution are biased and should be treated with caution. At the same time, the aggregate nature of dust grains hinted at by the scattering phase function poses another problem. It remains computationally daunting to compute scattering, absorption and emission properties of such grains in a consistent manner across all wavelengths from the UV to the millimeter. The use of Mie calculations or of tractable approximations is likely to remain a common strategy for the foreseeable future despite the shortcoming of such models. One possible improvement for future models is to make use of the analysis of cometary and zodiacal dust in our Solar System to inform the light interaction properties of dust in debris disks.

\section{Gas in Debris Disks}
\label{sec:gas_in_debris}

Understanding the frequency, amount, and composition of gas in debris disks is arguably the subfield of debris disk studies that has advanced most rapidly in recent years.  Here we discuss global statistical results related to demographic trends in gas content, summarize what is known about the composition of the atomic and molecular components of the gas disks, and review the current understanding of the origin of gas in debris disks.  

\subsection{Overview and demographics of gas-bearing debris disks}

The gas component of debris disk has long been poorly understood for lack of firm detections. One of the main limitations in this area is that while gas fluxes can be quite large in protoplanetary disks, the line sensitivity of single-dish telescopes and interferometers has been limited by relatively small collecting area, making gas surveys of debris disks around main sequence stars far more time intensive than dust surveys of similar-age systems. It is also challenging to detect cold gas in emission: once the inner disk has cleared, the remaining gas tends to have excitation temperatures of tens of K or less, implying that only the low-energy rotational transitions at millimeter wavelengths are sufficiently excited to be detectable, along with far-infrared fine-structure forbidden lines from ground state atomic species.  
Only recently has the sensitivity of far-infrared and millimeter-wave facilities (in particular {\it Herschel} and ALMA) become sufficient to detect gas emission from debris disks within minutes rather than hours of integration.  
%
However, while it is clear that the line fluxes are low compared to those of their protoplanetary counterparts, there is still considerable uncertainty in the total mass of gas in debris disk systems due to our lack of understanding of the molecular and atomic abundances and excitation and shielding conditions within the disk.  

Several heterogeneous methods have been used to probe the gas content of debris disks.  They can essentially be subdivided along two dimensions: atomic vs. molecular gas, and emission vs. absorption line studies.  Absorption line studies use the stellar spectrum as a backdrop against which to study the material along a single line of sight.  Absorption lines are powerful probes of the atomic and molecular content of debris disks, and are responsible for detecting the largest number of distinct species in circumstellar gas, including C, O, Na, Mg, Ca, Mn, Fe, Ni, Ti, Cr, and CO \citep[e.g.,][]{bra04}.  However, there are several important caveats: the utility of this method is restricted to a relatively narrow range of inclination angles such that the line of sight from the observer to the star actually passes through the circumstellar disk and is therefore only useful for the relatively small fraction of disks seen close to edge-on.  In addition, it can be difficult to distinguish circumstellar lines from interstellar lines in the same spectrum, unless the velocity of the star and the local ISM clouds along the line of sight are both well known and well separated in velocity space.  

Emission line studies, especially at high angular resolution, can also be used to study the composition and spatial distribution of material in the disk.  However, since they are less sensitive to small gas quantities, they are more limited in which species they can detect.  So far, the only molecular species detected in emission in debris disks\footnote{While there have been reports of other gas-phase molecular species in debris disks, they have so far not held up to further scrutiny. A gas-phase SiO feature was identified in the spectrum of HD\,172555 by \citet{lis09}, though absorption spectroscopy and additional analysis by \citet{wil16} indicate that the feature likely originates from solid-state rather than gas-phase material.  Similarly, a claimed detection of H$_3^+$ in the disk around HD\,141569 was later superseded by a more stringent upper limit \citep{got05}, as were claimed detections of H$_2$ in the $\beta$ Pic \citep{thi01,che07} and 49 Cet \citep{car08} disks.  There is also a likely detection of fluorescent H$_2$ emission from AU Mic \citep{fra07}, although its origin has not been definitively localized to the disk.} is CO, and with the exception of imaging studies in the extremely bright and nearby $\beta$ Pic system \citep{olo01,nil12}, the only atomic species that have been detected in emission in multiple disks are C and O.  Complicating matters further, emission and absorption lines can probe distinct components of the disk, with emission lines dominated by the large mass reservoirs at tens to hundreds of au from the star, while absorption lines tend to be weighted more strongly towards hot material orbiting close to the central star.  Absorption lines often include time-variable features related to putative ``falling evaporating bodies" (FEBs) in the innermost regions of the planetary system \citep[e.g.,][]{kie14,mil16}, as discussed in Section\,\ref{sec:time_domain}.

\begin{figure*}
\centering
\includegraphics[scale=0.375]{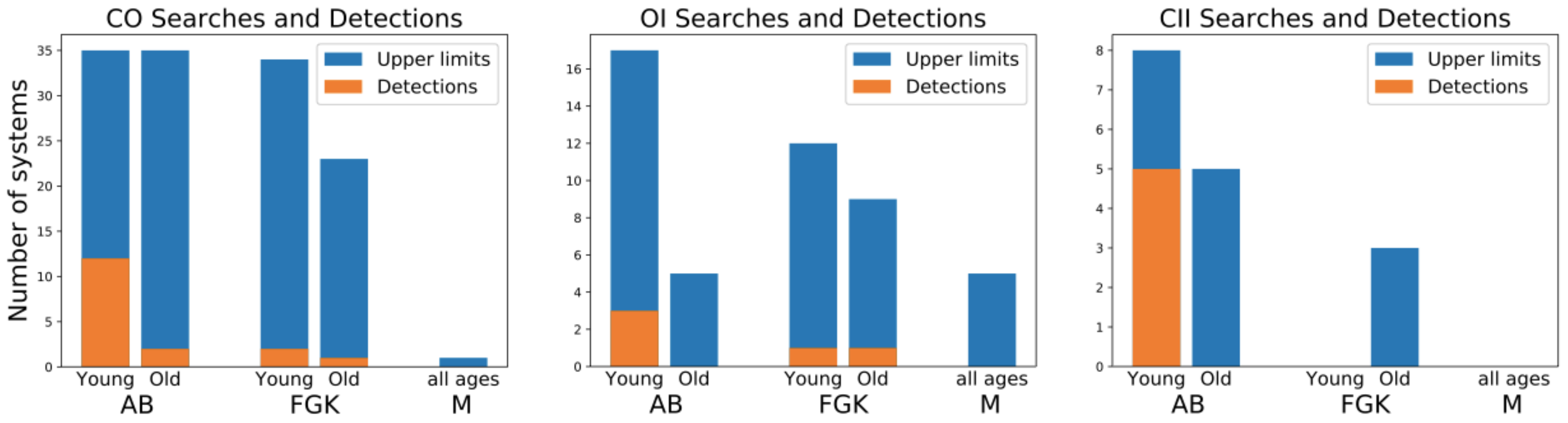}
\caption{
Histogram of all searches for (blue) and detections of (orange) gas emission from debris disks in millimeter-wavelength CO (left), [\ion{O}{1}] 63\,$\mu$m (center), and [\ion{C}{2}] 157\,$\mu$m (right), divided into bins on the basis of stellar age and spectral type. The dividing line between ``young" and ``old" systems is 50 Myr.  While the non-uniformity of the sensitivity and sample selection limit our ability to draw robust conclusions about the true incidence of gas-bearing systems, the available data provide suggestive evidence for greater incidence around younger stars and more massive stars, and highlight a need for more data on M star systems. The observations used in generating this figure are presented in Part\,A of the Supplementary Materials.
}
\label{fig:gas_hist}
\end{figure*}

Given the heterogeneity of detection methods and the non-uniformity of the sensitivity and sample selection of the various surveys conducted to date, our ability to derive robust demographic information is limited.  However, the available data reveal suggestive trends and areas of opportunity in understanding the prevalence of gas-bearing debris disks. \textbf{Figure\,\ref{fig:gas_hist}} presents a compilation of all of the searches for -- and detections of -- CO, [\ion{C}{2}], and [\ion{O}{1}] emission from debris disk systems, subdivided by the age and spectral type of the central star.  The data are drawn from all available surveys for CO emission in debris disks, limited to systems for which the fractional infrared excess luminosity is reported to be $<8\times10^{-3}$. 

Several commonalities emerge from this figure.  For all three species, the highest detection rate occurs among young ($<$50\,Myr) intermediate-mass stars.  Similarly, not only are the detection rates higher for young AB stars than older AB stars, but at least for the two species with sufficient numbers to examine the question, in both cases the AB star detection rates are higher than the FGK star detection rates.  The trend of higher gas detection rates around young intermediate mass stars was previously noted by \citet{lie16}, who surveyed 23 debris disk host stars in the 10\,Myr-old Sco-Cen region and detected strong CO emission from three of seven intermediate-mass stars but none of the 16 FGK stars.  

Interestingly, the FGK star detection rates appear to be flatter with age than those of the AB stars, although of course subject to limitations in the non-uniformity of the sample.  It is difficult to draw any conclusions about the prevalence of gas in M star debris disks, since only one disk has been surveyed in CO and none in [\ion{C}{2}]; while a few have been surveyed in [\ion{O}{1}], the number is so small that even if the prevalence of gas-bearing debris disks were similar to those around more massive stars, we would not expect to have detected any.  

The highest fractional gas detection rate occurs for young A stars surveyed in the [\ion{C}{2}] line.  While this might be a sample selection effect (if, for example, [\ion{C}{2}] searches were targeted preferentially at previously known gas-bearing debris disks), at least one of the five detections ($\eta$\,Tel) occurs in a system with no previous CO detection despite a sensitive search for CO(3-2) emission \citep{moo15}.  Interestingly, the [\ion{C}{2}] non-detections in this young A star category all come from disks with at least one other gas species detection: HD\,131835 and HD\,21997 (detected in CO but not [\ion{O}{1}]), and HD\,172555 (detected in [\ion{O}{1}] but not CO).  Detailed analysis of detections and nondetections within each system have the potential to reveal information about the composition of the gas; for example, the nondetection of [\ion{O}{1}] in the 49\,Ceti system despite abundant CO and [\ion{C}{2}] emission likely points to an anomalously high C/O ratio \citep{rob13}.  

\begin{figure}
    \centering
    \includegraphics[scale=0.7]{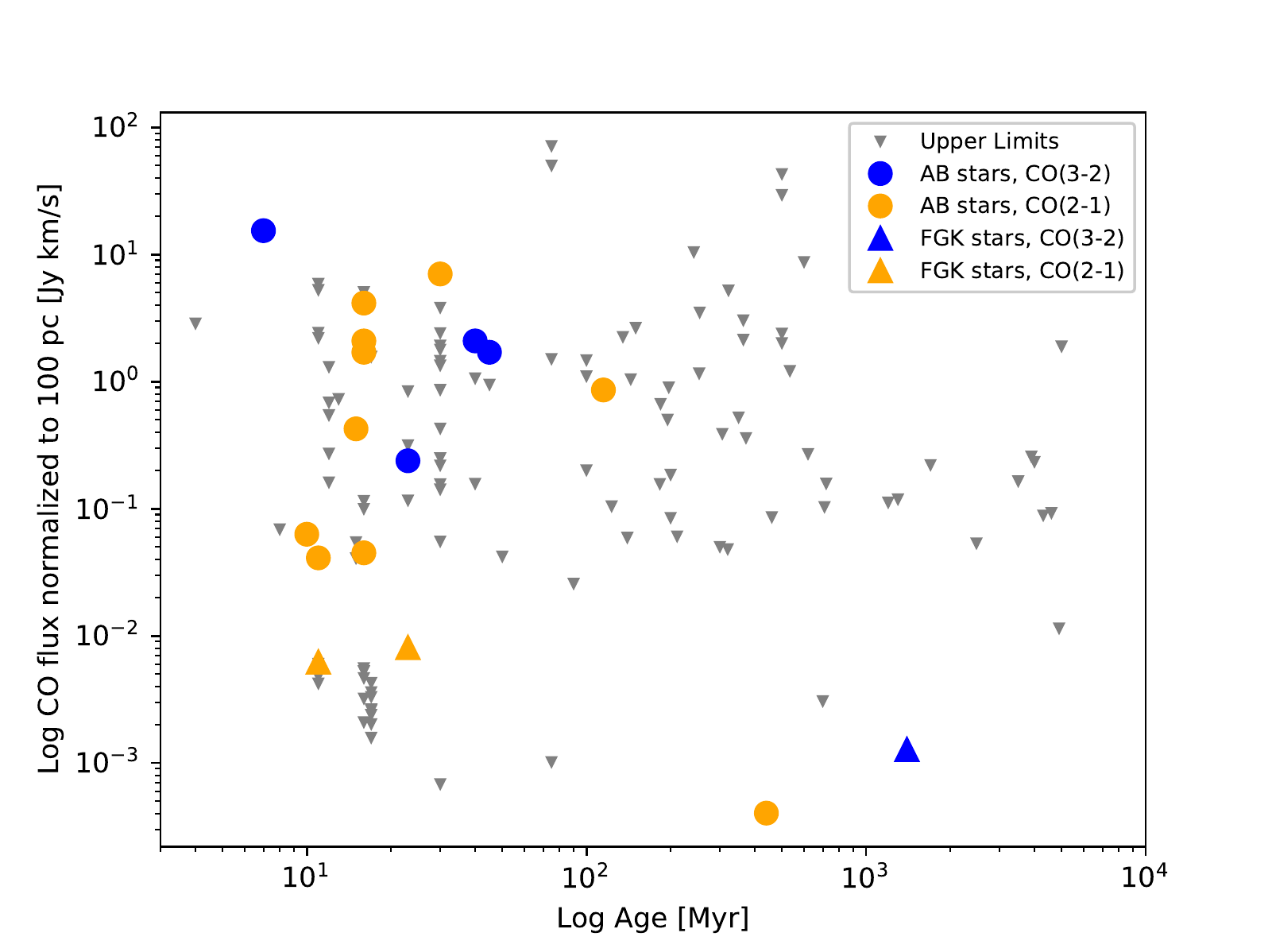}
    \caption{
    Flux of CO upper limits (gray triangles) and detections (colored symbols), normalized to a distance of 100\,pc, as a function of stellar age in Myr.  The stellar spectral type of each detection is indicated by the symbol (circles for AB stars and triangles for FGK stars), while the rotational transition is indicated by the color (blue for CO(3-2) and orange for CO(2-1)). The observations used in generating this figure are presented in Part\,A of the Supplementary Materials.
    }
    \label{fig:co_plot}
\end{figure}

In addition to the detection rates, it is also instructive to examine the detected fluxes in comparison with the range of upper limits.  Inspired by Figure\,7 from \citet{mat15}, \textbf{Figure\,\ref{fig:co_plot}} plots the flux of each gas detection or upper limit, normalized to a distance of 100\,pc, as a function of stellar age, with symbols dividing the detections by stellar spectral type and colors by the rotational transition of the detected species.  For systems where more than one CO transition was detected, we plot only the transition with the highest signal-to-noise detection.  

The difference in normalized flux between the brightest and faintest CO detection spans nearly 4 orders of magnitude, indicating that the range of CO flux is intrinsically broad.  Between those extremes, there is no obvious concentration at any particular flux level, although there is a gap of nearly two orders of magnitude between the faintest CO detection around an AB star \citep[Fomalhaut; see][]{mat17b} and the next-faintest AB star CO disk.  Since Fomalhaut is one of only a few stars that has been searched for CO to such exquisite sensitivity, it is possible (perhaps likely) that many other systems might harbor comparably faint CO disks.  

The three FGK star detections are all clustered within a relatively narrow range of values between $10^{-3}$ and $10^{-2}$ Jy\,km/s, which is about two orders of magnitude lower than the median value for the AB star detections.  This gap in flux suggests that the CO disks around FGK stars may be intrinsically fainter on average than those around AB stars.  The difference is unlikely to be due to stellar temperature alone, since the gap is far larger than the expected factor of 2-3 difference in brightness for an optically thick disk due to the difference in temperature between a typical AB star and a typical FGK star given a CO disk of comparable mass.  

While these broad demographic trends are suggestive, important recent insights into the nature and origin of the gaseous component of debris disks have come from detailed observations and modeling of individual objects.  In the sections below, we discuss the current state of our understanding of the gas composition, its total mass and corresponding implications for its lifetime, and the spatial distribution of material.  We conclude with a summary of the current state of understanding of the origin of the gaseous component in gas-bearing debris disks.  

\subsection{Composition of Atomic Gas}


Since CO is the only molecular species that has been detected, most of our insight into the composition of the gas comes from atomic lines, so far primarily observed through absorption rather than emission spectroscopy.  Multiple-line data and modeling exist for only a handful of systems; the richest data set currently exists for $\beta$\,Pic, followed closely by HD\,141569 and 49\,Ceti.  Here we review the current understanding of the composition of gas in these systems, with notes on some other systems of interest that provide information about either unique aspects of composition or the variation of composition between systems.  

\subsubsection{$\beta$\,Pic}

The unusual presence of Ca and Na absorption in the stellar spectrum was first noted by \citet{sle75}, predating the discovery of debris disks by almost a decade.  The absorption spectrum of $\beta$\,Pic includes both a ``stable" component at the velocity of the star, as well as a redshifted time-variable component that has been linked to the presence of FEBs in the innermost regions of the disk. \citet{bra04} detected and spatially resolved the disk in resonantly scattered emission from 88 different atomic lines and demonstrated that the stable \ion{Na}{1} absorption component originates from the extended disk and suggested that the other stable spectral components probably did as well.  Efforts to characterize the disk composition have therefore focused on the stable disk component rather than the time-variable FEB features, which are discussed in Section \ref{sec:time_domain}.  

Much of the current understanding of the disk's bulk composition hinges on the abundance of C and the C/O ratio, partly due to the generally high cosmic abundance of C and its corresponding importance to the chemistry and dynamics of gas in debris disks, but also because the C/O ratio in particular has been demonstrated to affect the outcome of the planet formation process \citep[e.g.,][]{kuc05}.  An analysis of absorption spectroscopy of [\ion{C}{2}] and [\ion{O}{1}] by \citet{rob06} demonstrated that C was extremely overabundant, by more than an order of magnitude, relative to every other measured element, a conclusion later confirmed by {\it Herschel} observations of [\ion{C}{2}] emission \citep{cat14}.  When interpreted in the context of theoretical work by \citet{fer06}, the enhanced C abundance (by $>10\times$ Solar) explains the previously perplexing persistence of the atomic gas disk, which should be subject to strong radiation pressure that should rapidly remove it from the system.  The high ionization fraction of the abundant C gas causes coupling into a fluid due to Coulomb collisions, which can stabilize the gas by self-braking.  In addition to the extreme overabundance of C, \citet{rob06} also derived abundances for a number of different elements showing that lithophile (e.g., Mg, Al) elements have roughly solar abundance relative to each other, as do siderophile elements (e.g., Fe, Ni), though lithophile  elements are slightly underabundant relative to siderophiles.  

The other interesting feature uncovered by absorption spectroscopy was a C/O ratio that exceeds the solar value by a factor of 18 \citep{rob06}.  Detection of [\ion{O}{1}] 63\,$\mu$m emission by \citet{bra16}, however, implied a lower C/O ratio due to a higher inferred O column density.  Untangling the different values implied by the emission and absorption spectroscopy requires the spatial distribution of the material, since modeling suggests that a clumpy spatial distribution is required to achieve O densities high enough to excite the observed emission.  Another possibility is that some of the O is coming from the photodissociation of H$_2$O, which would suggest an H$_2$O/CO ratio of $\sim 1.5$ \citep{kra16b}.  The $\beta$\,Pic gas disk shows strong deviations from axisymmetry, with the atomic emission lines imaged by \citet{bra04} brighter along the NE limb of the disk while the CO and [\ion{C}{2}] appear to be brighter in the SW \citep{den14,cat14}.  These opposite asymmetries may be explained by a region of enhanced electron density to the SW, which would excite [\ion{O}{1}] emission but remove Na and Fe since they would experience more frequent recombination and therefore spend more time in their neutral form that is more susceptible to the effects of radiation pressure.  

Despite several sensitive searches, a stable H component has never been detected, although the variable component exhibits Ly\,$\alpha$ emission \citep{wil17}.  The limits are so sensitive that they rule out a Solar abundance relative to Fe, and therefore imply that the gas disk is unlikely to be remnant gas from the original disk or gas expelled by the star.  The measured column density in the infalling FEB component is also too high to originate from a CI carbonaceous chondrite meteorite, and therefore \citet{wil17} suggest that the origin of the hydrogen gas might lie in the photodissociation of water from evaporating comets.  

In general, it is clear from the atomic line studies both that C and O are overabundant relative to refractory elements, and that the C/O ratio exceeds that of the star.  
The updated O column density from \citet{bra16} somewhat weakens the \citet{rob06} argument for a vaporized parent body, and modeling by \citet{xie13} presents a more detailed theoretical model favoring the preferential removal of metallic elements due to differing radiation pressure forces relative to C and O.

\subsubsection{Atomic species in other systems}

The inventory of stable gas components observed in absorption along the line of sight to 49\,Ceti includes atomic lines of Mg, Fe, and Mn \citep{mal14} as well as C, O, Cl, S, Si, Al, and Fe \citep{rob14}.  
A preliminary analysis of abundances in the circumstellar gas component of the 49\,Ceti system indicates that this system, like $\beta$\,Pic, exhibits an extreme C overabundance relative to Fe \citep{rob14}.  The 49\,Ceti system also appears to require an enhanced C/O ratio several times that of the Solar System, which is more extreme than the C/O ratio derived for $\beta$\,Pic.  

The line inventory by \cite{rob14} also includes a non-detection of CO in absorption, which is surprising given the strong CO emission observed from the spatially extended circumstellar disk orbiting nearly edge-on to the line of sight at an inclination of $79.4^\circ \pm 0.4^\circ$ \citep{hug17}.  The lack of CO absorption constrains the scale height of the molecular gas disk to be smaller than that of the atomic gas disk and restricts the opening angle to $<10^\circ$ (or, equivalently, $h/r < 0.17$)
A vertically thin configuration for the molecular gas is consistent with the analysis by \citet{hug17} demonstrating that the vertical structure of the gas disk is spatially unresolved, which is inconsistent with the expected scale height for a gas with a hydrogen-dominated, ISM-like composition and instead suggests a high mean molecular weight gas.  It is not yet clear whether the differing scale heights of the atomic and molecular gas distribution arise from an atomic surface layer above a molecular midplane, or whether the atomic gas preferentially arises from a parent body population with a large range of inclinations; however imaging of the recently detected [\ion{C}{1}] emission from the system \citep{hig17} is likely to provide some clues.  

Despite its relatively modest inclination of $60^\circ \pm 3^\circ$ \citep{fla16}, the disk around HD\,141569 also exhibits stable absorption features in Mg, Fe, and Mn that are likely to arise from circumstellar rather than interstellar material \citep{mal14}, as well as atomic emission in [\ion{O}{1}] and [\ion{C}{2}] \citep{thi14} and molecular CO emission \citep{fla16,whi16}.  Like 49\,Ceti, modeling of the observed emission lines favors a disk that is flatter than hydrostatic equilibrium for a disk with Solar abundances \citep{thi14}. Mirroring $\beta$ Pic, the small grid of models explored was unable to reproduce the observed [\ion{O}{1}] 63\,$\mu$m line flux, although in this case the models overpredicted rather than underpredicted the flux.  

Information about the composition of gas in other disks is far more scant, and in most cases sophisticated modeling to integrate information from different observing methods in a given system has yet to be attempted, which presents an opportunity to integrate theoretical understanding of the phenomenon of gas-bearing debris disks.  For example, HD\,32297 exhibits a stable line-of-sight absorption component in Na I that is five times as strong as that of $\beta$ Pic \citep{red07}, along with CO and [\ion{C}{2}] emission indicating substantial molecular and atomic gas reservoirs \citep{don13}.  The disk around HD\,172555 exhibits a stable Ca II absorption feature, but only variable \ion{Na}{1} D lines, indicating a \ion{Na}{1} / \ion{Ca}{2} ratio that deviates significantly from ISM ratios in the same sense as the anomalously low \ion{Na}{1} / \ion{Ca}{2} ratio measured for the $\beta$ Pic disk \citep{kie14}.  It also exhibits [\ion{O}{1}] emission, but no [\ion{C}{2}] \citep{riv12}.  

In general, atomic line studies paint a picture of an atomic disk component that deviates significantly from Solar composition, with evidence of both enhanced C abundance and likely high C/O ratios in multiple systems, as well as a potentially depleted H reservoir.  

\subsection{Molecular Gas: Quantity, Spatial Distribution, and Composition}

While there is good evidence of anomalous abundances and high volatile content in the atomic component of gas in debris disks, the relationship between the atomic and molecular components is not yet well understood.  Constraints on the composition of the molecular gas are scarce and depend largely on the unknown abundance of H$_2$ in the disk.  Complicating matters, there is also mounting evidence that the presence of gas in debris disks is not a monolithic phenomenon, but rather that there are likely to be at least two classes of gas-bearing debris disks: some with gas of protoplanetary origin and some with gas of secondary origin, possibly with a continuum in between.  

\subsubsection{Spatial Distribution}

The spatial distribution of molecular gas in debris disks is one of a number of potential indicators of the origin and evolutionary status of the disk.  Generally speaking, there are three interesting properties of the spatial distribution that can provide clues to its origin: 
1) the degree of axisymmetry, since non-axisymmetric gas distributions are suggestive of second-generation origin given that they should shear out within a dynamical timescale; 2) the spatial distribution of dust versus gas, since if the gas origins are secondary, we expect them to match the distribution of debris dust; and 3) the spatial distribution of gas in debris disks versus protoplanetary disks, since the latter host gas density profiles that decrease outwards gradually over a wide range of radii \citep{wil11} and debris gas disks that do the same may indicate a common origin.

Apparently symmetric systems in CO emission (HD\,141569, HD\,110058, HD\,131835, HD\,138813, HD\,32297, 49\,Ceti, HD\,21997, HD\,181327, HD\,146897, HD\,156623) are so far more common than asymmetric systems ($\beta$\,Pic, Fomalhaut), although the current observations limit our ability to distinguish deviations from axisymmetry in most cases to within several tens of percent, and the issue of optical depth also complicates our ability to distinguish subtle contrast within the density distribution.  It is also worth noting that a few systems, while dominated by a symmetric CO distribution, exhibit significant minor asymmetries with no obvious explanation \citep[HD\,141569, 49\,Ceti,][]{whi16,hug17}.  There are a couple of cases for which the angular resolution or signal-to-noise ratio is insufficient to distinguish any degree of axisymmetry ($\eta$\,Crv, HD\,23642).  The asymmetric debris observed in $\beta$\,Pic is consistent with theoretical predictions for long-lived asymmetry at the sites of recent impacts between planetary embryos \citep{jac14}. 

As for the comparison of the spatial distribution between dust and gas, several different configurations emerge.  The asymmetric CO disks around $\beta$\,Pic and Fomalhaut both exhibit gas that traces closely the location of dust continuum emission in the system, enhancing the evidence for a common second-generation origin. The edge-on $\beta$\,Pic disk exhibits a weak asymmetry in dust continuum emission between the northeast and southwest disk limbs that is exaggerated in CO \citep{den14}, while the location of the CO gas in Fomalhaut's eccentric dust ring corresponds at least approximately with the location of pericenter \citep{mat17b}, although the dust emission otherwise demonstrates no deviations from axisymmetry \citep{mac17}.  Among the systems with generally axisymmetric gas disks, most that have been imaged at sufficiently high resolution seem to exhibit marked differences between the radial extent of the gas and dust emission \citep[HD\,21997, HD\,141569, 49\,Ceti, $\eta$\,Crv,][]{kos13,whi16,hug17,mar17}, although at least one - the only F star - exhibits cospatial gas and dust \citep[HD\,181327,][]{mar16}.  Due to the coarse angular resolution of the survey by \cite{lie16},
it is not yet clear how the extent of the gas and dust disks (i.e., HD\,110058, HD\,131835, HD\,138813, HD\,146897, HD\,156623) in the sample compare.  The overall picture that emerges is therefore one in which asymmetric gas disks are observed to closely trace the parent dust distributions, as expected for a common second-generation origin, while the relationship between gas and dust in the systems with symmetric (and, typically, extended) gas disks is more ambiguous.  

Finally, a comparison between the gas surface density profile observed in symmetric debris disks and that typical of protoplanetary disks is theoretically possible, although in practice is complicated by our lack of understanding of the excitation conditions and general lack of multiple tracers of gas in debris disks.  Observations of the 49\,Ceti disk with ALMA yield surprising evidence for a CO surface density distribution that increases with radius, which is the opposite of protoplanetary disk surface density profiles \citep{hug17}.  Our ability to measure the surface density profile, however, hinges on the unknown optical depth of the disk; while the CO(3-2)/CO(2-1) line ratio is consistent with marginally optically thin gas, a scenario in which the H$_2$ abundance is low leads to higher optical depths that might still be consistent with the line ratio, and might obscure the underlying surface density trend.  A similar analysis of the HD\,21997 gas disk yielded poor constraints on the variation of CO intensity with distance from the central star, due to low angular resolution \citep{kos13}.  SMA observations of the CO emission from the HD\,141569 disk provide a similarly poorly contrained radial distribution, although a surface density that decreases with radius is favored \citep{fla16}.  

\subsubsection{Quantity, and Therefore Lifetime, of the Gas}

The path from CO flux to CO mass has not been straightforward.  Since observations of molecular gas are time consuming, discovery surveys and initial reconnaissance have typically been limited to single lines in the CO rotational ladder, with follow-up in multiple lines and isotopologues only recently gathered for some systems.  The excitation and optical depth of the CO gas is therefore highly uncertain for most systems.  

Knowledge of the CO mass and excitation affects our understanding of the gaseous component of debris disks in a number of ways.  For example, dynamical interactions between gas and dust may shape the morphology of debris disks and cause features like narrow eccentric rings that would instead be attributed to the presence of planets \citep[e.g.,][]{lyr13}, although only when the gas masses are sufficiently large, in the range of M$_\mathrm{gas}$/M$_\mathrm{dust} \gtrsim 1$ \citep[e.g.,][]{tak01,bes07}.  

One major question is that of the lifetime of the molecular gas, an upper limit for which can generally be estimated by dividing the mass of the gas by the rate at which it is photodissociated.  The lifetime of the gas is an important indicator of protoplanetary vs. second-generation origin, since a lifetime shorter than the age of the star would be a clear sign that the material must not be primordial.  This knowledge is also important in determining the rate at which CO must be replenished to retain a second-generation gas disk for a significant length of time, which can aid in distinguishing which methods of gas production are reasonably consistent with the required rates.  The main problem is that the rate at which CO is photodissociated depends not only on the production rate of the relevant photons by the star, but also on shielding: self-shielding by CO if found in sufficient quantities, shielding by atomic carbon if the C/CO ratio is sufficiently high \citep{mat17}, and shielding by H$_2$, which depends on the generally unknown CO/H$_2$ abundance \citep[e.g.,][]{vis09}.  

Several lines of evidence point to a low H$_2$ abundance in debris disk systems.  One is the typically low excitation temperatures indicated by multi-transition line measurements \citep[e.g.][]{kos13,fla16,mat17}.  These low excitation temperatures appear common, though not universal (49\,Ceti's disk-averaged excitation temperature is 44\,K, which is not clearly subthermal).  Another is the previously mentioned lack of CO absorption along the line of sight to 49\,Ceti, combined with the spatially unresolved vertical structure of its highly inclined CO disk, which indicates a molecular gas scale height smaller than that predicted for a primarily H$_2$ disk and suggests that the mean molecular weight is substantially larger than an ISM-based value \citep{hug17}.  Taking the excitation temperature measurement further with the help of a non-LTE line analysis, \citet{mat17} demonstrate that the line ratios in the $\beta$\,Pic system are incompatible with an ISM-like CO/H$_2$ ratio, and derive a CO+CO$_2$ ice abundance of at most $\sim$6\% in the molecular gas disk.  

Overall, the gas masses derived for these disks are low, of order a lunar mass or less for the CO component, and the lifetimes are short, typically $< 1000$ yr. 









\subsection{Origin}

Modeling efforts to understand individual disks have recently begun to converge into a more comprehensive theory of the origin and nature of gas in debris disks.  

There are a number of different proposed second-generation mechanisms for producing the observed gaseous component of debris disks.  These generally include photodesorption \citep{gri07}, collisional vaporization of icy dust grains \citep{cze07}, either at the site of a resonant point with an unseen planet or as the result of a recent major collision between planetary embryos \citep{jac14}, collisional desorption from comets \citep{zuc12}, or sublimation of comets \citep{beu90}.  While it is not possible to completely disambiguate these various scenarios based on existing observations, there are hints that particular scenarios may be favored for different systems.  For example, the highly asymmetric nature of the gas around $\beta$ Pic and Fomalhaut suggests either a recent collision or series of collisions due to orbital configurations that enhance the likelihood of collisions in a certain area of the disk \citep{den14,mat17b}.  There are some systems for which the cometary sublimation rate required to sustain the observed gas mass is uncomfortably large and requires the gas to be produced in a location where dust is not detected \citep[e.g. HD\,21997,][]{kos13}.  But overall, these are generally viable scenarios for liberating dust from its solid form on the surfaces of dust grains or the interiors of comets and sustaining a rate comparable to that needed to produce the observed CO line fluxes.  

An important recent attempt at unifying observations of CO, C, and O, which together comprise the dominant mass constituents of debris disks (with most of the C and O likely generated from photodissociation of CO), comes from \citet{kra17b}.  They use a semi-analytical model that assumes an input CO rate tied to the mass loss rate in the steady-state collisional cascade and released from exocomets with Solar System composition, and then allow the gas to evolve by photodissociation and viscous spreading.  Their line flux predictions are consistent with most of the detections of C, O, and CO lines in debris disks.  Systems with CO fluxes that exceed the predicted values by more than two orders of magnitude are interpreted as likely long-lived protoplanetary systems (HD\,21997, HD\,131835, HD\,138813).  Several CO nondetections in gas-bearing systems ($\eta$\,Tel, HD\,172555) are explained by this model as due to dust temperatures too high to retain volatiles or high stellar luminosity that shortens the CO lifetime.  The perplexing presence of C lines around intermediate-mass stars that should experience high levels of radiation pressure is explained for the first time by appealing to efficient self-shielding from the low levels of CO present in the disk.  

According to the \citet{kra17b} model, most gas-bearing debris disks can be comfortably interpreted within a second-generation scenario.  However, this model does not clearly explain the much higher prevalence of gas-bearing disks around intermediate-mass stars relative to solar mass stars.  The reliance on fractional excess luminosity to predict the CO flux is also somewhat at odds with the observational trend that millimeter continuum flux does not appear to be strongly correlated with the presence of CO at 10\,Myr \citep{lie16}. 

Overall, some gas-bearing debris disks are likely protoplanetary remnants, like HD\,21997, and some are clearly second-generation, like $\beta$\,Pic and Fomalhaut.  For other systems, while most can be plausibly explained by a second-generation scenario \citep{kra17b}, there are also some at younger ages and higher CO fluxes that might plausibly include a protoplanetary component as well.  



It seems clear that an unexplored dimension of the molecular gas component of debris disks is their chemistry.  Molecular line surveys therefore have the potential both to disambiguate the origin of individual disks, and also to potentially reveal the chemistry of exocomets.  Now that more edge-on gas-bearing debris disks have been identified, there are also opportunities for expanding the available data on the atomic component through a combination of absorption and emission line studies.  Since a large part of the uncertainty in the composition and origin of the disk comes from uncertainty in the molecular hydrogen abundance, it would seem fruitful to follow up on the detection of fluorescent H$_2$ emission from AU Mic in other nearby systems.  And finally, filling out the sample of objects observed in [\ion{O}{1}] emission would be useful both for studying the C/O ratio and also for hinting at the H$_2$O/CO abundance like in $\beta$\,Pic, although it is not possible to make such a measurement with currently available instrumentation.  The surprising discovery of significant quantities of molecular gas in debris disks is rapidly approaching a stage where we can draw conclusions about its role in late-stage planet formation and the composition of ices in distant solar systems.

\section{Variability in Debris Disks}
\label{sec:time_domain}

The study of source variability is an exciting and expanding area of debris disk research that to date has focused on a few intriguing sources of many ages. Below, we focus on variability observed toward main sequence debris hosts. Typically, observations of debris disks are only sensitive to the products of collisional evolution, namely the dust and gas, whereas several recent detections are sensitive to the parent planetesimal population. 

\subsection{Detections in Stellar Spectra}

The observation of time variable phenomena from debris disks is not new. The variation of features in spectra were observed around $\beta$\,Pic by \cite{beu90}; subsequent studies have confirmed ongoing detections of such transiting FEBs, or exocomets, in many A star systems \citep[e.g.,][and references therin]{eir16,mont17}.
In edge-on systems, such bodies can be expected to cross the line of sight to the star. In a comprehensive analysis of over 1000 spectra, \cite{kie14b} identified two different comet families within the $\beta$\,Pic disk by examining 493 individual cometary clouds. They found that one class of objects produced weak absorption features consistent with old, volatile-depleted comets in a mean motion resonance with a planet \citep{beu96}.  A second class produced strong narrow features consistent with younger, more volatile-rich bodies on similar orbits with a very narrow range of longitude of periastron, consistent with a family of comets formed by the distruption of one or more larger bodies. The authors liken these to the Kreutz family of comets in our Solar system, a group of sun-grazing comets with periods of $\sim 2300$ yr. 


\subsection{Variable Light Curves of Circumstellar Dust}

Over the past decade, variation in continuum debris disk emission over time has been detected from several sources in the form of sudden and rapid declines or increases in brightness. Disks observed to be abnormally bright in emission have been coined ``extreme" debris disks. 
Such disks may have higher fractional dust luminosities than are typical of debris disks, i.e., $f \ge 10^{-2}$ \citep{meng15}.
They are also rare, detected around just 1\% of 250 stars aged 30-130 Myr surveyed in the infrared \citep{bal09}.
These extreme debris disks are candidates for recent major collisional events that could evolve on orbital timescales, i.e., months to decades, which has stimulated a strong interest in monitoring of disks in the infrared to isolate these events. 

Monitoring of six such disks with {\it Spitzer} demonstrated that four show evidence of variable emission at 3.6 and 4.5 \micron\ \citep{meng15}. An additional very rapidly decaying disk around the 10 Myr debris disk host TYC\,8241\,2652\,1 is reported by \cite{mel12}, who attribute the decline of a factor of 30 in brightness over two years
to a collisional avalanche or runaway accretion. \cite{ost13} propose consideration of coronal mass ejections as a mechanism for removing dust in such sources, but the timescales of its efficacy are in fact too rapid, days rather than years as observed to date. 


The most well-characterized extreme debris disk is reported by \cite{meng14}. The star, ID8, showed a sudden increase in 3.6 and 4.5 \micron\ emission followed by a year-long decay.  At an age of 35 Myr, ID8 is in the right range for terrestrial planet formation.  Optical observations of the star over the same period showed no variation. The light curves at 3.6 and 4.5 \micron\ show a significant amount of variation, which the authors attribute to a new impact in the disk in 2013 -- a period when the source was not observable with {\it Spitzer}.  Based on the rise time of the light curve, the authors posit that the dust light curve and brightness increase records the aftermath of a violent collision in the terrestrial zone between two large bodies, likely 100-1000\,km in size. Such a collision would create an initially silica-rich vapour, from which solids would condense out with a minimum size of 100 \micron\ based on the decay rate.  
\cite{meng15} note that the largest objects fueling the collisional cascade are likely mm-sized bodies produced from the impact, which means that a much more rapid decay phase of 100\,days to 10\,yr is expected.  This decay time is much more rapid than the timescales expected for a typical debris disk evolving via a collisional cascade originating from 1-100\,km parent bodies \citep{wya02}. Analysis of the light curve further places the impacted body on a period of $\sim 71$\,days, or 0.33\,au from the star. 


This rapid timescale for depletion reconciles the issues of the low incidence of observed extreme debris disks with the {\it Kepler} data, which indicate that short period ($<85$ days) terrestrial planets are common around Sun-like stars \citep{fre13}. Theories  support the idea that collisions during the terrestrial planet formation phase should be common \citep{stew12}. If many systems are forming terrestrial planets, we might expect to see many extreme debris disks, but the rapid depletion time means the extreme debris disk emission is only briefly observable.  How frequently the collisions occur is then the limiting factor in comparing the presence of a planetary system with that of an extreme debris disk.

 
In the near-infrared, \cite{ert16} note that only one of the seven targeted sources for H-band excess in their study showed evidence for variability: HD\,7788. They note, however, that since their detections are typically near their sensitivity limits, the study of variability remains challenging at these wavelengths and corroborating observations are needed. Other systems show that the H-band emission persists for years, meaning the hot dust close to the stars isn't eroding on very rapid timescales.

\subsection{Variability in Stellar Light Curves}

While it is easier in our own solar system to observe the planetesimals (comets and asteroids) that are the parent bodies that produce the dust in the debris disk, in other systems, we have traditionally been restricted to observing the products of collisions: dust, and now secondary gas as well.  However, there are now several cases in the literature of distinct, stellar light curves revealing the presence of complex circumstellar material around nearby stars.  These transit curves suggest the presence of planetesimal-scale bodies in orbit around their parent stars. 

The star KIC 8462852, also known as TYC\,3162-655-1 or ``Boyajian's Star", is unique in the {\it Kepler} stellar spectral catalogue, showing complex, non-periodic behavior in its light curve with dimming events up to 20\% of the stellar brightness.  \cite{boy16} provide a comprehensive analysis of the light curve as well as ancillary data to provide insight into the system. The variations observed are of astrophysical origin, and companions and stellar variation are ruled out as the source. The star shows no evidence of an IR excess, and given its distance (400\,pc) current technology is unable to detect the modest amounts of dust seen in nearby debris-bearing systems, such as Fomalhaut.  \cite{boy18} use additional data to constrain the material to be optically thin with the majority of the material comprised of small ($\le 1$ \micron\ bodies) consistent with their original favored interpretation of the data as a debris swarm from the breakup of one or more massive comets passing in front of the star, with an original body, or bodies, having at least 0.3\% the mass of Ceres.  \cite{ball17} instead model the system using a large ($5 R_{Jup}$) ringed body shepherding a system of Trojan asteroids clustered at the L5 region, with the former producing the first, deep transit and the smaller bodies producing the period of multiple, shallow dimmings 700\,days later. \cite{wri16b} provide a summary of the families of ideas that have been put forth to explain the emission, which in addition to the ideas above, include artificial sources, although the constraints on optical depth and increased reddening during short-term dimming events reported by \cite{boy18} strongly favor obscuration by circumstellar dust as the source of the variations in the light curve. 



\cite{rap18} report the detection of transits in the continuum emission of two {\it Kepler} targets. The authors report six transits over the {\it Kepler} observing period of four years toward KIC\,3542116 and a single transit event toward KIC\,11084727. Both stars are early F type stars and considerably fainter for spectral monitoring compared to the A stars discussed above. The search of the {\it Kepler} database was done visually, but the authors note that more such extended transits, showing dips at the $\ge 0.1$\% level and lasting for hours to days, should be easily assessed in data from the upcoming TESS mission.  These signatures could indicate the presence of obscuring dust and planetesimals in the host systems, enabling many more systems like KIC\,8462852 to be detected. 

\subsection{Directly Imaged Features}

One of the enticing promises of debris disk science is that substructure imaged in disks could reveal the presence of unseen planets. The $\beta$\,Pic warp, while not time variable, was one of the first clear cases of disk structure revealing the presence of an underlying planet. Hence, searches for structure have traditionally focused on imaging of clumps or asymmetries in disks, which can then be modeled to reveal the underlying planetary architecture \citep{wya08,mat14c}. 
In general, the orbital timescales of Kuiper Belt analogues are too long to hope to observe rotation of any substructures in real time through direct imaging, although features in warm dust belts on much shorter orbital periods may be imaged in future, using infrared facilities with high angular resolution. 

Time variable features have in fact been detected at 10\,au scales in images of the AU\,Mic debris disk.  \cite{boc15} report the detection of five time-varying scattered light features moving outward in projection on the south east side of the AU\,Mic disk. They find that these features persist over 1-4 years and that their projected speeds in some cases exceed escape velocity, suggesting they are unbound. \cite{boc15} further determined that the features are radially offset in each epoch, with the furthest projected features having the highest velocities. Changes in disk colour have been measured where the fast-moving features are detected, suggesting that their passage is changing the distribution of sub-micron grains at those locations \citep{lom17}. 

\cite{sez17} model the features in the AU\,Mic disk as arising from proper motion of the dust itself.  They suggest the dust is generated either by resonance with a parent body (i.e., a planet on a Keplerian orbit), or at the location of a recent giant collision that has generated a swarm of smaller bodies \citep{jac14}. These bodies then release the dust emission in separate events, which may or may not be grouped in time. In both scenarios, the best fit models place the dust inside the radius of AU\,Mic's planetesimal belt \citep[40\,au,][]{mac13}, even as close as 8\,au in the case of an orbiting planet. \cite{sez17} note that the arches seen above the disk are composed of very small grains that experience very strong stellar wind relative to gravitational forces. 

\begin{textbox}[t]\section{AVALANCHE CASCADE}
A dust avalanche or avalanche cascade typically refers to the collision of grains on hyperbolic trajectories with larger parent bodies still bound within the disk, leading to the creation of an exponential increase in the density of impactor-sized grains. The newly created grains, subject to the same forces, are rapidly accelerated and collide with more bound material in the disk. This chain reaction results in an exponential amplification of dust material on unbound trajectories in the disk \citep{gri07}. 
\end{textbox}

\cite{chi17} explain the observed features as clouds of sub-micron sized dust produced in an avalanche cascade (see Sidebar), 
precipitated by the disruption of a $< 400$ km sized body (mass $<10^{-4}$ $M_\oplus$) in the AU\,Mic birth ring less than $3 \times 10^4$ years ago and resulting in the creation of a secondary debris ring. The avalanche is launched from the intersection of the birth and secondary rings (a relatively small intersection of only a few au in size), providing a natural cadence to the launch events, and disk rotation carries the avalanche cascade around the disk to the southeast, as observed, providing new mass into the cascade. Because the clouds are composed of sub-micron sized grains, they experience a strong ram force, exceeding 20 times the stellar gravity.  \cite{chi17} predict not only the color changes seen by \cite{lom17}, but also the potential to image the secondary ring if it has sufficient mass. They also note that the vertical height of the features may vary with the stellar magnetic cycle.  
In conclusion, the detection of rapid variations in brightness in debris disks, and now time variable features in disk images, could have origins in giant impacts and avalanche cascades, which alter disks on timescales that can be observed and monitored to reveal the internal dynamics of debris disks. It is worth noting that both \cite{sez17} and \cite{chi17} present scenarios to explain the time-variable emission from the AU\,Mic disk without requiring a perturbing planet: time variable features do not rely on planet-related dynamics.

\section{Planet-Disk Connections}
\label{sec:planet-disk_interaction}

Given the role played by debris disks in planetary formation and evolution, one may expect some tangible signs such as correlations between the physical properties of debris disks and planetary systems. To date, there are statistically robust correlations between the presence of planets and the presence of debris around stars (Section\,\ref{sec:planet_disk_demographics}), but there are some strong suggestions in the existing data that disks are more frequent and brighter in systems with planets below Jovian masses \citep{mat14c}.  In addition, case studies of systems where planets and dust coexist can be extremely valuable in revealing the dynamics of planetary systems. Such studies are still in their early stages, since the ability to image planets and debris has been increasing at a rapid rate over the past decade.  There are essentially two subsets of systems that have been studied using imaging techniques to understand interactions between disks and planets: systems where an RV planet (or planets) at small radii coexist with debris at large radii (Section\,\ref{sec:rv_planets}), and systems where both directly imaged planets and directly imaged debris dust coexist in the same system (Section\,\ref{sec:direct_obs}).  

\subsection{Disk Correlations with Planetary Systems}
\label{sec:planet_disk_demographics}

Assessing the demographics of systems with debris disks and planets is challenging. The strongest overlap in observationally favored targets for both debris disks and directly imaged planets lies in young systems. Currently, thermal imaging of exoplanets is limited to young systems and the rates of detection are higher for young debris disks as well (see Section \ref{sec:demographics}). 
There is tentative evidence for a higher occurrence rate of debris disks in systems hosting directly imaged planets \citep{mesh17}, albeit with the caveats of small number statistics and potential selection biases.  For example, some imaging surveys target disk hosts as more favorable for planetary detections \citep{bow16}.
The vast majority of known planets have been detected by methods via the transit method using {\it Kepler}; however, they are typically too far away for faint debris disks to be detectable. 

Theories may predict correlations or anticorrelations between debris disks and planets, depending  primarily on the masses and orbits of the planets. For example, \cite{ray12} predict that orbital instabilities in systems of giant planets will produce an anticorrelation between eccentric Jupiters and debris disks, but a correlation between terrestrial planets and debris disks. 
Large statistical samples, constructed to be as bias-free as possible, are the key to establishing robust statistically relations such as these. Recent analyses of {\it Spitzer} \citep{kos09,bry09}, {\it Herschel} \citep{mor15} and {\it WISE} \citep{mal17} data have not established statistically significant correlations of planets and debris, however. Searches of planet host samples for debris characteristics have not found strong correlations either \citep{wit15}. \cite{mor12} searched 591 planetary systems around main sequence and evolved stars for evidence of infrared excess with {\it WISE} and found an excess rate of 2.6\%, but the rate dropped to 1\% if only main sequence stars were considered. 
Smaller samples and analysis of systems in which planets are known do consistently suggest that the presence of debris correlates with the presence of a low mass planet (or planets) in the system, but anticorrelates with the presence of a planet of Jupiter mass or higher \citep[i.e.,][]{mar14}, consistent with the findings of \cite{mal15}.
This apparent contradiction stems from the nature of the samples: construction of large, bias-free samples is challenging \cite[e.g.,][]{mor15}, and the sample of stars that has been uniformly searched for debris and exoplanets is too small to yield a definitive correlation. 

\cite{wit15} suggest a weak trend in brightness of debris emission as a function of planetary mass. 
By contrast, \cite{mor15} find no statistically significant correlation between the fraction of low- or high-mass planet host stars with disks -- or with the brightness of disks when present -- compared to stars with no evidence for planetary companions. In addition, the presence of single or multiple planets in a system appears to have no impact on the detection fraction or brightness of associated debris disks \citep{mor15}; we note however that \cite{mal12} find that the presence of planets is correlated with a lower dust luminosity in that study.  

Several studies have looked for differences between the population of stars hosting debris disks and those with no evidence of debris. Known metallicity relationships for planetary systems suggest a relationship between disk incidence or brightness and stellar metallicity.  Many statistical studies have reaffirmed the established correlation of giant planets with higher stellar metallicities \citep[e.g.,][]{mal12}. Early studies suggested that there was no corresponding correlation of debris disks with metallicity \citep{bei06}. Several more recent studies have pointed out that there could be a deficit of stars with discs at very low metallicities (i.e., $-0. 50 <  [\mathrm{Fe/ H}] < -0. 20$) with respect to stars without detected discs \citep[][and references therein]{mon16b}. 
\cite{mal15} investigated whether any solar type (FGK) stars hosting debris show any chemical peculiarities that could relate to the planet formation process. They found no significant differences in metallicities or abundances between stars that host debris disks and those which show no evidence of debris or planetary companions.  One class of stars that does appear to exhibit a relation between stellar abundance anomalies and debris disks are the $\lambda$\,Boo class. \cite{dra16b} suggest the stellar chemical signatures exhibited by $\lambda$\,Boo stars are harbingers for other dynamical activity in such systems, related to the deposition of material in the stellar atmosphere and the presence of a bright debris disk, such as planetary perturbations.

\subsection{RV Planet systems with Debris Disks at Large Radial Separations}
\label{sec:rv_planets}

A number of imaged debris disks are seen in systems with detected RV planets at much smaller radii (generally less than a few au compared to radii of tens to hundreds of au for the debris).  In general, the relatively young stellar ages and the presence of bright dust at large radii suggest that many of these systems are stirred by planets rather than any self-stirring mechanism \citep{moo15b}.  While the approximate characteristics of the putative planets can be inferred via the RV method, our knowledge is typically limited to a minimum mass (or an inferred mass using the inclination derived from the aspect ratio of the debris disk, assuming that the disk and planet(s) are coplanar) and orbital period / semimajor axis. 

At the time of writing, fewer than a dozen systems with RV planets also have had spatially resolved observations of their outer disks.  The most prominent examples include q$^1$\,Eri, $\epsilon$\,Eri and GJ\,581 \citep{lis10, cha16, les12}.
There is a catalog of stars with both debris disks and planets currently maintained by Universidad Autonoma de Madrid \footnote{http://svo2.cab.inta-csic.es/vocats/debris2/}, which should serve as a resource for updates to this list.

In most currently known systems, the outer disk is only marginally resolved, and the inner edge of the debris ring is not spatially resolved at all.  Its location is typically only constrained by SED models, which depend on the SED-fitted temperature (and less well-constrained size and composition) of the dust grains. 
While there is some indication that the disks surrounding planets with larger minimum masses may have a larger radial extent \citep{ken15}, the numbers of systems sampled are far too small to draw any firm conclusions.
At this stage, the most that can be said is that there is a tendency for the derived inner radius of the debris disk (typically tens to hundreds of au) to be too large to be reasonably sculpted by the known planet(s) within the age of the system (typically tens to hundreds of Myr), implying either that there are dynamical effects not taken into account by the chaotic zone defined by \cite{wis80} or that additional planets are required to clear the region interior to the edge of the debris disk. One possible exception to this trend is HD\,69830, which is host to three Neptune-mass planets and an asteroid belt analog resolved with mid-IR interferometry \citep[e.g.,][]{smi09}.  

\subsection{Direct Observations of Planet-Disk Interaction}
\label{sec:direct_obs}

Several systems have now been demonstrated to host both a directly imaged planet(ary system), some with measured orbital properties, as well as a debris disk that has been imaged at high angular resolution. Interactions between the disk and known planet(s) can now be directly observed, and the properties of the disk can, when interpreted through an appropriate theoretical framework, place dynamical constraints on planet properties like mass. Here we review the known systems and current efforts to understand their properties.  


The masses of directly imaged planets are highly uncertain and estimated primarily based on uncalibrated theoretical models of hot- and cold-start evolution that predict the luminosity (or temperature) of the planet as a function of time.  The mass derived from the observed flux therefore depends not only on the poorly understood initial conditions of the planetary system, but also on the stellar age, which can be highly uncertain, especially for isolated young A stars.  Observations of the interaction between the planet(s) and debris disk(s) have the potential to help calibrate the uncertain flux-based planetary mass estimates, since the distance between the planet and the disk is a sensitive function of the mass of the planet (and its orbital properties).  For example, \citet{mal98} derive the width of a planet's ``chaotic zone,'' which is expected to be cleared of dust, as $\Delta a \simeq 1.4 a_p (M_p/M_*)^{2/7}$, where $\Delta a$ is the difference between the planet and disk semimajor axis, $a_p$ is the semimajor axis of the planet, $M_p$ is the mass of the planet, and $M_*$ is the mass of the star.  The width of the chaotic zone can also depend on the eccentricity of the bodies \citep[e.g.,][]{chi09,mus12} as well as time-variable orbital parameters in multi-planet systems \citep[e.g.,][]{mor10}.  The relationship has also been evaluated for different timescales and outcomes and extended into the brown dwarf regime \citep{mor15b}.  Planets and disks can interact in even more complex ways, and the long-term stability of the planetary system may depend on the underlying mass of the planetesimal disk producing the debris \citep{moo13b}.  

The HR\,8799 system is often considered to be a scaled-up Solar System analog, since it contains four giant planets orbiting between two debris belts, similar to the architecture of the outer Solar System although with larger radii and masses \citep[e.g.,][]{su09}.  Due to the large semimajor axes and orbital periods of the planets, there is still some uncertainty on their orbital parameters as well as the degree of coplanarity between the star, disk, and planetary system.  Given the existing data, there are stable solutions for the planetary orbits involving a 1:2:4:8 mean-motion resonance between the four observed planets \citep{god14}.  While the semimajor axis of the planetary orbits depends on the inclination, estimates for which range from 25$^\circ$ to 40$^\circ$ depending on the methodology \citep{wri11,mat14b,kon16b,wer17}, 
recent ALMA imaging resolves the inner edge of the dust disk and permits an estimate of the system geometry \citep{boo16}.  With the addition of short-baseline data from the Submillimeter Array, \citet{wil18} place the first dynamical limit on the mass of planet b, yielding a value of $5.8^{+7.9}_{-3.1}$\,M$_\mathrm{Jup}$.  Since the chaotic zone extent is so steeply dependent on the planet mass, increasing the precision of the inner edge measurement by even a modest amount will lead to improved constraints on the planet mass.  


In many ways $\beta$\,Pic is the best example we have of a well-understood interaction between planet and debris disk.  High-resolution imaging of its debris disk has been conducted across a wide range of wavelengths \citep[for recent examples, see e.g.][]{van10,lag12,den14}, so that the geometry of both large and small grains is well understood, and the planet's orbital properties are reasonably well constrained \citep[e.g.,][]{wan16}.  After the discovery of the planet \citep{lag09}, two numerical models demonstrated that planet b was primarily or solely responsible for producing the scattered light warp \citep{daw11}, and demonstrated that the orbital properties of planet b could plausibly explain not only the x-shaped morphology observed in scattered light but also the central clearing observed in thermal emission at millimeter wavelengths and potentially also the presence of a bright clump of CO emission on one side of the disk \citep{nes15}.  Difficulties in placing constraints on the planet mass via these numerical simulations arise due to the uncertainty in the age of the system as well as the unknown timing of the planet's formation or scattering onto its current orbit. 


HD\,95086 hosts a system similar to that of HR\,8799 in several ways: it has a two-temperature SED indicating inner and outer debris belts \citep{su15} and at least one directly imaged giant planet orbiting between the belts \citep{ram13,gal14}.  While the orbital parameters of planet b seem to be inconsistent with its ability to clear the gap inferred from the SED measurement \citep{ram16}, recent ALMA imaging has placed more stringent constraints on the inner edge of the debris belt \citep{su17}.  The inner edge of the debris belt is farther than the standard chaotic zone approximation would predict, leaving three possibilities: (1) the orbit of HD\,95086\,b may be more eccentric than suggested by either the ALMA or astrometry measurements, (2) the disk and planet may not be coplanar, or (3) there may be a second planet orbiting exterior to HD\,95086\,b but interior to the inner edge of the outer debris belt.  While an additional planet is possible, its properties are fairly well restricted by the existing observations, and it would need to be low-mass (0.2-1.5 M$_{Jup}$) and low-eccentricity \citep{su17}.  


Even within the small sample of known stars with directly imaged disks and planets, HD\,106906 stands out as unique because its perturber is external rather than internal.  After the discovery of the companion object HD\,106906\,b, which orbits at a projected separation of $\sim$650\,au \citep{bai14}, imaging of the disk revealed an extremely asymmetric dust distribution \citep[][and references therein]{wu16}.  The ability to study planet-disk interactions in this system has provided a unique opportunity to investigate the formation mechanism of wide-separation companions, since some formation scenarios are ruled out by the timescales and geometry of the system when interpreted in the context of numerical simulations.  Since external perturbers can be extremely destructive to debris dust, the timescale of evolution of the debris disk around this 13\,Myr-old star can constrain the timescale of the process that created the external perturber.  \citet{jil15} use this observation to argue for a scenario in which the current configuration resulted from planet-planet scattering from the inner disk, although their model predicts that large-amplitude oscillations should be excited through a mechanism similar to Kozai-Lidov.  \citet{nes17} conduct N-body simulations demonstrating that a long-term exterior orbit resulting from an {\it in situ} formation mechanism can reproduce the observed asymmetries without vertically exciting the dust.  


Finally, upcoming systems of interest in this category include two stars orbited by a brown dwarf companion located interior to a debris ring.  The recently discovered companions to HD\,206893 \citep{mil17} and HR\,2562 \citep{kon16} are both directly imaged brown dwarfs orbiting interior to a debris ring inferred from SED fitting.  Neither disk has yet been imaged at high resolution and fidelity (though both are accessible to ALMA), and it is expected that future imaging will provide useful constraints on the masses of the brown dwarf companions.  




\begin{summary}[SUMMARY POINTS]

\begin{enumerate}
\item Debris disks are a common phenomenon around main sequence stars, with current detection rates at $\sim 25$\% despite the limited sensitivity of instrumentation (for Kuiper Belt analogues, a few $\times$ Solar System levels for even the nearest stars, and orders of magnitude from Solar System levels for exozodis) indicating that this figure is unquestionably a lower limit.  Debris disks are more commonly detected around early-type and younger stars, and they frequently show evidence of two dust belts, like the Solar System.
\item High-resolution imaging of structures in outer debris disks reveals a rich diversity of structures including narrow and broad rings, gaps, haloes, wings, warps, clumps, arcs, spiral arms, and eccentricity.  There is no obvious trend in disk radial extent with stellar spectral type or age.  Most of the observed structures can be explained by the presence of planets, but for most structures there is also an alternative proposed theoretical mechanism that does not require the presence of planets.  
\item The combined analysis of a debris disk's SED with high-resolution (polarized) scattered light images and thermal emission maps is a powerful method to constrain the properties of its dust. Despite some shortcomings in current models, a coherent picture arises in which the grain size distribution is characterized by a power law similar to that predicted from collisional models, albeit with a minimum grain size that is typically a few times larger than the blowout size. Beyond the ubiquitous silicates, constraints on dust composition remain weak. Remarkably, dust in most debris disks appears to be characterized by a nearly universal scattering phase function that also matches that observed for dust populations in the Solar System, possibly because most dust grains share a similar aggregate shape.
\item Many debris disk systems harbor detectable amounts of atomic and/or molecular gas.  Absorption spectroscopy of a few key edge-on systems reveals volatile-rich gas in a wide variety of atomic tracers, and there is evidence of an enhanced C/O ratio in at least some systems.  Emission spectroscopy reveals that CO gas is common, especially around young A stars.  The quantities of molecular gas found in most systems are likely insufficient to strongly affect the dust dynamics or planet formation potential.  While the origin of the gas is still a matter of discussion, there is increasing evidence that the sample is not homogeneous.  The molecular gas in some systems is clearly second-generation like the dust, while in other systems the disk is likely to be a ``hybrid'' with second-generation dust coexisting with at least some primordial gas.  
\item Detection of time variable features in debris disk SEDs, images and light curves on timescales of days to years provide a window into ongoing dynamical processes in the disks and potentially the planetesimal belts as well.  
\item Planet-disk interaction is now directly observable in a handful of systems that contain both a directly imaged planet(ary system) and a spatially resolved debris disk.  These rare systems provide valuable opportunities to place dynamical constraints on the masses of planets, and thereby to calibrate models of planetary atmospheres. 
\end{enumerate}
\end{summary}

\begin{issues}[FUTURE ISSUES]
\begin{enumerate}
\item {\it M star exploration:}  Most studies of debris disk structure and composition have so far focused on F, G, K, and A stars.  While the incidence of M star debris disks appears to be low, the known disks include some of the most iconic systems including AU\,Mic, and their gas and dust properties are poorly understood. Given then high frequency of terrestrial planets around M dwarfs, understanding their debris disks is a high priority.
\item {\it Physics of Debris Disk Morphology:} Multiwavelength imaging is beginning to untangle the underlying physical mechanisms sculpting debris disk structure.   Dynamically-induced disk structures from planets or stellar flybys should affect even the largest grains, whereas those caused by the ISM or radiation effects tend to produce the largest effect on the smallest grains.  As limits on gas emission improve, gas becomes a less plausible mechanism for sculpting narrow and eccentric rings in many systems. Imaging of structures across the electromagnetic spectrum, particularly comparison of ALMA thermal imaging with high-contrast scattered light imaging, will be key for moving from categorization of structure to conceptualization of the underlying physics.  
\item {\it The connection between hot, warm and cold dust belts:} While multiple components are often detected in debris disk systems, identifying correlations between them has not yet delivered a fully coherent picture. Understanding the relative importance of in situ dust production and dust migration, as well as the physical mechanisms explaining the latter, remains an open question. Future observations sensitive enough to detect dust at intermediate radii between separate belts will help in clarifying this issue.
\item {\it Variability:} While studies of gas absorption variability have a long and fascinating history -- revealing the dynamics and composition of falling evaporating bodies -- studies of the variability of dust, both directly in emission and indirectly through stellar variability, are in their infancy.  Currently there are a handful of known instances, each of which has multiple proposed explanations ranging from planetary dynamics to collisional avalanches.  While large variations in integrated infrared light are rare, they are important for understanding stochastic events.  Furthermore, the recent detection of fast-moving scattered light features in the dust around AU\,Mic suggests that at least in some cases systems previously assumed to be static might be variable on more subtle spatial and flux contrast scales.  Debris disk variability is an area that is wide open for discovery and modeling, particularly in the approaching epoch of {\it JWST}.  
\item {\it Gas statistics and chemistry:} As the number of gas detections in debris disks increases, the opportunity for characterization of atomic and molecular abundances of the gas also grows.  Clear opportunities include surveys designed to measure incidence of gas emission in a statistically meaningful way, in addition to detection of molecules other than CO and the corresponding opportunity to characterize the composition of exocometary gas.  It would also be beneficial to amass rich data sets comparable to that of $\beta$\,Pic for a sample of several different objects so that modeling of both the atomic and molecular components can begin to explore the diversity of exosolar gas composition and better distinguish between primordial and secondary gas origins. 
\end{enumerate}
\end{issues}

\section*{DISCLOSURE STATEMENT}
The authors are not aware of any affiliations, memberships, funding, or financial holdings that might be perceived as affecting the objectivity of this review. 

\section*{ACKNOWLEDGMENTS}
The authors wish to thank the following people for providing feedback and commentary on the article: Christine Chen, Kevin Flaherty, Paul Kalas, Grant Kennedy, Sasha Krivov, Luca Matra, Aki Roberge, Kate Su and Ewine van Dishoeck. The authors thank Eugene Chiang and Mark Wyatt for points of clarification. The authors are also grateful to the following people who agreed to share data used in the preparation of the figures in this review: Dan Apai, John Carpenter, Cail Daley, Bill Dent, Carsten Dominik, Jane Greaves, Paul Kalas, Markus Kasper, Mihoko Konishi, Meredith MacGregor, Sebastian Marino, Julien Milli, Johan Olofsson, Glenn Schneider.
A.M.H. gratefully acknowledges support from NSF grant AST-1412647, and G.D. from NSF grants AST-1413718 and AST-1616479.  

%


\appendix{}



\newcommand{\cmark}{\ding{51}}%
\newcommand{\xmark}{\ding{55}}%
\renewcommand{\micron}{$\upmu$m}

{\bf \large A. Debris disks database used in this review} \\

Several figures presented in this review are based on the aggregation of many published results, both from survey-type and single-target studies. From these studies, we have assembled a database of 152 debris disks systems for which at least one of the following types of observation is currently available:
\begin{itemize}
\item[--] spectral data probing one or more (sub)millimeter CO rovibrational emission lines ($\uplambda_\mathrm{c}=$\,870\,\micron, 1.3mm, 2.6mm);
\item[--] spectral data probing the far-infrared [\ion{C}{2}] emission line ($\uplambda_\mathrm{c}=$\,157\,\micron);
\item[--] spectral data probing the far-infrared [\ion{O}{1}] emission line ($\uplambda_\mathrm{c}=$\,63\,\micron);
\item[--] well-resolved visible and/or near-infrared scattered light images ($\uplambda \leq$\,4\,\micron);
\item[--] well-resolved mid-infrared continuum thermal emission maps (10\,\micron\,$\leq \uplambda \leq$\,25\,\micron);
\item[--] well-resolved far-infrared continuum thermal emission maps (70\,\micron\,$\leq \uplambda \leq$\,160\,\micron);
\item[--] well-resolved (sub)millimeter continuum thermal emission maps ($\uplambda \geq$\,800\,\micron);
\item[--] estimated Henyey-Greenstein phase function $g$ parameters from modeling of scattered light images.
\end{itemize}

The database is presented in \textbf{Table\,\ref{tab:database}}, with all numbered references listed below. For each system, the target's spectral type and distance, as well as the best estimate of its age, are provided. In some cases, there are significant uncertainties on these quantities (stellar ages are notoriously difficult to establish) and they should be used with caution. The database is also available in electronic format from the editor's website.\\

\begin{landscape}
\begin{longtable}{ccccc|ccc|cccc|c}
\caption{Gas and Dust Data Compilation For Debris Disks\label{tab:database}}\\
\hline
Target & Alias & Sp.T. & Dist. & Age & \multicolumn{3}{c|}{Gas Emission Lines$^\mathrm{a}$} & \multicolumn{4}{c|}{Continuum Images$^\mathrm{b}$} & Dust\\
&&& (pc) & (Myr) & CO & [\ion{C}{2}] & [\ion{O}{1}] & OIR & MIR & FIR & (sub)mm & $g^\mathrm{ac}$\\
\hline
\endfirsthead
\multicolumn{13}{c}%
{\tablename\ \thetable\ -- \textit{Continued from previous page}} \\
\hline
Target & Alias & Sp.T. & Dist. & Age & \multicolumn{3}{c|}{Gas Emission Lines$^\mathrm{a}$} & \multicolumn{4}{c|}{Continuum Images$^\mathrm{b}$} & Dust\\
&&& (pc) & (Myr) & CO & [\ion{C}{2}] & [\ion{O}{1}] & OIR & MIR & FIR & (sub)mm & $g^\mathrm{c}$\\
\hline
\endhead
\hline \multicolumn{13}{r}{\textit{Continued on next page}} \\
\endfoot
\hline
\endlastfoot
HD\,105 & HIP\,490 & G0V & 39 & 30 & \xmark$^{30,73}$ & -- & \xmark$^{18}$ & -- & -- & -- & -- & -- \\
HD\,377 & HIP\,682 & G2V & 39 & 180 & \xmark$^{30,73}$ & -- & \xmark$^\dagger$ & \cmark$^{9}$ & -- & -- & -- & \cmark$^{9}$ \\
HD\,627 & HIP\,876 & B7V & 400 & 75 & \xmark$^{17}$ & -- & -- & -- & -- & -- & -- & -- \\
HD\,2772 & $\uplambda$\,Cas & B8V & 109 & 180 & \xmark$^{79}$ & -- & -- & -- & -- & -- & -- & -- \\
HD\,3003 & $\upbeta^3$\,Tuc & A0V & 46 & 45 & \xmark$^{30}$ & -- & \xmark$^{18}$ & -- & -- & -- & -- & -- \\
HD\,3670 & CD\,-53\,130 & F5V & 76 & 30 & \xmark$^{71}$ & -- & \xmark$^\dagger$ & -- & -- & -- & -- & -- \\
HD\,6028 & HIP\,4844 & A3V & 91 & 500 & \xmark$^{17}$ & -- & -- & -- & -- & -- & -- & -- \\
HD\,9672 & 49\,Cet & A1V & 59 & 40 & \cmark$^{32,34}$ & \cmark$^{18,32}$ & \xmark$^{18}$ & \cmark$^{10}$ & -- & \cmark$^{33,85}$ & \cmark$^{34}$ & \cmark$^{10}$ \\
HD\,10476 & 107\,Psc & K1V & 8 & 4900 & \xmark$^{27}$ & -- & --  & -- & -- & -- & -- & -- \\
HD\,10700 & $\uptau$\,Cet & G8V & 3.7 & 8200 & -- & -- & \xmark$^\dagger$ & -- & -- & -- & \cmark$^{54}$ & -- \\
HD\,10939 & q$^2$\,Eri & A1V & 62 & 350 & \xmark$^{73}$ & \xmark$^{29}$ & \xmark$^{29}$ & -- & -- & -- & -- & -- \\
HD\,12039 & DK\,Cet & G4V & 42 & 30 & \xmark$^{30}$ & -- & -- & -- & -- & -- & -- & -- \\
HD\,14055 & $\upgamma$\,Tri & A1V & 36 & 300 & \xmark$^{79}$ & -- & --  & -- & -- & -- & -- & -- \\
HD\,15115 & HIP\,11360 & F2 & 45 & 12 & \xmark$^{71}$ & -- & \xmark$^\dagger$ & \cmark$^{65,90}$ & -- & -- & -- & \cmark$^{65,86,87}$ \\
HD\,15745 & HIP\,11847 & F0 & 64 & 12 & \xmark$^{41}$ & -- & -- & \cmark$^{90}$ & -- & \cmark$^{33}$ & -- & -- \\
HD\,16743 & HIP\,12361 & F0/2III & 59 & 30 & \xmark$^{73}$ & -- & \xmark$^\dagger$ & -- & -- & -- & -- & -- \\
HD\,17081 & HIP\,12770 & B7IV & 135 & 135 & \xmark$^{17}$ & -- & -- & -- & -- & -- & -- & -- \\
HD\,17206 & $\uptau^1$\,Eri & F6V & 14 & 710 & \xmark$^{27}$ & -- & -- & -- & -- & -- & -- & -- \\
HD\,17848 & $\nu$\,Hor & A2V & 51 & 100 & \xmark$^{30,73,79}$ & -- & -- & -- & -- & -- & -- & -- \\
HD\,21563 & HR\,1053 & A4V & 180 & 1000 & \xmark$^{30}$ & -- & -- & -- & -- & -- & -- & -- \\
HD\,21620 & HIP\,16424 & A0V & 140 & 250 & \xmark$^{79}$ & -- & -- & -- & -- & -- & -- & -- \\
HD\,21997 & HR\,1082 & A3IV/V & 72 & 45 & \cmark$^{45,71}$ & \xmark$^{29}$ & \xmark$^{29}$ & -- & -- & \cmark$^{74}$ & \cmark$^{72}$ & -- \\
HD\,22049 & $\upepsilon$\,Eri & K2V & 3.2 & 75 & \xmark$^{50}$ & -- & \xmark$^\dagger$ & -- & -- & \cmark$^{28}$ & \cmark$^{8}$ & -- \\
HD\,23362 & HIP\,17473 & K5III & 310 & 2200 & \xmark$^{17}$ & -- & -- & -- & -- & -- & -- & -- \\
HD\,23642 & V1229\,Tau & A0V & 110 & 115 & \cmark$^{79}$ & -- & -- & -- & -- & -- & -- & -- \\
HD\,24966 & HIP\,18437 & A0V & 104 & 360 & \xmark$^{79}$ & -- & -- & -- & -- & -- & -- & -- \\
HD\,27290 & $\upgamma$\,Dor & F4III & 16 & 620 & \xmark$^{27}$ & -- & -- & -- & -- & -- & -- & -- \\
HD\,28001 & CD\,-57\,894 & A4V & 310 & 500 & \xmark$^{30}$ & -- & -- & -- & -- & -- & -- & -- \\
HD\,30422 & EX\,Eri & A3IV & 58 & 13 & \xmark$^{79}$ & -- & -- & -- & -- & -- & -- & -- \\
HD\,30447 & HIP\,22226 & F3V & 80 & 30 & \xmark$^{41,71}$ & -- & -- & -- & -- & -- & -- & -- \\
HD\,31295 & $\uppi^1$\,Ori & A0V & 37 & 120 & \xmark$^{79}$ & -- & -- & -- & -- & -- & -- & -- \\
HD\,32297 & HIP\,23451 & A0V & 112 & 30 & \cmark$^{29}$ & \cmark$^{29}$ & \xmark$^{29}$ & \cmark$^{13,39}$ & \cmark$^{24}$ & -- & -- & \cmark$^{4,6}$ \\
HD\,33636 & HIP\,24205 & G0V & 29 & 2500 & -- & \xmark$^\dagger$ & \xmark$^\dagger$ & -- & -- & -- & -- & -- \\
HD\,35650 & HIP\,25283 & K6V & 18 &  150 & -- & -- & -- & \cmark$^{9}$ & -- & -- & -- & -- \\
HD\,35817 & HIP\,25613 & F5 & 155 & 5000 & -- & -- & \cmark$^\dagger$ & -- & -- & -- & -- & -- \\
HD\,35841 & BD\,-22\,1109 & F5V & 96 & 30 & \xmark$^{71}$ & -- & \xmark$^\dagger$ & -- & -- & -- & -- & -- \\
HD\,35850 & AF\,Lep & F8V & 27 & 12 & \xmark$^{79}$ & -- & -- & -- & -- & -- & -- & -- \\
HD\,36546 & HIP\,26062 & A0 & 114 &  6 & -- & -- & -- & \cmark$^{15,78}$ & -- & -- & -- & \cmark$^{15}$ \\
HD\,38206 & HIP\,26966 & A0V & 75 & 30 & \xmark$^{73,79}$ & -- & \xmark$^{29}$ & -- & -- & -- & -- & -- \\
HD\,38207 & BD\,-20\,1162 & F2V & 93 & 530 & \xmark$^{71}$ & -- & \xmark & -- & -- & -- & -- & -- \\
HD\,38678 & $\upzeta$\,Lep & A2IV/V & 22 & 370 & \xmark$^{18}$ & \xmark$^{18}$ & \xmark$^{18}$ & -- & -- & -- & -- & -- \\
HD\,39060 & $\upbeta$\,Pic & A6V & 19 & 23 & \cmark$^{32}$ & \cmark$^{29}$ & \cmark$^{29}$ & \cmark$^{5,36,66}$ & -- & -- & \cmark$^{5,19}$ & \cmark$^{2,68}$ \\
HD\,40136 & $\upeta$\,Lep & F1III & 15 & 1300 & \xmark$^{27}$ & -- & -- & -- & -- & -- & -- & -- \\
HD\,42111 & HIP\,29151 & B9V & 170 & 150 & \xmark$^{79}$ & -- & -- & -- & -- & -- & -- & -- \\
HD\,48682 & HIP\,32480 & G0V & 16 & 2500 & \xmark$^{17}$ & -- & -- & -- & -- & -- & -- & -- \\
HD\,50554 & HIP\,33212 & F8V & 31 & 390 & -- & \xmark$^\dagger$ & \xmark$^\dagger$ & -- & -- & -- & -- & -- \\
HD\,52265 & HIP\,33719 & G0V & 28 & 2400 & -- & \xmark$^\dagger$ & \xmark$^\dagger$ & -- & -- & -- & -- & -- \\
HD\,53143 & LTT\,2715 & G9V & 18 & 1000 & -- & -- & -- & \cmark$^{90}$ & -- & -- & -- & -- \\
HD\,54341 & CD\,-43\,2931 & A0V & 93 & 360 & \xmark$^{79}$ & -- & -- & -- & -- & -- & -- & -- \\
HD\,56986 & $\updelta$\,Gem & F1IV/V & 16 & 720 & \xmark$^{27}$ & -- & -- & -- & -- & -- & -- & -- \\
HD\,58946 & $\uprho$\,Gem & F0V & 16 & 1200 & \xmark$^{27}$ & -- & -- &  -- & -- & -- & -- & -- \\
HD\,61005 & HIP\,36948 & G3/5V & 35 & 40 & \xmark$^{30,73}$ & -- & \xmark$^\dagger$ & \cmark$^{90}$ & -- & \cmark$^{76}$ & \cmark$^{77}$ & \cmark$^{77}$ \\
HD\,71043 & HIP\,41081 & A0V & 73 & 30 & \xmark$^{79}$ & -- & -- & -- & -- & -- & -- & -- \\
HD\,71155 & 30\,Mon & A0V & 38 & 310 & \xmark$^{79}$ & -- & -- & -- & -- & -- & -- & -- \\
HD\,75732 & 55\,Cnc & G8V & 14 & 30 & \xmark$^{27}$ & -- & -- & -- & -- & -- & -- & -- \\
HD\,78072 & BD\,-08\,2577 & F6V & 95 & 5000 & \xmark$^{79}$ & -- & -- & -- & -- & -- & -- & -- \\
HD\,78154 & $\upsigma^2$\,UMa & F7IV/V & 18 & 3900 & \xmark$^{27}$ & -- & -- & -- & -- & -- & -- & -- \\
HD\,81515 & HIP\,46182 & A5V & 106 & 500 & \xmark$^{17}$ & -- & -- & -- & -- & -- & -- & -- \\
HD\,82943 & HIP\,47007 & F9V & 27 & 3000 & -- & -- & -- & -- & -- & \cmark$^{42}$ & -- & -- \\
HD\,85672 & HIP\,48541 & A0V & 93 & 30 & \xmark$^{41}$ & -- & -- & -- & -- & -- & -- & -- \\
HD\,92945 & V419\,Hya & K1V & 21 & 210 & \xmark$^{73}$ & -- & -- & \cmark$^{90}$ & -- & -- & -- & \cmark$^{25}$ \\
HD\,95086 & HIP\,53524 & A8III & 90 & 17 & \xmark$^{49,73}$ & -- & --  & -- & -- & -- & \cmark$^{96}$ & -- \\
HD\,95418 & $\upbeta$\,UMa & A1V & 24 & 320 & \xmark$^{103}$ & -- & -- & -- & -- & -- & -- & -- \\
HD\,98363 & HIP\,55188 & A2V & 124 & 15 & \xmark$^{75}$ & -- & -- & -- & -- & -- & -- & -- \\
HD\,102365 & GJ\,442 & G5V & 9.2 & 4300 & \xmark$^{27}$ & -- & -- & -- & -- & -- & -- & -- \\
HD\,102647 & $\upbeta$\,Leo & A3V & 11 & 50 & \xmark$^{17}$ & -- & -- & -- & -- & -- & -- & -- \\
HD\,105211 & $\upeta$\,Cru & F2V & 20 & 1400 & -- & -- & -- & -- & -- & \cmark$^{31}$ & -- & -- \\
HD\,105452 & $\upalpha$\,Crv & F2III/IV & 14 & 460 & \xmark$^{27}$ & -- & -- & -- & -- & -- & -- & -- \\
HD\,106906 & HIP\,59960 & F5V & 92 & 17 & \xmark$^{49,71}$ & -- & -- & \cmark$^{47}$ & -- & -- & -- & \cmark$^{47}$ \\
HD\,107146 & HIP\,60074 & G2 & 29 & 100 & \xmark$^{41}$ & -- & -- & \cmark$^{90}$ & -- & -- & \cmark$^{81}$ & \cmark$^{3}$ \\
HD\,109085 & $\eta$\,Crv & F2V & 18 & 1400 & \cmark$^{57}$ & -- & -- & -- & -- & \cmark$^{48}$ & \cmark$^{20}$ & -- \\
HD\,109573 & HR\,4796A & A0 & 73 & 8 & \xmark$^{30,17,50}$ & \xmark$^{18}$ & \xmark$^{83}$ & \cmark$^{69,97,92}$ & \cmark$^{70}$ & -- & \cmark$^{43}$ & \cmark$^{16,89,97}$ \\
HD\,109832 & HIP\,61684 & A9V & 112 & 15 & \xmark$^{75}$ & -- & -- & -- & -- & -- & -- & -- \\
HD\,110058 & HIP\,61782 & A0V & 107 & 10 & \cmark$^{49}$ & -- & --  & -- & -- & -- & -- & -- \\
HD\,110411 & $\uprho$\,Vir & A0V & 36 & 90 & \xmark$^{73}$ & -- & -- & -- & -- & -- & -- & -- \\
HD\,111161 & HIP\,62842 & A3III/IV & 125 & 17 & \xmark$^{49}$ & -- & -- & -- & -- & -- & -- & -- \\
HD\,111520 & HIP\,62657 & F5/6V & 109 & 17 & \xmark$^{49}$ & -- & -- & -- & -- & -- & -- & -- \\
HD\,112810 & HIP\,63439 & F3/5IV/V & 140 & 17 & \xmark$^{49}$ & -- & -- & -- & -- & -- & -- & -- \\
HD\,113556 & HIP\,63886 & F2V & 107 & 17 & \xmark$^{49}$ & -- & -- & -- & -- & -- & -- & -- \\
HD\,113766 & HIP\,63975 & F4V & 125 & 17 & \xmark$^{49,71}$ & -- & -- & -- & -- & -- & -- & -- \\
HD\,114082 & HIP\,64184 & F3V & 85 & 17 & \xmark$^{49,71}$ & -- & \xmark$^\dagger$ & \cmark$^{99}$ & -- & -- & -- & \cmark$^{99}$ \\
HD\,115600 & HIP\,64995 & F2IV/V & 110 & 17 & \xmark$^{49,71}$ & -- & -- & \cmark$^{14}$ & -- & -- & -- & \cmark$^{14}$ \\
HD\,115617 & 61\,Vir & G6V & 8.6 & 4600 & \xmark$^{27}$ & -- & -- & -- & -- & -- & -- & -- \\
HD\,117214 & HIP\,65875 & F6V & 110 & 17 & \xmark$^{49,71}$ & -- & -- & -- & -- & -- & -- & -- \\
HD\,120326 & HIP\,67497 & F0 & 107 & 16 & -- & -- & -- & \cmark$^{7,78}$ & -- & -- & -- & \cmark$^{7}$ \\
HD\,121191 & CD\,-52\,5860 & A5IV/V & 130 & 17 & \cmark$^{75}$ & -- & -- & -- & -- & -- & -- & -- \\
HD\,121617 & CD\,-46\,8973 & A1V & 128 & 17 & \cmark$^{75}$ & -- & -- & -- & -- & -- & \cmark$^{75}$ & -- \\
HD\,121847 & 47\,Hya & B8V & 104 & 75 & \xmark$^{17}$ & -- & -- & -- & -- & -- & -- & -- \\
HD\,123356 & BD\,-20\,3960 & G1V & 41 & 4000 & \xmark$^{17}$ & -- & -- & -- & -- & -- & -- & -- \\
HD\,123247 & HIP\,69011 & B9.5V & 101 & 11 & \xmark$^{17}$ & -- & -- & -- & -- & -- & -- & -- \\
HD\,128167 & $\upsigma$\,Boo & F2V & 16 & 1700 & \xmark$^{27}$ & -- & -- & -- & -- & -- & -- & -- \\
HD\,129590 & HIP\,72070 & G1V & 135 & 16 & \xmark$^{49}$ & -- & -- & \cmark$^{63}$ & -- & -- & \cmark$^{49}$ & \cmark$^{63}$ \\
HD\,131488 & CD\,-40\,9088 & A1V & 150 & 16 & \cmark$^{75}$ & -- & -- & -- & -- & -- & \cmark$^{75}$ & -- \\
HD\,131835 & HIP\,73145 & A2IV & 125 & 16 & \cmark$^{30,49,73}$ & \xmark$^{73}$ & \xmark$^{73}$ & \cmark$^{22}$ & \cmark$^{35}$ & -- & \cmark$^{49}$ & \cmark$^{22}$ \\
HD\,134888 & HIP\,74499 & F3/5V & 90 & 16 & \xmark$^{49}$ & -- & -- & -- & -- & -- & -- & -- \\
HD\,135953 & HIP\,74959 & F5V & 135 & 16 & \xmark$^{49}$ & -- & -- & -- & -- & -- & -- & -- \\
HD\,136246 & HIP\,75077 & A1V & 145 & 16 & \xmark$^{79}$ & -- & -- & -- & -- & -- & -- & -- \\
HD\,138813 & HIP\,76310 & A0V & 151 & 11 & \cmark$^{49,59}$ & -- & \xmark$^{59}$ & -- & -- & -- & -- & -- \\
HD\,139365 & HIP\,76600 & B2.5V & 135 & 11 & \xmark$^{17}$ & -- & -- & -- & -- & -- & -- & -- \\
HD\,139664 & g\,Lup & F5 & 17 & 300 & -- & -- & -- & \cmark$^{90}$ & -- & -- & -- & -- \\
HD\,141569 & HIP\,77542 & A0Ve & 99 & 7 & \cmark$^{17,101}$ & \cmark$^{29}$ & \cmark$^{29}$ & \cmark$^{12,44,64}$ & \cmark$^{58}$ & -- & \cmark$^{101}$ & \cmark$^{44,100}$ \\
HD\,142114 & HIP\,77840 & B2.5V & 130 & 11 & \xmark$^{17}$ & -- & -- & -- & -- & -- & -- & -- \\
HD\,142165 & HIP\,77858 & B5V & 125 & 30 & \xmark$^{17}$ & -- & -- & -- & -- & -- & -- & -- \\
HD\,142315 & HIP\,77911 & B9V & 150 & 11 & \xmark$^{49,59}$ & -- & \xmark$^{59}$ & -- & -- & -- & -- & -- \\
HD\,142446 & HIP\,78043 & F3V & 145 & 16 & \xmark$^{49}$ & -- & -- & -- & -- & -- & -- & -- \\
HD\,142860 & $\upgamma$\,Ser & F6V & 12 & 3500 & \xmark$^{27}$ & -- & -- & -- & -- & -- & -- & -- \\
HD\,143018 & $\uppi$\,Sco & B1V & 140 & 11 & \xmark$^{17}$ & -- & -- & -- & -- & -- & -- & -- \\
HD\,143675 & HIP\,78641 & A5IV/V & 113 & 16 & \xmark$^{75}$ & -- & -- & -- & -- & -- & -- & -- \\
HD\,145263 & HIP\,79288 & F0V & 150 & 11 & \xmark$^{49}$ & -- & -- & -- & -- & -- & -- & -- \\
HD\,145560 & HIP\,79516 & F5V & 135 & 16 & \xmark$^{49}$ & -- & -- & -- & -- & -- & \cmark$^{49}$ & -- \\
HD\,145631 & HIP\,79439 & B9V & 130 & 11 & -- & -- & \xmark$^{18}$ & -- & -- & -- & -- & -- \\
HD\,145880 & HIP\,79631 & B9.5V & 128 & 16 & \xmark$^{75}$ & -- & -- & -- & -- & -- & -- & -- \\
HD\,146181 & HIP\,79742 & F6V & 145 & 16 & \xmark$^{49}$ & -- & -- & -- & -- & -- & -- & -- \\
HD\,146606 & HIP\,79878 & A0V & 130 & 15 & -- & -- & \xmark$^{59}$ & -- & -- & -- & -- & -- \\
HD\,146897 & HIP\,79977 & F2/3V & 125 & 11 & \cmark$^{49}$ & -- & -- & \cmark$^{21,98}$ & -- & -- & -- & -- \\
HD\,147137 & HIP\,80088 & A9V & 140 & 11 & \xmark$^{49}$ & -- & \xmark$^{59}$ & -- & -- & -- & -- & -- \\
HD\,156623 & HIP\,84881 & A0V & 118 & 16 & \cmark$^{49}$ & -- & -- & -- & -- & -- & -- & -- \\
HD\,157587 & HIP\,85224 & F5V & 107 & 3000 & -- & -- & -- & \cmark$^{67,78}$ & -- & -- & -- & -- \\
HD\,158352 & HR\,6507 & A7V & 60 & 600 & \xmark$^{30}$ & \xmark$^{29}$ & \xmark$^{29}$ & -- & -- & -- & -- & -- \\
HD\,159082 & HIP\,85826 & B9.5V & 135 & 200 & \xmark$^{79}$ & -- & -- & -- & -- & -- & -- & -- \\
HD\,160305 & HIP\,86598 & F8/G0V & 73 & 23 & \xmark$^{73}$ & -- & -- & -- & -- & -- & -- & -- \\
HD\,161868 & $\upgamma$\,Oph & A1V & 32 & 200 & \xmark$^{73}$ & \xmark$^{29}$ & -- & -- & -- & -- & -- & -- \\
HD\,164249 & HIP\,88399 & F5V & 48 & 12 & \xmark$^{71,79}$ & \xmark$^{84}$ & \xmark$^{84}$ & -- & -- & -- & -- & -- \\
HD\,164577 & HR\,6723 & A2V & 39 & 250 & \xmark$^{30}$ & -- & -- & -- & -- & -- & -- & -- \\
HD\,166191 & HIP\,89046 & F3/5V & 102 & 4 & \xmark$^{79}$ & -- & -- & -- & -- & -- & -- & -- \\
HD\,170773 & HR\,6948 & F5V & 37 & 200 & -- & -- & -- & -- & -- & \cmark$^{73}$ & -- & -- \\
HD\,172167 & Vega & A0V & 7.7 & 700 & \xmark$^{103}$ & \xmark$^{29}$ & \xmark$^{29}$ & -- & -- & \cmark$^{93,95}$ & -- & -- \\
HD\,172555 & HIP\,92024 & A7V & 29 & 23 & \xmark$^{30,71,79}$ & \xmark$^{84}$ & \cmark$^{82}$ & -- & -- & -- & -- & -- \\
HD\,178253 & $\upalpha$\,CrA & A2V & 39 & 250 & \xmark$^{30}$ & -- & -- & -- & -- & -- & -- & -- \\
HD\,181296 & $\upeta$\,Tel & A0V & 48 & 23 & \xmark$^{30,73,79}$ & \cmark$^{84}$ & \xmark$^{84}$& -- & -- & -- & -- & -- \\
HD\,181327 & HIP\,95270 & F5/6V & 52 & 23 & \cmark$^{56}$ & \xmark$^{84}$ & \xmark$^{84}$ & \cmark$^{94,80}$ & -- & -- & \cmark$^{56}$ & \cmark$^{88}$ \\
HD\,182681 & HIP\,95619 & B8/9V & 70 & 145 & \xmark$^{73,79}$ & -- & -- & -- & -- & -- & -- & -- \\
HD\,182919 & 5\,Vul & A0V & 73 & 200 & \xmark$^{73}$ & -- & -- & -- & -- & -- & -- & -- \\
HD\,183324 & c\,Aql & A0V & 61 & 140 & \xmark$^{73}$ & -- & -- & -- & -- & -- & -- & -- \\
HD\,184800 & CD\,-51\,12185 & A8/9V & 280 & 500 & \xmark$^{30}$ & -- & -- & -- & -- & -- & -- & -- \\
HD\,191089 & HIP\,99273 & F5V & 52 & 23 & \xmark$^{71}$ & -- & \xmark$^\dagger$ & -- & \cmark$^{11}$ & -- & -- & -- \\
HD\,192758 & CD\,-43\,13915 & F0V & 62 & 40 & \xmark$^{71}$ & -- & \xmark$^\dagger$ & -- & -- & -- & -- & -- \\
HD\,197481 & AU\,Mic & M1V & 9.9 & 23 & \xmark$^{51}$ & -- & \xmark$^\dagger$ & \cmark$^{23,38}$ & -- & \cmark$^{62}$ & \cmark$^{53}$ & \cmark$^{23,26,46}$ \\
HD\,202206 & HIP\,104903 & G6V & 46 & 570 & -- & -- & \xmark$^\dagger$ & -- & -- & -- & -- & -- \\
HD\,202628 & LTT\,8444 & G5V & 24 & 2300 & -- & -- & -- & \cmark$^{91}$ & -- & -- & -- & \cmark$^{91}$ \\
HD\,202917 & HIP\,105388 & G5V & 43 & 30 & \xmark$^{73}$ & -- & -- & \cmark$^{91}$ & -- & --& -- & \cmark$^{91}$ \\
HD\,207129 & LTT\,8704 & G2V & 16 & 2350 & -- & -- & -- & \cmark$^{91}$ & -- & \cmark$^{52}$   & -- & \cmark$^{91}$ \\
HD\,212676 & BD\,+53\,2870 & B9V & 670 & 75 & \xmark$^{17}$ & -- & -- & -- & -- & -- & -- & -- \\
HD\,216956 & Fomalhaut & A4V & 7.7 & 440 & \cmark$^{60}$ & \xmark$^{15}$ & \xmark$^{15}$ & \cmark$^{40}$ & -- & \cmark$^{1}$ & \cmark$^{55}$ & \cmark$^{39}$ \\
HD\,218396 & HR\,8799 & A5V & 40 & 30 & \xmark$^{79}$ & -- & \xmark$^\dagger$ & -- & -- & \cmark$^{61}$ & \cmark$^{102}$ & -- \\
HD\,220825 & HIP\,115738 & A2V & 47 & 200 & \xmark$^{79}$ & -- & -- & -- & -- & -- & -- & -- \\
HD\,221853 & HIP\,116431 & F0 & 68 & 100 & \xmark$^{41,71}$ & -- & -- & -- & -- & -- & -- & -- \\
HD\,225200 & HIP\,345 & A0V & 130 & 320 & \xmark$^{79}$ & -- & -- & -- & -- & -- & -- & -- \\
TWA\,2 & CD\,-29\,8887 & M2 & 42 & 10 & -- & -- & \xmark$^{83}$ & -- & -- & -- & -- & -- \\
TWA\,7 & CE\,Ant & M3 & 35 & 10 & -- & -- & \xmark$^{83}$ & \cmark$^{9}$ & \xmark$^{83}$ & -- & -- & \cmark$^{9}$ \\
TWA\,10 & V1252\,Cen & M2 & 62 & 10 & -- & -- & \xmark$^{83}$ & -- & -- & -- & -- & -- \\
TWA\,25 & V1249\,Cen & M0.5 & 54 & 10 & -- & -- & -- & \cmark$^{9}$ & -- & -- & -- & \cmark$^{9}$ \\
\end{longtable}
\noindent Note: Throughout the table, a long-dash symbol indicates that no relevant data is available.\\
$^\mathrm{a}$ Indicated here is whether a given emission line has been detected (\cmark) or whether an upper limit has been set (\xmark). CO: either of the 1-0 (2601\,\micron), 2-1 (1300\,\micron) and 3-2 (867\,\micron) rotational transitions is considered whenever available; [\ion{C}{2}]: 158\,\micron\ transition; [\ion{O}{1}]: 63\,\micron\ transition. Entries with a $\dagger$ symbol are objects with no published [\ion{C}{2}] or [\ion{O}{1}] line information but for which we have visually inspected archival {\it Herschel} spectra.\\
$^\mathrm{b}$ Objects with a \cmark\ symbol are those for which observations exist in which the disk inner radius is well defined. Some of these debris disks have been partially resolved in other studies that are not listed here due to their limited resolution. OIR: optical and near-infrared scattered light imaging (up to 4\,\micron); MIR: mid-infrared thermal imaging (10--25\,\micron); FIR: far-infrared thermal imaging (70--160\,\micron); (sub)mm: (sub-)millimeter thermal imaging (800--1400\,\micron).\\
$^\mathrm{c}$ Objects with a \cmark\ symbol are those for which a scattered light image has been modeled using the Henyey-Greenstein phase function and a best fitting asymmetry parameter has been published.\\
\end{landscape}

\begin{enumerate}
\setlength\itemsep{-0.25em}
\item  {Acke} B, {Min} M, {Dominik} C, {et~al.} 2012. {\it Astron. Astrophys.} 540:A125
\item  {Ahmic} M, {Croll} B, \& {Artymowicz} P. 2009. {\it Ap. J.} 705:529
\item  Ardila DR, Golimowski DA, Krist JE, et al. 2004. {\it Ap. J. Lett.} 617:L147
\item  Asensio-Torres R, Janson M., Hashimoto J, et al. 2016. {\it Astron. Astrophys.} 593:A73
\item  {Ballering} NP, {Su} KYL, {Rieke} GH, {G{\'a}sp{\'a}r} A. 2016. {\it Ap. J.} 823:108
\item  Boccaletti A, Augereau JC, Lagrange AM, et al. 2012. {\it Astron. Astrophys.} 544:A85
\item  {Bonnefoy} M, {Milli} J, {M{\'e}nard} F, {et~al.} 2017. {\it Astron. Astrophys.} 597:L7
\item  {Booth} M, {Dent} WRF, {Jord{\'a}n} A, {et~al.} 2017. {\it MNRAS} 469:3200
\item  Choquet E, Perrin MD, Chen CH, et al. 2016. {\it Ap. J. Lett.} 817:L2
\item  Choquet E, Milli J, Wahhaj Z, et al. 2017. {\it Ap. J. Lett} 834:L12
\item  Churcher L, Wyatt M, Smith R. 2011. {\it MNRAS} 410:2
\item  Clampin M, Krist JE, Ardila RD, et al. 2003. {\it Astron. J.} 126:385
\item  Currie T, Rodigas TJ, Debes J, et al. 2012. {\it Ap. J.} 757:28
\item  Currie T, Lisse CM, Kuchner M, et al. 2015. {\it Ap. J. Lett.} 807:L7
\item  {Currie} T, {Guyon} O, {Tamura} M, {et~al.} 2017. {\it Ap. J. Lett.} 836:L15
\item  {Debes} JH, {Weinberger} AJ, {Schneider} G. 2008. {\it Ap. J. Lett.} 673:L191
\item  Dent WRF, Greaves JS, Coulson IM. 2005. {\it MNRAS} 359:663
\item  Dent WRF, Thi WF, Kamp I, et al. 2013. {\it Publ. Astron. Soc. Pac.} 125:477
\item  {Dent} WRF, {Wyatt} MC, {Roberge} A, {et~al.} 2014. {\it Science} 343:1490
\item  Duch\^ene G, Arriaga P, Wyatt M, et al. 2014. {\it Ap. J.} 784:148
\item  Engler N, Schmid HM, Thalman C, et al. 2017. {\it Astron. Astrophys.} 607:A90
\item  {Feldt} M, {Olofsson} J, {Boccaletti} A, {et~al.} 2017. {\it Astron. Astrophys.} 601:A7
\item  Fitzgerald MP, Kalas PG, Duch\^ene G, et al. 2007a. {\it Ap. J.} 670:536
\item  Fitzgerald MP, Kalas PG, Graham JR. 2007b. {\it Ap. J.} 670:557
\item  Golimowski DA, Krist JE, Stapelfeldt KR, et al. 2011. {\it Astron. J.} 142:30
\item  {Graham} JR, {Kalas} PG, {Matthews} BC. 2007. {\it Ap. J.} 654:595
\item  Greaves JS, Coulson IM, Holland WS. 2000. {\it MNRAS Lett.} 312:L1
\item  {Greaves} JS, {Sibthorpe} B, {Acke} B, {et~al.} 2014. {\it Ap. J. Lett.} 791:L11
\item  Greaves JS, Holland WS, Matthews BC, et al. 2016. {\it MNRAS} 461:3910
\item  Hales AS, De Gregorio-Monsalvo I, Montesinos B, et al. {\it Astron. J.} 148:47
\item  Hengst S, Marshall JP, Horner J, Marsden SC. 2016. {\it MNRAS} 468:4725
\item  {Higuchi} AE, {Sato} A, {Tsukagoshi} T, {et~al.} 2017. {\it Ap. J. Lett.} 839:L14
\item  {Holland} WS, {Matthews} BC, {Kennedy} GM, {et~al.} 2017. {]it MNRAS} 470:3606
\item  {Hughes} AM, {Lieman-Sifry} J, {Flaherty} KM, {et~al.} 2017. {\it Ap. J.} 839:86
\item  Hung LW, Fitzgerald MP, Chen CH, et al. 2015. {\it Ap. J.} 802:138
\item  Kalas P, Jewitt D. 1995. {\it Astron. J.} 110:794
\item  Kalas P. 2005. {\it Ap. J. Lett.} 635:L169
\item  Kalas P, Liu MC, Matthews BC. 2004. {\it Science} 303:1990
\item  {Kalas} P, {Graham} JR, {Clampin} M. 2005. {\it Nature} 435:1067
\item  {Kalas} P, {Graham} JR, {Fitzgerald} MP, {Clampin} M. 2013. {\it Ap. J.} 775:56
\item  Kastner JH, Hily-Blant P, Sacco GG, Forveille T, Zuckerman B. 2010. {\it Ap. J. Lett.} 723:L248
\item  {Kennedy} GM, {Wyatt} MC, {Bryden} G, {Wittenmyer} R, {Sibthorpe} B. 2013. {\it MNRAS} 436:898
\item  {Kennedy} GM, {Marino} S, {Matr\`{a}} L, et al. 2018. {\it MNRAS} in press (arxiv:1801.05429)
\item  {Konishi} M, {Grady} CA, {Schneider} G, {et~al.} 2016. {\it Ap. J. Lett.} 818:L23
\item  {K{\'o}sp{\'a}l} {\'A}, {Mo{\'o}r} A, {Juh{\'a}sz} A, {et~al.} 2013. {\it Ap. J.} 776:77
\item  Krist JE, Ardila DA, Golimowski DA, et al. 2005. {\it Astron. J.} 129:1008
\item  Lagrange AM, Langlois M, Gratton R, et al. 2016. {\it Astron. Astrophys. Lett.} 586:L8
\item  {Lebreton} J, {Beichman} C, {Bryden} G, {et~al.} 2016. {\it Ap. J.} 817:165
\item  {Lieman-Sifry} J, {Hughes} AM, {Carpenter} JM, {et~al.} 2016. {\it Ap. J.} 828:25
\item  Liseau R. 1999. {\it Astron. Astrophys.} 348:133
\item  Liu MC, Matthews BC, Williams JP, Kalas PG. 2004. {\it Ap. J.} 608:526
\item  Lohne T, Augereau JC, Ertel, S, et al. 2012. {\it Astron. Astrophys.} 537:A110
\item  {MacGregor} MA, {Wilner} DJ, {Rosenfeld} KA, {et~al.} 2013. {\it Ap. J. Lett.} 762:L21
\item  MacGregor MA, Lawler AM, Wilner DJ, et al. 2016. {\it Ap. J.} 828:113
\item  {MacGregor} MA, {Matra} L, {Kalas} P, {et~al.} 2017. {\it MNRAS} 842:8
\item  Marino S, Matr\`a L, Stark C, et al. 2016. {\it MNRAS} 460:2933
\item  {Marino} S, {Wyatt} MC, {Pani\'c} O, {et~al.} 2017. {\it MNRAS} 465:2595
\item  Marsh KA, Silverstone MD, Becklin EE, et al. 2002. {\it Ap. J.} 573:425
\item  Mathews GS, Pinte C, Duch\^ene G, Williams JP, M\'enard F. 2013. {\it Astron. Astrophys.} 558:A66
\item  {Matr{\`a}} L, {MacGregor} MA, {Kalas} P, {et~al.} 2017. {\it MNRAS} 842:9
\item  {Matthews} B, {Kennedy} G, {Sibthorpe} B, {et~al.} 2014. {\it Ap. J.} 780:97
\item  {Matthews} BC, {Kennedy} G, {Sibthorpe} B, {et~al.} 2015. {\it Ap. J.} 811:100
\item  Matthews E, Hinkley S, Vigan A, et al. 2017. {\it Ap. J. Lett.} 843:L12
\item  Mawet D, Choquet E, Absil O, et al. 2017. {\it Astron. J.} 153:44
\item  Mazoyer J, Boccaletti A, Augereau JC, et al. 2014. {\it Astron. Astrophys.} 569:A29
\item  Millar-Blanchaer MA, Graham JR, Pueyo L, et al. 2015. {\it Ap. J.} 811:18
\item  Millar-Blanchaer MA, Wang JJ, Kalas P, et al. 2016. {\it Astron. J.} 152:128
\item  Milli J, Lagrange AM, Mawet D, et al. 2014. {\it Astron. Astrophys.} 566:A91
\item  {Milli} J, {Vigan} A, {Mouillet} D, {et~al.} 2017. {\it Astron. Astrophys.} 599:A108
\item  Moerchen MM, Churcher LJ, Telesco CM, et al. 2011. {\it Astron. Astrophys.} 526:A34
\item  Mo{\'o}r A, {\'A}brah{\'a}m P, Juh{\'a}sz A, et al. 2011. {\it Ap. J. Lett.} 740:L7
\item  Mo{\'o}r A, Juh{\'a}sz A, {K{\'o}sp{\'a}l}, et al. 2013. {\it Ap. J. Lett.} 777:L25
\item  Mo{\'o}r A, Henning T, Juh{\'a}sz A, et al. 2015. {\it Ap. J.} 814:42
\item  Mo{\'o}r A, K{\'o}sp{\'a}l {\'A}, {\'A}brah{\'a}m P, et al. 2015. {\it MNRAS} 447:577
\item  Mo{\'o}r A, Cur{\'e} M, K{\'o}sp{\'a}l {\'A}, et al. 2017. {\it Ap. J.} 849:123
\item  {Morales} FY, {Bryden} G, {Werner} MW, {Stapelfeldt} KR. 2016. {\it Ap. J.} 831:97
\item  {Olofsson} J, {Samland} M, {Avenhaus} H, {et~al.} 2016. {\it Astron. Astrophys.} 591:A108
\item  Padgett D, Stapelfeldt K. 2016. {\it Intern. Astron. Union Proc.} 314:175
\item  P\'ericaut J, Di Folce E, Dutrey A, Guilloteau S, Pi\'etu V. 2017., {\it Astron. Astrophys.} 600:A62
\item  Ren B, Pueyo L, Zhu GB, Debes J, Duch\^ene G. 2017, {\it Ap. J.} in press (arxiv:1712:10317)
\item  {Ricci} L, {Carpenter} JM, {Fu} B, {et~al.} 2015. {\it Ap. J.} 798:124
\item  {Riviere-Marichalar} P, {Barrado} D, {Augereau} JC, {et~al.} 2012. {\it Astron. Astrophys.} 546:L8
\item  Riviere-Marichalar P, Pinte C, Barrado D, et al. 2013. {\it Astron. Astrophys.} 555:A67
\item  {Riviere-Marichalar} P, {Barrado} D, {Montesinos} B, {et~al.} 2014. {\it Astron. Astrophys.} 565:A68
\item  {Roberge} A, {Kamp} I, {Montesinos} B, {et~al.} 2013. {\it Ap. J.} 771:69
\item  Rodigas TJ, Hinz PM, Leisenring J, et al. 2012. {\it Ap. J.} 752:57
\item  {Sai} S, {Itoh} Y, {Fukagawa} M, {Shibai} H, {Sumi} T. 2015. {\it Publ. Astron. Soc. Jap.} 67:20
\item  {Schneider} G, {Silverstone} MD, {Hines} DC, {et~al.} 2006. {\it Ap. J.} 650:414
\item  Schneider G, Weinberger AJ, Becklin EE, Debes JH, Smith BA. 2009. {\it Astron. J.} 137:53
\item  {Schneider} G, {Grady} CA, {Hines} DC, {et~al.} 2014. {\it Astron. J.} 148:59
\item  {Schneider} G, {Grady} CA, {Stark} CC, {et~al.} 2016. {\it Astron. J.} 152:64
\item  {Schneider} G, {Debes} JH, {Grady} CA, {et~al.} 2018. {\it Astron. J.} in press (arxiv:1712.08599)
\item  Sibthorpe B, Vandenbussche B, Greaves JS, et al. 2010. {\it Astron. Astrophys. Lett.} 518:L130
\item  {Stark} CC, {Schneider} G, {Weinberger} AJ, {et~al.} 2014. {\it Ap. J.} 789:58
\item  {Su} KYL, {Rieke} GH, {Misselt} KA, {et~al.} 2005. {\it Ap. J.} 628:487
\item  {Su} KYL, {MacGregor} MA, {Booth} M, {et~al.} 2017. {\it Astron. J.} 154:225
\item  {Thalmann} C, {Janson} M, {Buenzli} E, {et~al.} 2011. {\it Ap. J. Lett.} 743:L6
\item  Thalmann C, Janson M, Buenzli E, et al. 2013. {\it Ap. J. Lett.} 763:L29
\item  Wahhaj Z, Milli J, Kennedy G, et al. 2016. {\it Astron. Astrophys. Lett.} 596:L4
\item  Weinberger AJ, Becklin EE, Schneider G, et al. 1999. {\it Ap. J. Lett.} 525:L53
\item  {White} JA, {Boley} AC, {Hughes} AM, {et~al.} 2016. {\it Ap. J.} 829:6
\item  Wilner D, MacGregor MA, Andrews SM, et al. 2018. {\it Ap. J.} submitted
\item  Yamashita T, Handa T, Omodaka T, et al. 1993. {\it Ap. J. Lett.} 402:L65
\end{enumerate}

\pagebreak

{\bf \large B. Henyey-Greenstein phase function fits to scattered light images of debris disks}\\

Scattered light images of debris disks provide insight about dust properties via the phase function, i.e., the distribution of scattered light as a function of scattering angle. The most robust method for interpretion of this phase function is to sample it across all possible scattering angles and to then compare it to other dust populations or to models of dust populations. The state-of-the-art of this approach is discussed in this review. Unfortunately, very few debris disks are amenable to such an analysis. Unfavorable viewing geometries (specifically inclination), the small angular extent of the disk, and substantial artefacts associated with the subtraction of residual starlight conspire to limit our ability to infer with precision the underlying phase function. 

A commonly employed method to circumvent these limitations and obtain a constraint on the phase function consists of modeling the scattered light image of a debris disk under the assumption that the scattering follows the Henyey-Greenstein phase function. The result of this model fit is a single parameter, $g$, that measures the degree of forward scattering associated with the dust residing in the debris disk. While this approach is not anchored in a physical description of scattering by dust grains, the simplicity of a one-parameter analytical fit is appealing, and it has been used for 25 disks (some with multiple estimates based on different datasets). The database in Section\,A of these Supplementary Materials provides a complete list of relevant references.

\begin{figure}
\centering
\includegraphics[scale=0.7]{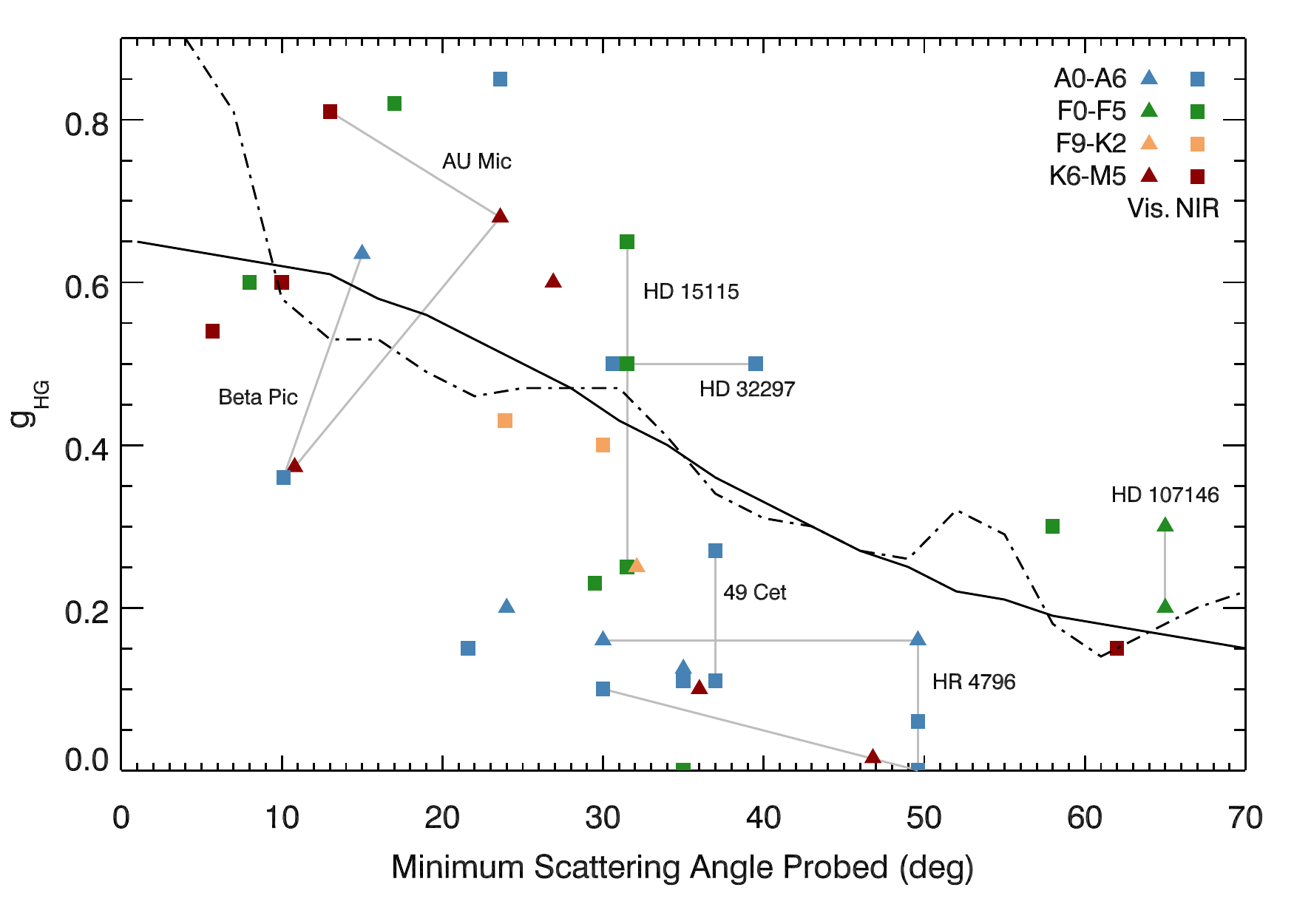}
\caption{Estimated Henyey-Greenstein $g$ parameter based on model fitting of scattered light images of debris disks plotted as a function of the smallest scattering angle probed in the data. Triangle and squares distinguish results based on visible and near-infrared images, while the spectral type of each target is color-coded. Objects with more than one estimated $g$ values are connected by gray segments and labelled. The dot-dashed and solid curves show the result of fitting an Henyey-Greenstein phase function to the observed phase function of Saturn's rings and to a model consisting of a two-parameter Henyey-Greenstein phase function characterized by $g_1 = 0.7$ and $g_2 = -0.5$ that is a qualitatively good fit to observations of cometary and Saturn's ring dust.\label{fig:g_hg}}
\end{figure}

\textbf{Figure\,\ref{fig:g_hg}} presents all estimated $g$ values from the literature. Despite a significant scatter between objects and even within different observations of the same object, \textbf{Figure\,\ref{fig:g_hg}} reveals a clear correlation between the fitted $g$ value and the range of scattering angles probed by the data: the closer the data probe to the forward scattering peak, the higher the $g$ value. This is qualitatively consistent with a phase function that is rather flat (low $g$ value) at scattering angles in the $\approx$ 50--130$^\circ$\ range, but with a sharp forward scattering peak that is sampled when data are available at smaller scattering angle. Indeed, this is the shape of the scattering phase function observed both in debris disks and in Solar System dust population, in cases where it can be measured directly, as discussed in the review (see Section\,C of these Supplementary Materials). To illustrate this, we fit a single-parameter Henyey-Greenstein phase function to the observed phase function in Saturn's G ring (Hedman \& Stark 2015), as well as to a composite model that is a weighted average of two Henyey-Greenstein phase functions (assuming $g_1 = 0.7$ and $g_2 = -0.5$ and using a 3:1 weight ratio). Such a composite model is often proposed to reproduce observed Solar System phase functions. By varying the range of scattering angle included in the fit, we recover the qualitative behavior of a lower $g$ value when a smaller range of scattering angles is available. 

On the other hand, \textbf{Figure\,\ref{fig:g_hg}} shows that the fitted $g$ parameters are not significantly correlated with the spectral type of the star. Such a correlation could have been expected, since hotter, higher-luminosity stars are expected to have a larger blow-out size. Since large grains are expected to be more forward-scattering, this could have led to higher $g$ values. Yet, no correlation is apparent, suggesting that either the grain size distribution is not uniquely set by the stellar luminosity, or that the Henyey-Greenstein formalism is inadequate to infer the (minimum) grain size.

While the Henyey-Greenstein parametrization can be useful as an easy-to-implement formalism, we caution against over-interpreting the results of an image fit based on it. Nonetheless, the trends observed in \textbf{Figure\,\ref{fig:g_hg}} suggest that the scattering phase function in most, if not all, debris disks share a common shape, which expands the conclusion reached on the basis of the empirically-determined phase functions discussed in this review.

\vspace*{0.5cm}

{\bf \large C. Empirical scattering phase functions in the Solar System} \\

The scattering phase function of several dust populations in the Solar System has been analyzed from ground-based and {\it in situ} measurements. Contrary to debris disks surrounding other stars, these populations offer the advantage that they can be studied from a varying range of viewing angles. For one, the orbital motion around the Sun of the Earth and other objects in the Solar Systems allows measurement of broad ranges of scattering angles with ground-based observations. In addition, the phase function can be efficiently probed through imaging data from a spacecraft orbiting a Solar System body, which again provides the desired variable viewing geometry.

\begin{figure}
\centering
\includegraphics[scale=0.7]{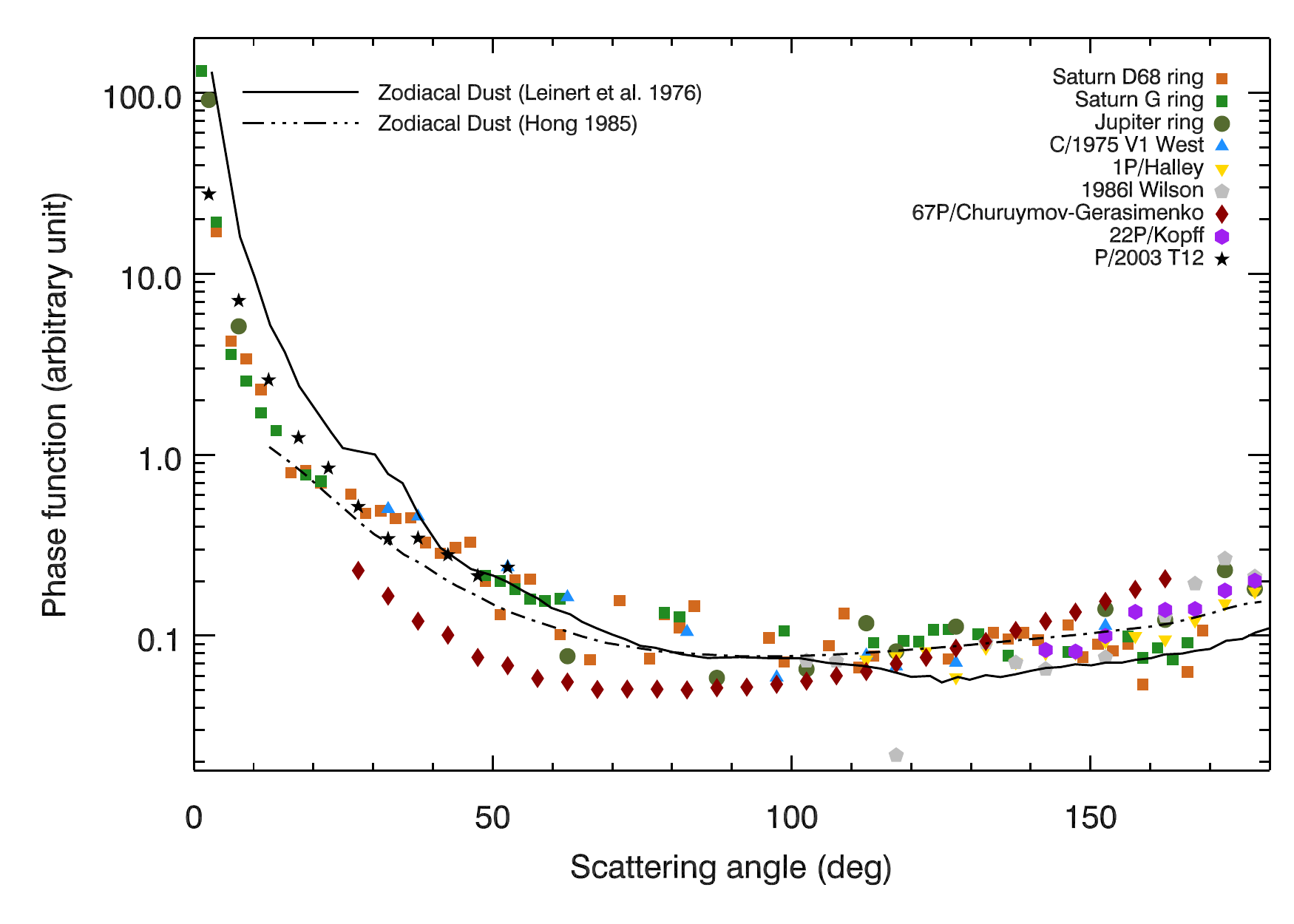}
\caption{Observed scattering phase function for different dust populations in the Solar System. Solid and dashed curves are two estimates of the phase function for the zodiacal dust population while color symbols represent planetary rings and cometary dust populations. Data used in this figure are from Bertini et al. (2017, comet 67P), Hanner \& Newburn (1989, comet 1986l), Hedman \& Stark (2015, Saturn's rings), Hong (1985), Hui (2013, comet P/2003 T12), Leinert et al. (1976), Moreno et al. (2012, comet 22P), Schleicher et al. (1998, comet 1P), Throop et al. (2004, Jupiter's ring), Zubko et al. (2014, comet C/1975 V1).
\label{fig:pf_solsyst}}
\end{figure}

The zodiacal dust population is the closest to the Earth. It is populated via mass loss from or disintegration of asteroids and comets in the inner Solar System. Scattering off that dust is responsible for the zodiacal light.  However, this dust population is spatially very extended and contains a number of discrete substructures. Since any light of sight through the zodiacal light effectively integrates scattering along an extended column of dust, it is necessary to determine the spatial structure of the zodiacal dust in order to estimate its scattering phase function. Unfortunately, our location inside of this disk makes this endeavor difficult. So far, this separation has been performed under the assumption of a simple power law surface density profile by Leinert et al. (1976) and Hong (1985), leading to tyhe phase functions shown in \textbf{Figure\,\ref{fig:pf_solsyst}}. These efforts effectively assume that the effects of small-scale substructures cancel out in the integral along the line of sight. As an indication of uncertainty, Leinert et al. (1976) estimated 4 separate phase functions, each under different assumptions for the surface density profile of the zodiacal disk, suggesting that we only understand the phase function to within a factor of 2 or so. Specifically, the phase function could have a more marked back-scattering peak associated with a somewhat weaker forward-scattering one. On the other hand, rather than a direct empirical estimate of the zodiacal phase function, Hong (1985) provides a 3-parameter composite Henyey-Greenstein phase function. Nonetheless, the general shape of both estimates of the phase function are in good agreement, suggesting that the results are robust.

Another well-studied dust population in the Solar System is the dust tail of comets. In this case, the relative orbital motions of the Earth and comet allow to probe a large range of scattering angles. This approach, however, relies on the assumption that the dust population remains in steady-state for timescales of weeks to months so that observed differences in surface brightness can be ascribed to the scattering phase function. For comets, however, this assumption is often invalid, as the mass loss is generally not constant and grains of different sizes have different drift timescales within the tail, so that any location in the tail is characterized by a time-dependent dust population. Nonetheless, these factors can be modeled, leading to estimates of the scattering phase function in cometary tails. Very few comets, however, provide a full evaluation of the phase function. Comets whose orbit remain outside of the Earth's orbit only probe scattering angles larger than 90$^\circ$, while comets that come much closer to the Sun allow us to probe small scattering angles. Interestingly, almost all phase functions for cometary dust agree with one another over common ranges of scattering angles, suggesting that they all share the same phase function, as illustrated in \textbf{Figure\,\ref{fig:pf_solsyst}}. The latter is also in good agreement with that observed for the zodiacal light. While this may be unsurprising if comets are the main contributor to zodiacal dust, size sorting through Poynting-Robertson drag and radiation pressure could have introduced significant differences between these phase functions.

There is one remarkable exception among comets: comet 67P/Churuymov-Gerasimenko, which has been studied by the {\it Rosetta} spacecraft. With an orbital period of just a few hours, time variability of the dust population in this comet is not a concern. It remains unclear whether the distinct phase function observed from this comet's tail is due to the comet being of different origin (it is a Jupiter-family comet), or to observations taken at a special time. The data presented by Bertini et al. (2017) were taken around periastron, a period characterized by a markedly different grain size distribution in this comet's tail (Fulle et al. 2016). Interestingly, the phase function for comet 67P is distinct from other comets in a qualitatively similar manner as the HR\,4796 disk is from other debris disks, with the minimum in the phase function occurring at a scattering angle of $\approx$70$^\circ$\ instead of $\approx$90$^\circ$, as observed for other comets. Whether this is purely coincidental remains to be determined.

The last Solar System dust population for which the scattering phase function has been estimated is planetary rings. The Saturn ring system has been mapped in depth by the {\it Cassini} mission, while a combination of ground-based, {\it Galileo} and {\it Voyager} observations characterized Jupiter's ring. While analysis of the thickest and densest of Saturn's rings (A, B, C) is complicated by optical depth issues, Hedman \& Stark (2015) studied the G and D ring, respectively at the outer and inner edges of the main rings. Both rings share a nearly identical phase function (see \textbf{Figure\,\ref{fig:pf_solsyst}}), which is also a good match to that of the zodiacal and cometary dust. Jupiter's ring also appears to have the same characteristic phase function, although it is not as well sampled as Saturn's ring (Throop et al. 2004). This is rather surprising since the composition and size distribution of grains in planetary rings are believed to be markedly different from both asteroidal and cometary dust, although we note that the more tenuous rings around Saturn are characterized by much smaller dust grains than others parts of the ring system (Hedman \& Stark 2015).
\vspace*{0.5cm}

{\bf \large References} \\

\noindent Bertini I, La Forgia F, Tubiana C, et al. 2017. {\it MNRAS} 469:404 \\
 Fulle M, Marzari F, Della Corte V, et al. 2016. {\it Ap. J.} 821:19 \\
 Hanner MS, Newburn RL. 1989. {\it Astron. J.} 97:245 \\
 Hedman MM, Stark CC. 2015. {\it Ap. J.} 811:67 \\
 Hong SS. 1985. {\it Astron. Astrophys.} 146:67 \\
 Hui MT. 2013. {\it MNRAS} 436:1564 \\
 Leinert C, Link H, Pitz E, Giese R.H. 1976. {\it Astron. Astrophys.} 47:221 \\
 Moreno F, Pozuelos F, Aceituno F, et al. 2012. {\it Ap. J.} 752:136 \\
 Schleicher DG, Miller RL, Birch PV. 1998. {\it Icarus} 132:397 \\
 Throop HB, Porco CC, West RA, et al. 2004. {\it Icarus} 172:59 \\
 Zubko E, Muinonen K, Videen G, Kiselev NN. 2014. {\it MNRAS} 440:2928

\end{document}